# REPORT OF THE

# DARK ENERGY TASK FORCE

Andreas Albrecht, University of California, Davis
Gary Bernstein, University of Pennsylvania
Robert Cahn, Lawrence Berkeley National Laboratory
Wendy L. Freedman, Carnegie Observatories
Jacqueline Hewitt, Massachusetts Institute of Technology
Wayne Hu, University of Chicago
John Huth, Harvard University
Marc Kamionkowski, California Institute of Technology
Edward W. Kolb, Fermi National Accelerator Laboratory and The University of Chicago
Lloyd Knox, University of California, Davis
John C. Mather, Goddard Space Flight Center
Suzanne Staggs, Princeton University
Nicholas B. Suntzeff, Texas A&M University

Dark energy appears to be the dominant component of the physical Universe, yet there is no persuasive theoretical explanation for its existence or magnitude. The acceleration of the Universe is, along with dark matter, the observed phenomenon that most directly demonstrates that our theories of fundamental particles and gravity are either incorrect or incomplete. Most experts believe that nothing short of a revolution in our understanding of fundamental physics will be required to achieve a full understanding of the cosmic acceleration. For these reasons, the nature of dark energy ranks among the very most compelling of all outstanding problems in physical science. These circumstances demand an ambitious observational program to determine the dark energy properties as well as possible.

The Dark Energy Task Force (DETF) was established by the Astronomy and Astrophysics Advisory Committee (AAAC) and the High Energy Physics Advisory Panel (HEPAP) as a joint sub-committee to advise the Department of Energy, the National Aeronautics and Space Administration, and the National Science Foundation on future dark energy research.

# I. Executive Summary

Over the last several years scientists have accumulated conclusive evidence that the Universe is expanding ever more rapidly. Within the framework of the standard cosmological model, this implies that 70% of the universe is composed of a new, mysterious dark energy, which unlike any known form of matter or energy, counters the attractive force of gravity. Dark energy ranks as one of the most important discoveries in cosmology, with profound implications for astronomy, high-energy theory, general relativity, and string theory.

One possible explanation for dark energy may be Einstein's famous cosmological constant. Alternatively, dark energy may be an exotic form of matter called quintessence, or the acceleration of the Universe may even signify the breakdown of Einstein's Theory of General Relativity. With any of these options, there are significant implications for fundamental physics. The problem of understanding the dark energy is called out prominently in major policy documents such as the *Quantum Universe Report* and *Connecting Quarks with the Cosmos*, and it is no surprise that it is featured as number one in *Science* magazine's list of the top ten science problems of our time.

To date, there are no compelling theoretical explanations for the dark energy. In the absence of useful theoretical guidance, observational exploration must be the focus of our efforts to understand what the Universe is made of.

Although there is currently conclusive observational evidence for the existence of dark energy, we know very little about its basic properties. It is not at present possible, even with the latest results from ground and space observations, to determine whether a cosmological constant, a dynamical fluid, or a modification of general relativity is the correct explanation. We cannot yet even say whether dark energy evolves with time.

Fortunately, the extraordinary scientific challenge of the dark energy has generated outstanding ideas for an observational program that can greatly impact our understanding. A properly executed dark energy program should have as its goals to
1. Determine as well as possible whether the accelerating expansion is consistent with a cosmological constant.
2. Measure as well as possible any time evolution of the dark energy.
3. Search for a possible failure of general relativity through comparison of the effect of dark energy on cosmic expansion with the effect of dark energy on the growth of cosmological structures like galaxies or galaxy clusters.

To recommend a program to reach these goals, the Dark Energy Task Force first requested input from the community. The community responded with fifty impressive white papers outlining current and future research programs on dark energy. Second, using these submissions and our own expertise, we performed extensive calculations so different approaches could be compared side-by-side in a standardized and quantitative manner. We then developed a quantitative "figure of merit" that is sensitive to the properties of dark energy, including its evolution with time.  Our extensive findings are based on these calculations.

Using our figure of merit, we evaluated ongoing and future dark energy studies in four areas represented in the white papers. These are based on observations of Baryon Acoustic Oscillations, Galaxy Clusters, Supernova, and Weak Lensing.

One of our main findings is that no single technique can answer the outstanding questions about dark energy: combinations of at least two of these techniques must be used to fully realize the promise of future observations.  Already there are proposals for major, long-term (Stage IV[1]) projects incorporating these techniques that have the promise of increasing our figure of merit by a factor of ten beyond the level it will reach with the conclusion of current experiments.  What is urgently needed is a commitment to fund a program comprised of a selection of these projects.  The selection should be made on the basis of critical evaluations of their costs, benefits, and risks.

Success in reaching our ultimate goal will depend on the development of dark-energy science.  This is in its infancy.  Smaller, faster programs (Stage III[1]) are needed to provide the experience on which the long-term projects can build.  These projects can reduce systematic uncertainties that could otherwise impede the larger projects, and at the same time make important advances in our knowledge of dark energy.

We recommend that the agencies work together to support a balanced program that contains from the outset support for both the long-term projects and the smaller projects that will have more immediate returns.  We call for a *coordinated* program to attack one of the most profound questions in the physical sciences.   Our report provides a quantitative basis for prioritizing near-term and long-term projects.

We are very fortunate that a wide range of new observations are possible that can drive significant progress in this field.  Many researchers from both particle physics and astronomy are being drawn to these remarkable opportunities.  It is a rare moment in the history of science when such clear steps can be taken to address such a profound problem.

---

[1] In this Report we describe dark-energy research in *Stages*: Stage I represents dark-energy projects that have been completed; Stage II represents ongoing projects relevant to dark-energy; Stage III comprises near-term, medium-cost, currently proposed projects; Stage IV comprises a Large Survey Telescope (LST), and/or the Square Kilometer Array (SKA), and/or a Joint Dark Energy (Space) Mission (JDEM).

Our recommendations are based on the results of our modeling. They are discussed in detail in Section V. In summary, they are

I. We strongly recommend that there be an aggressive program to explore dark energy as fully as possible, since it challenges our understanding of fundamental physical laws and the nature of the cosmos.

II. We recommend that the dark energy program have multiple techniques at every stage, at least one of which is a probe sensitive to the growth of cosmological structure in the form of galaxies and clusters of galaxies.

III. We recommend that the dark energy program include a combination of techniques from one or more Stage III projects designed to achieve, in combination, at least a <u>factor of three</u> gain over Stage II in the DETF figure of merit, based on critical appraisals of likely statistical and systematic uncertainties.

IV. We recommend that the dark energy program include a combination of techniques from one or more Stage IV projects designed to achieve, in combination, at least a <u>factor of ten</u> gain over Stage II in the DETF figure of merit, based on critical appraisals of likely statistical and systematic uncertainties. Because JDEM, LST, and SKA all offer promising avenues to greatly improved understanding of dark energy, we recommend continued research and development investments to optimize the programs and to address remaining technical questions and systematic-error risks.

V. We recommend that high priority for near-term funding should be given as well to projects that will improve our understanding of the dominant systematic effects in dark energy measurements and, wherever possible, reduce them, even if they do not immediately increase the DETF figure of merit.

VI. We recommend that the community and the funding agencies develop a coherent program of experiments designed to meet the goals and criteria set out in these recommendations.

# II. Dark Energy in Context

1. Conclusive evidence from supernovae and other observations shows that the expansion of the Universe, rather than slowing because of gravity, is increasingly rapid. Within the standard cosmological framework, this must be due to a substance that behaves as if it has negative pressure. This substance has been termed "dark energy." Experiments indicate that dark energy accounts for about 70% of the mass-energy in the Universe.

2. One possibility is that the Universe is permeated by an energy density, constant in time and uniform in space. Such a "cosmological constant" (Lambda: $\Lambda$) was originally postulated by Einstein, but later rejected when the expansion of the Universe was first detected. General arguments from the scale of particle interactions, however, suggest that if $\Lambda$ is not zero, it should be very large, larger by a truly enormous factor than what is measured. If dark energy is due to a cosmological constant, its ratio of pressure to energy density (its equation of state) is $w = P/\rho = -1$ at all times.

3. Another possibility is that the dark energy is some kind of dynamical fluid, not previously known to physics. In this case the equation of state of the fluid would likely not be constant, but would vary with time, or equivalently with redshift $z$ or with $a = (1+z)^{-1}$, the scale factor (or size) of the Universe relative to its current scale or size. Different theories of dynamical dark energy are distinguished through their differing predictions for the evolution of the equation of state.

4. The impact of dark energy (whether dynamical or a constant) on cosmological observations can be expressed in terms $w(a) = P(a)/\rho(a)$, which is to be measured through its influence on the large-scale structure and dynamics of the Universe.

5. An alternative explanation of the accelerating expansion of the Universe is that general relativity or the standard cosmological model is incorrect. We are driven to consider this prospect by potentially deep problems with the other options. A cosmological constant leaves unresolved one of the great mysteries of quantum gravity and particle physics: If the cosmological constant is not zero, it would be expected to be $10^{120}$ times larger than is observed. A dynamical fluid picture usually predicts new particles with masses thirty-five orders of magnitude smaller than the electron mass. Such a small mass could imply the existence of a new observable long-range force in nature in addition to gravity and electromagnetism. Regardless of which (if any) of these options are realized, exploration of the acceleration of the Universe's expansion will profoundly change our understanding of the composition and nature of the Universe.

6. It is not at present possible, even with the latest results from ground and space observations, to determine whether a cosmological constant, a dynamical fluid, or a modification of general relativity is the correct explanation of the observed accelerating Universe.

7. Dark energy appears to be the dominant component of the physical Universe, yet there is no persuasive theoretical explanation for its existence or magnitude. The acceleration of the Universe is, along with dark matter, the observed phenomenon that most directly demonstrates that our theories of fundamental particles and gravity are either incorrect or incomplete.  Most experts believe that nothing short of a revolution in our understanding of fundamental physics will be required to achieve a full understanding of the cosmic acceleration.  For these reasons, the nature of dark energy ranks among the very most compelling of all outstanding problems in physical science. These circumstances demand an ambitious observational program to determine the dark energy properties as well as possible.

# III. Goals and Methodology for Studying Dark Energy

1. The goal is to determine the very nature of the dark energy that causes the Universe to accelerate and seems to comprise most of the mass-energy of the Universe.

2. Toward this goal, our observational program must
   a. Determine as well as possible whether the accelerating expansion is consistent with being due to a cosmological constant.
   b. If the acceleration is not due to a cosmological constant, probe the underlying dynamics by measuring as well as possible the time evolution of the dark energy by determining the function $w(a)$.
   c. Search for a possible failure of general relativity through comparison of the effect of dark energy on cosmic expansion with the effect of dark energy on the growth of cosmological structures like galaxies or galaxy clusters.

3. Since $w(a)$ is a continuous function with an infinite number of values at infinitesimally separated points, $w(a)$ must be modeled using just a few parameters whose values are determined by fitting to observations. No single parameterization can represent all possibilities for $w(a)$. We choose to parameterize the equation of state as $w(a) = w_0 + (1-a)w_a$, where $w_0$ is the present value of $w$ and where $w_a$ parameterizes the evolution of $w(a)$. This simple parameterization is most useful if dark energy is important at late times and insignificant at early times.

4. The goals of a dark energy observational program may be reached through measurement of the expansion history of the Universe [traditionally measured by luminosity distance vs. redshift, angular-diameter distance vs. redshift, expansion rate vs. redshift, and volume element vs. redshift], and through measurement of the growth rate of structure, which is suppressed during epochs when the dark energy dominates. All these measurements of dark energy properties can be expressed in terms of the value of the dark energy density today, $w_0$, and its evolution, $w_a$. If the accelerating expansion is due instead to a failure of general relativity, this could be revealed by finding discrepancies between the values of $w(a)$ inferred from these two types of data.

5. In order to quantify progress in measuring the properties of dark energy we define a dark-energy "figure of merit" formed from a combination of the uncertainties in $w_0$ and $w_a$.

> The **DETF figure of merit** is the reciprocal of the area of the error ellipse enclosing the 95% confidence limit in the $w_0$–$w_a$ plane. Larger figure of merit indicates greater accuracy.

The one-dimensional errors in $w_0$ and $w_a$ are correlated, and their product is not a good indication of the power of a particular experiment. This is why the DETF figure of merit is defined as the area contained within the 95% confidence limit contours in the $w_0$–$w_a$ plane (not the simple product of one-dimensional uncertainties in $w_0$ and $w_a$). In Section VII we discuss the DETF figure of merit. We also discuss the utility of defining a pivot value of $w$, defined as $w_p$.

The pivot value of $w$ is the value at the redshift for which $w$ is best constrained by a particular experiment; its variance is equal to the variance of $w$ in a model assuming $w_a = 0$. We demonstrate that the figure of merit is the inverse of the product of uncertainties in $w_p$ and $w_a$. The error in $w_p$ reflects the ability of a single experiment or a combination of experiments to test whether dark energy equation of state is consistent $w = -1$; i.e., a cosmological constant.

6. The DETF dark-energy parameterization of $w(a)$ and the associated figure of merit serve as a robust, quantitative guide to the ability of an experimental program to constrain a large, but not exhaustive, set of dark-energy models. Since the nature of dark energy is so poorly understood, no single figure of merit is appropriate for every eventuality. Particular experiments may excel at testing dark-energy models that are poorly described by our parameterization and their utility may not be reflected in out figure of merit. However, potential shortcomings of the choice of any figure of merit must be evaluated in the larger context, which includes the critical need to make side-by-side comparisons and specific choices to move the field forward. In our judgment there is no better choice of a figure of merit available at this time. We expect continuing theoretical and experimental advances in our understanding of dark energy will allow us to explore other figures of merit. We recognize that developments may eventually lead to recognition by the community that some new measure better meets the overall needs of the field.

7. We have made extensive use of statistical (Fisher-Matrix) techniques incorporating information about cosmic microwave background (CMB) and Hubble's constant ($H_0$) to predict the future performance of possible dark-energy projects, and combinations of these projects.

8. Our considerations for a dark-energy program follow developments in "Stages:"
   a. Stage I represents what is now known.
   b. Stage II represents the anticipated state of knowledge upon completion of ongoing projects that are relevant to dark-energy.
   c. Stage III comprises near-term, medium-cost, currently proposed projects.
   d. Stage IV comprises a Large Survey Telescope (LST), and/or the Square Kilometer Array (SKA), and/or a Joint Dark Energy (Space) Mission (JDEM).

9. Just as dark-energy science has far-reaching implications for other fields of physics, advances and discoveries in other fields of physics may point the way toward understanding the nature of dark energy; for instance, any observational evidence for modifications of General Relativity.

# IV. Findings of the Dark-Energy Task Force

1. Four observational techniques dominate the White Papers received by the task force. In alphabetical order:

    a. **Baryon Acoustic Oscillations (BAO)** are observed in large-scale surveys of the spatial distribution of galaxies. The BAO technique is sensitive to dark energy through its effect on the angular-diameter distance vs. redshift relation and through its effect on the time evolution of the expansion rate.

    b. **Galaxy Cluster (CL)** surveys measure the spatial density and distribution of galaxy clusters. The CL technique is sensitive to dark energy through its effect on a combination of the angular-diameter distance vs. redshift relation, the time evolution of the expansion rate, and the growth rate of structure.

    c. **Supernova (SN)** surveys use Type Ia supernovae as standard candles to determine the luminosity distance vs. redshift relation. The SN technique is sensitive to dark energy through its effect on this relation.

    d. **Weak Lensing (WL)** surveys measure the distortion of background images due to the bending of light as it passes by galaxies or clusters of galaxies. The WL technique is sensitive to dark energy through its effect on the angular-diameter distance vs. redshift relation and the growth rate of structure.

    Other techniques discussed in White Papers, such as using $\gamma$-ray bursts or gravitational waves from coalescing binaries as standard candles, merit further investigation. At this time, they have not yet been practically implemented, so it is difficult to predict how they might be part of a dark energy program. We do note that if dark energy dominance is a recent cosmological phenomenon, very high-redshift ($z \gg 1$) probes will be of limited utility.

2. Different techniques have different strengths and weaknesses and are sensitive in different ways to the dark energy properties and to other cosmological parameters.

3. Each of the four techniques can be pursued by multiple observational approaches, *e.g.,* radio, visible, near-infrared (NIR), and/or x-ray observations, and a single experiment can study dark energy with multiple techniques. Individual missions need not necessarily cover multiple techniques; combinations of projects can achieve the same overall goals.

4. The techniques are at different levels of maturity:

    a. The **BAO** technique has only recently been established. It is less affected by astrophysical uncertainties than other techniques.

    b. The **CL** technique has the statistical potential to exceed the BAO and SN techniques but at present has the largest systematic errors. Its eventual accuracy is currently very difficult to predict and its ultimate utility as a dark energy technique can only be determined through the development of

techniques that control systematics due to non-linear astrophysical processes.

c. The **SN** technique is *at present* the most powerful and best proven technique for studying dark energy. If redshifts are determined by multiband photometry, the power of the supernova technique depends critically on the accuracy achieved for photo-$z$'s. **(Multiband photometry measures the intensity of the object in several colors. A redshift determined by multiband photometry is called photometric redshift, or a <u>photo-$z$</u>.)** If spectroscopically measured redshifts are used, the power of the experiment as reflected in the DETF figure of merit is much better known, with the outcome depending on the uncertainties in supernova evolution and in the astronomical flux calibration.

d. The **WL** technique is also an emerging technique. Its eventual accuracy will also be limited by systematic errors that are difficult to predict. *If* the systematic errors are at or below the level asserted by the proponents, it is likely to be the most powerful individual Stage-IV technique and also the most powerful component in a multi-technique program.

5. A program that includes multiple techniques at Stage IV can provide an order of magnitude increase in the DETF figure of merit. This would be a major advance in our understanding of dark energy. A program that includes multiple techniques at Stage III can provide a factor of three increase in the DETF figure of merit. This would be a valuable advance in our understanding of dark energy. In the absence of a persuasive theoretical explanation for dark energy, we must be guided by ever more precise observations.

6. We find that no single observational technique is sufficiently powerful and well established that it is guaranteed to achieve by itself an order of magnitude increase in the DETF figure of merit. Combinations of the principal techniques have substantially more statistical power, much greater ability to discriminate among dark energy models, and more robustness to systematic errors than any single technique. The case for multiple techniques is supported as well by the critical need for confirmation of results from any single method. (The results for various model combinations can be found at the end of Section IX.)

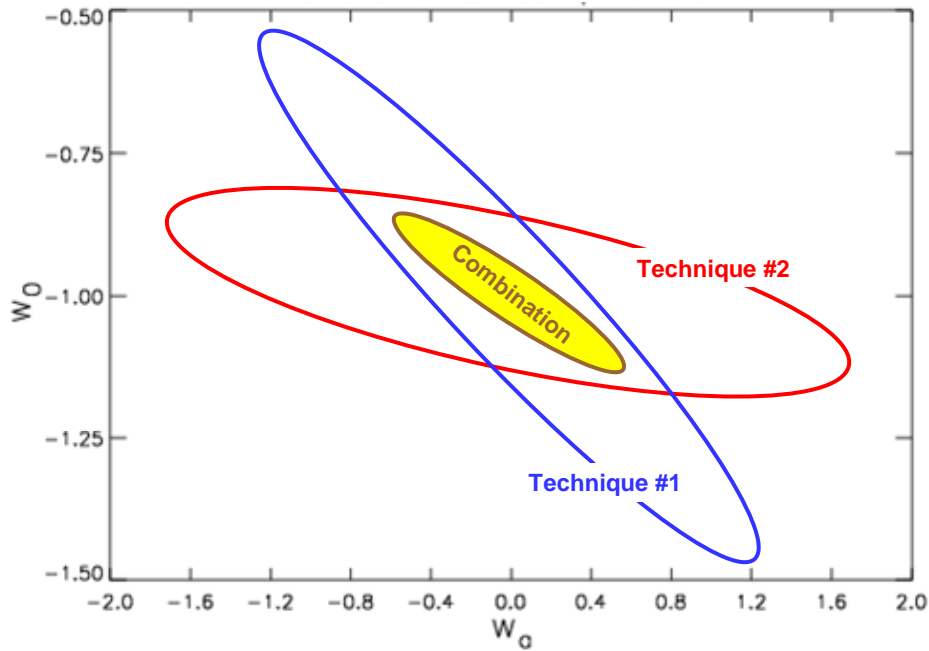

*Illustration of the power of combining techniques. Technique #1 and Technique #2 have roughly equal DETF figure of merit. When results are combined, the DETF figure of merit is substantially improved.*

7.  Results on structure growth, obtainable from weak lensing or cluster observations, provide additional information not obtainable from other techniques. In particular, they allow for a consistency test of the basic paradigm: spatially constant dark energy plus general relativity.

8.  In our modeling we assume constraints on $H_0$ from current data and constraints on other cosmological parameters expected to come from further measurement of CMB temperature and polarization anisotropies.
    a.  These data, though insensitive to $w(a)$ on their own, contribute to our knowledge of $w(a)$ when combined with any of the dark energy techniques we have considered.
    b.  Increased precision in a particular cosmological parameter may improve dark-energy constraints from a single technique. Increased precision is valuable for the important task of comparing dark energy results from different techniques.

9.  Increased precision in cosmological parameters tends not to improve significantly the overall DETF figure of merit obtained from a multi-technique program. Indeed, a multi-technique program would itself provide powerful new constraints on cosmological parameters within the context of our parametric dark-energy model.

10. Setting the spatial curvature of the Universe to zero greatly strengthens the dark-energy constraints from supernovae, but has a modest impact on the other techniques once a dark-energy parameterization is selected. When techniques are combined, setting the spatial curvature of the Universe to zero makes little difference to constraints on parameterized dark energy, because the curvature is one of the parameters well determined by a multi-technique approach.

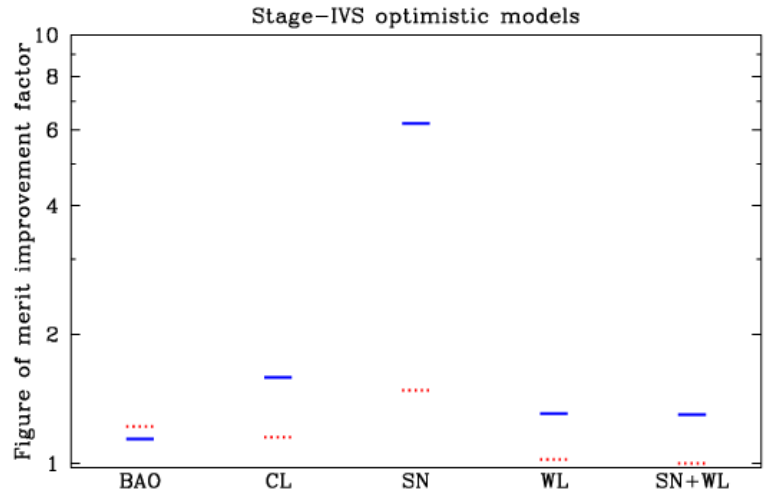

*Illustration of the sensitivity of dark energy constraints to prior assumptions about cosmological parameters in the case of Stage IV space-based measurements with optimistic systematic errors. The solid lines indicate the factor by which the DETF figure of merit increases with the assumption that the spatial curvature of the Universe vanishes. There is a marked improvement in the power of the SN technique with the assumption that the spatial curvature vanishes. However, if the SN technique is combined with other techniques, e.g., the WL technique, the improvement is modest. The dotted lines indicate the factor by which the DETF figure of merit increases with the assumption that the uncertainty in the Hubble constant is $4 \text{ km s}^{-1} \text{ Mpc}^{-1}$ compared to the present uncertainty of $8 \text{ km s}^{-1} \text{ Mpc}^{-1}$. Reducing the uncertainty in $H_0$ makes at most a 50% improvement on individual techniques at the Stage IV level. Space experiments are illustrated here but results from ground Stage IV experiments are similar.*

11. Optical, NIR, and x-ray experiments with very large numbers of astronomical targets will rely on photometrically determined redshifts. The ultimate accuracy that can be attained for photo-$z$'s is likely to determine the power of such measurements. (Radio HI (neutral hydrogen) surveys produce precise redshifts as part of the survey.)

12. Our inability to forecast systematic error levels reliably is the biggest impediment to judging the future capabilities of the techniques. Assessments of effectiveness could be made more reliably with:
    a. For **BAO–** Theoretical investigations of how far into the non-linear regime the data can be modeled with sufficient reliability and further understanding of galaxy bias on the galaxy power spectrum.

   b. For **CL**– Combined lensing, Sunyaev-Zeldovich, and x-ray observations of large numbers of galaxy clusters to constrain the relationship between galaxy cluster mass and observables.

   c. For **SN**– Detailed spectroscopic and photometric observations of about 500 nearby supernovae to study the variety of peak explosion magnitudes and any associated observational signatures of effects of evolution, metallicity, or reddening, as well as improvements in the system of photometric calibrations.

   d. For **WL**– Spectroscopic observations and multi-band imaging of tens to hundreds of thousands of galaxies out to high redshifts and faint magnitudes in order to calibrate the photometric redshift technique and understand its limitations.  It is also necessary to establish how well corrections can be made for the intrinsic shapes and alignments of galaxies, the effects of optics, (from the ground) the atmosphere, and the anisotropies in the point-spread function.

13. Six types of Stage-III projects have been considered.  They include:

   a. a BAO survey on a 4-m class telescope using photo-$z$'s.

   b. a BAO survey on an 8-m class telescope employing spectroscopy.

   c. a CL survey on a 4-m class telescope obtaining optical photo-$z$'s for clusters detected in ground-based SZ surveys.

   d. a SN survey on a 4-m class telescope using photo-$z$'s.

   e. a SN survey on a 4-m class telescope employing spectroscopy from an 8-m class telescope.

   f. a WL survey on a 4-m class telescope using photo-$z$'s.

These projects are typically projected by proponents to cost in the range of tens of millions of dollars. (Cost projections were not independently checked by the DETF.)

14. Our findings regarding Stage-III projects are

   a. Only an incremental increase in knowledge of dark-energy parameters is likely to result from a Stage-III BAO project using photo-$z$'s.  The primary benefit from a Stage-III BAO photo-$z$ project would be in exploring systematic photo-$z$ uncertainties.

   b. A modest increase in knowledge of dark-energy parameters is likely to result from a Stage-III SN project using photo-$z$'s.  Such a survey would be valuable if it were to establish the viability of photometric determination of supernova redshifts, types, and evolutionary effects.

   c. A modest increase in knowledge of dark-energy parameters is likely to result from any single Stage-III CL, WL, spectroscopic BAO, or spectroscopic SN survey.

   d. The SN, CL, or WL techniques could, individually, produce factor of two improvements in the DETF figure of merit, if the systematic errors are close to what the proponents claim.

   e. If executed in combination, Stage-III projects would increase the DETF figure of merit by a factor in the range of approximately three to five, with the large degree of uncertainty due to uncertain forecasts of systematic errors.

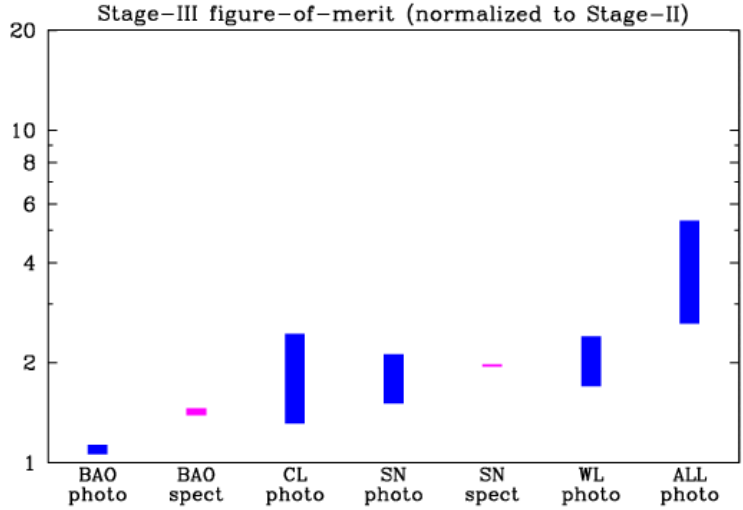

*Illustration of the potential improvement in the DETF figure of merit arising from Stage III projects. The improvement is given for the different techniques individually, along with various combinations of techniques. In the figure 'photo' and 'spect' refers to photometric and spectroscopic surveys, respectively. Each bar extends from the expectation with pessimistic systematics up to the expectation with optimistic systematics. "ALL photo" combines photometric survey results from BAO, CL, SN, and WL.*

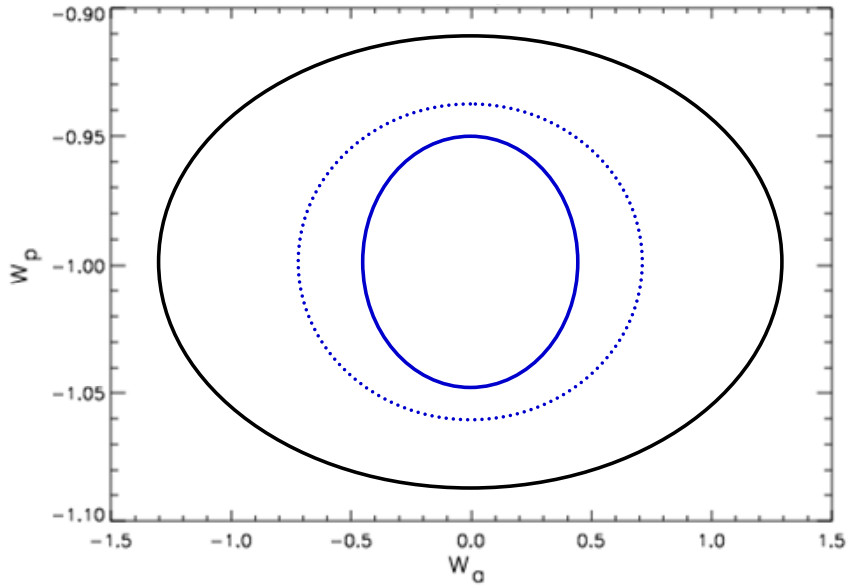

*Illustration of the potential improvement in the DETF figure of merit arising from Stage III projects in the $w_a$–$w_p$ plane. The DETF figure of merit is the reciprocal of the area enclosed by the contours. The outer contour corresponds to Stage II, and the inner contours correspond to pessimistic and optimistic ALL-photo. All contours are 95% C.L.*

15. Four types of Stage-IV projects have been considered
    a. an optical Large Survey Telescope (LST), using one or more of the four techniques.
    b. an optical/NIR Joint Dark Energy Mission (JDEM) satellite, using one or more of the four techniques.
    c. an x-ray JDEM satellite, which would study dark energy by the cluster technique.
    d. a radio Square Kilometer Array, which could probe dark energy by WL and/or BAO techniques through a hemisphere-scale survey of 21-cm and continuum emission. The very large range of frequencies currently demanded by the SKA specifications would likely require more than one type of antenna element. Our analysis is relevant to a lower frequency system, specifically to frequencies below 1.5 GHz.

    Each of these projects is projected by proponents to cost in the \$0.3-1B range, but dark energy is not the only (in some cases not even the primary) science that would be done by these projects. (Cost projections were not independently checked by the DETF.) According to the white papers received by the Task Force, the technical capabilities needed to execute LST and JDEM are largely in hand. (The Task Force is not constituted to undertake a study of the technical issues.)

16. Each of the Stage IV projects considered (LST, JDEM, and SKA) offers compelling potential for advancing our knowledge of dark energy as part of a multi-technique program.

17. The Stage IV experiments have different risk profiles:
    a. The SKA would likely have very low systematic errors, but it needs technical advances to reduce its costs and risk. Particularly important is the development of wide-field imaging techniques that will enable large surveys. The effectiveness of an SKA survey for dark energy would also depend on the number of galaxies it could detect, which is uncertain.
    b. An optical/NIR JDEM can mitigate systematic errors because it would likely obtain a wider spectrum of diagnostic data for SN, CL, and WL than possible from the ground, and it has no systematics associated with atmospheric influence, though it would incur the usual risks and costs of a space-based mission.
    c. LST would have higher systematic-error risk than an optical/NIR JDEM, but could in many respects match the power of JDEM if systematic errors, especially if those due to photo-$z$ measurements, are small. An LST Stage IV program could be effective only if photo-$z$ uncertainties on very large samples of galaxies can be made smaller than what has been achieved to date.

18. A mix of techniques is essential for a fully effective Stage IV program. The technique mix may be comprised of elements of a ground-based program, or elements of a space-based program, or a combination of elements from ground- and space-based programs. No unique mix of techniques is optimal (aside from doing them all), but the absence of weak lensing would be the most damaging provided this technique proves as effective as projections suggest.

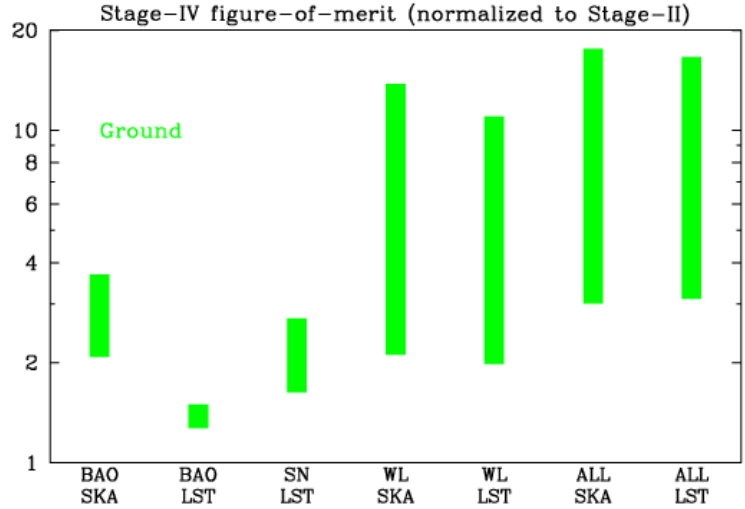

*Illustration of the potential improvement in the DETF figure of merit arising from Stage IV ground-based projects. The bars extend from the pessimistic to the optimistic projections in each case.*

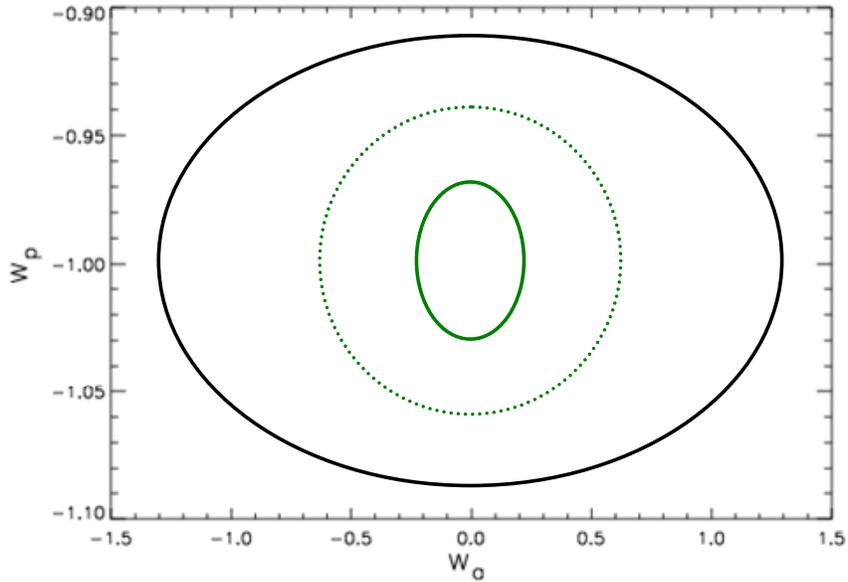

*Illustration of the potential improvement in the DETF figure of merit arising from Stage IV ground-based projects in the $w_a$–$w_p$ plane. The DETF figure of merit is the reciprocal of the area enclosed by the contours. The outer contour corresponds to Stage II, and the inner contours correspond to pessimistic and optimistic ALL-LST. (ALL-SKA would result in similar contours.) All contours are 95% C.L.*

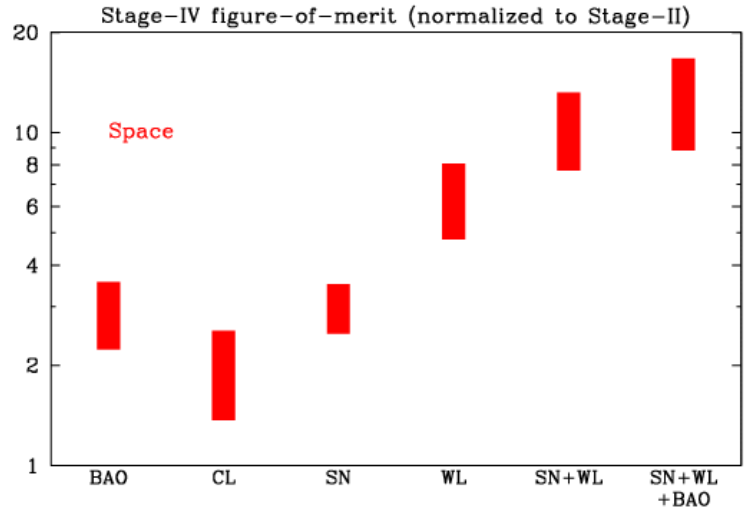

*Illustration of the potential improvement in the DETF figure of merit arising from Stage IV space-based projects. The bars extend from the pessimistic to the optimistic projections in each case. The final two error bars illustrate the improvement available from combining techniques; other combinations of techniques may be superior or more cost-effective. CL results are from an x-ray satellite; the others results from an optical/NIR satellite.*

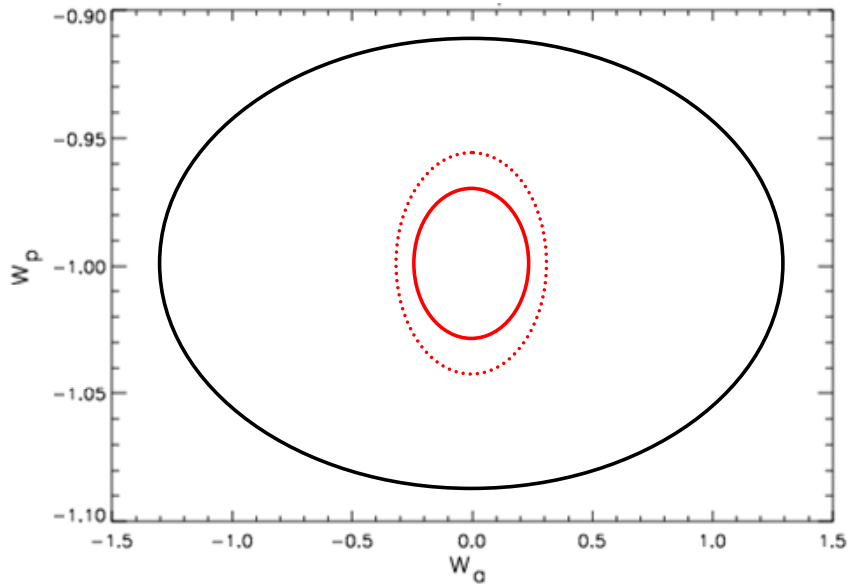

*Illustration of the potential improvement in the DETF figure of merit arising from Stage IV space-based projects in the $w_a$–$w_p$ plane. The DETF figure of merit is the reciprocal of the area enclosed by the contours. The outer contour corresponds to Stage II, and the inner contours correspond to pessimistic and optimistic BAO+SN+WL. All contours are 95% C.L.*

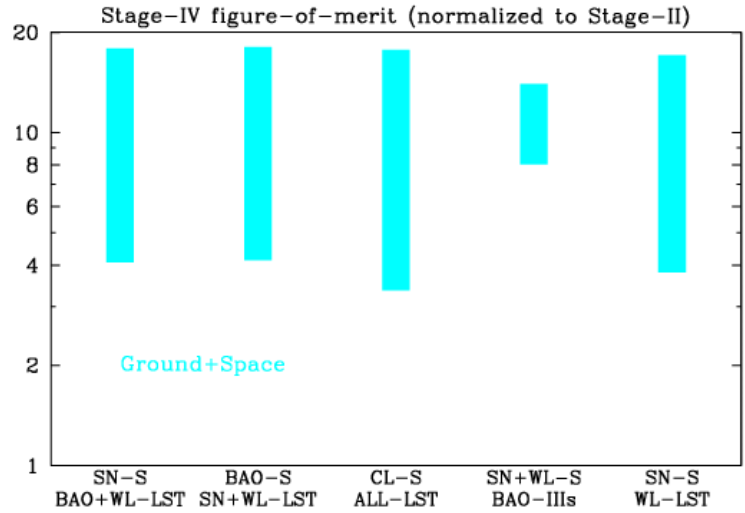

*This figure illustrates the potential improvement in the DETF figure of merit arising from a combination of Stage IV space-based and ground-based projects. The bars extend from the pessimistic to the optimistic projections in each case. This is by no means an exhaustive search of possible ground/space combinations, just a representative sampling to illustrate that uncertainties on each combination are as large as the differences among them.*

# V. Recommendations of the Dark Energy Task Force

Among the outstanding problems in physical science, the nature of dark energy ranks among the very most compelling

> **I. We strongly recommend that there be an aggressive program to explore dark energy as fully as possible, since it challenges our understanding of fundamental physical laws and the nature of the cosmos.**

---

We model advances in dark energy science in Stages. Stage I represents what is now known. Stage II represents the anticipated state of knowledge upon completion of ongoing dark energy projects. Stage III comprises near-term, medium-cost, currently proposed projects. Stage IV comprises a Large Survey Telescope (LST), and/or the Square Kilometer Array (SKA), and/or a Joint Dark Energy (Space) Mission (JDEM).

There are four primary observational techniques for studying dark energy: Baryon Acoustic Oscillations, Clusters, Supernovae, and Weak Lensing. We find that no single observational technique alone is sufficiently powerful and well established that we can be certain it will adequately address the question of dark energy. We also find that combinations of techniques are much more powerful than individual techniques. In addition, we find that techniques sensitive to growth of cosmological structure have the potential of testing the possibility that the acceleration is caused by a modification of general relativity. Finally, multiple techniques are valuable not just for their improvement of the figure of merit but for the protection they provide against modeling errors, either in the dark energy or the observables.

> **II. We recommend that the dark energy program have multiple techniques at every stage, at least one of which is a probe sensitive to the growth of cosmological structure in the form of galaxies and clusters of galaxies.**

---

To quantify our empirical knowledge of dark energy we form a figure of merit from a product of observational uncertainties in parameters that describe the evolution of dark energy. The <u>DETF figure of merit</u> is the reciprocal of the area of the error ellipse enclosing the 95% confidence limit in the $w_0$–$w_a$ plane. Larger figure of merit indicates greater accuracy. (The DETF figure of merit is discussed in detail in Section VII.)

> **III. We recommend that the dark energy program include a combination of techniques from one or more Stage III projects designed to achieve, in combination, at least a <u>factor of three</u> gain over Stage II in the DETF figure of merit, based on critical appraisals of likely statistical and systematic uncertainties.**

Our modeling indicates that a Stage III program can, in principle, reach this goal. Moreover, such a program would help to determine systematic uncertainties and would provide experience valuable to Stage IV planning and execution using the same techniques. Significant progress understanding Stage IV systematic error levels should be made as soon as possible. As much as possible these goals should be integrated with the Stage III projects.

> **IV. We recommend that the dark energy program include a combination of techniques from one or more Stage IV projects designed to achieve, in combination, at least a <u>factor of ten</u> gain over Stage II in the DETF figure of merit, based on critical appraisals of likely statistical and systematic uncertainties. Because JDEM, LST, and SKA all offer promising avenues to greatly improved understanding of dark energy, we recommend continued research and development investments to optimize the programs and to address remaining technical questions and systematic-error risks.**

Our modeling suggests that there are several combinations of Stage IV projects and techniques capable, in principle, of reaching a factor of ten increase, by a ground-based program, a space-based program, or a combination of ground-based and space-based programs. Further improvements in our understanding of systematic error levels are required to determine with confidence the overall and relative effectiveness of specific combinations of Stage IV projects. Findings 12 and 17 discuss this issue in detail.

---

> **V. We recommend that high priority for near-term funding should be given as well to projects that will improve our understanding of the dominant systematic effects in dark energy measurements and, wherever possible, reduce them, even if they do not immediately increase the DETF figure of merit.**

Among the projects that can contribute to this goal are

A. Improving knowledge of the precision and reliability attainable from near-infrared and visible photometric redshifts for both galaxies and supernovae, through statistically significant samples of spectroscopic measurements for a wide range in redshift. The precision with which photometric redshifts can be measured will impact many dark energy measurements. They are particularly critical for large-scale weak lensing surveys, and they bound the potential of baryon-oscillation and supernova surveys that forego spectroscopy. There must be a robust program to develop the precision that will be required for experiments in Stages III and IV.

B. Demonstrating weak-lensing observations with low shear-measurement errors. Future weak-lensing surveys will demand measurements of gravitational shear, in the presence of optical and atmospheric distortions, that exceed currently demonstrated accuracy. Development of the lensing methodology and testing on large volumes of real and simulated image data are required.

C. Obtaining high-precision spectra and light curves of a large ensemble of Type Ia SNe in the ultraviolet/visible/near-infrared to constrain, for example, systematic effects due to reddening, metallicity, evolution, and photometric/spectroscopic calibrations.

D. Establishing a high-precision photometric and spectrophotometric calibration system in the ultraviolet, visible, and near-infrared. Precision photometric redshifts, $K$-corrections, and luminosity distances cannot be achieved until the fundamental calibration system is significantly improved.

E. Obtaining better estimation of the galaxy population that would be detectable in 21 cm by a SKA at high redshifts ($2 > z > 0.5$). Current plausible models show considerable differences in the evolution of the HI luminosity function. This is the primary uncertainty in our predictions of the performance of an SKA galaxy survey as it determines the size and redshift distribution of the galaxy sample.

F. Better characterization of cluster mass-observable relations through joint x-ray, SZ, and weak lensing studies and also via numerical simulations including the effects of cooling, star-formation, and active galactic nuclei.

G. Supporting theoretical work on non-linear gravitational growth and its impacts on baryon acoustic oscillation measurements, weak lensing error statistics, cluster mass observables, simulations, and development of analysis techniques.

———————————

Because the dark energy program will employ a variety of techniques, a number of experiments, and three funding agencies, management of the program poses special challenges.

**VI. We recommend that the community and the funding agencies develop a coherent program of experiments designed to meet the goals and criteria set out in these recommendations.**

We propose a number of guidelines for the development of the program:

1. Individual proposals should not be reviewed in isolation.   Decisions on projects should take into account how they fit into the overall dark-energy program.
    - In judging Stage III proposals, in addition to contributing toward a factor of three increase in the DETF figure of merit, significant weight should be placed on their capacity to enhance the efforts to develop an optimal Stage IV program.   In this regard, the timing of experiments is an issue. That is, Stage III experiments will be of most value if they inform the planning and/or execution of the Stage IV program.
    - In ranking proposed projects, the precision gain in an individual technique is not necessarily the most important factor. When considered in conjunction with other techniques, significant gains in precision for a single technique may not be as valuable as more modest advances in another technique.  In evaluating projects, there is considerable opportunity for trade-offs between different techniques.
    - Projects that combine multiple techniques are desirable. While multiple techniques are crucial, the order of magnitude gain in the dark energy figure of merit is unlikely to require that all four techniques be pursued through Stage IV.  As detailed in our report, combinations of three (or possibly even two) techniques probably can achieve the stated goal.

2. It is incumbent on proponents of Stage III and IV projects to demonstrate that they will be able to limit systematic uncertainties well enough to achieve the claims they make for improving the measurements of dark energy parameters.
    - In modeling projected performance of Stage III and Stage IV projects, the DETF concluded that systematic uncertainties will ultimately determine the accuracy of our knowledge of dark energy.   Critical assessment of the potential systematic uncertainties is a necessary step in the evaluation of these projects.

3. Potential gains from the Stage IV facilities beyond their dark energy studies should be taken into account.
    - Each of the Stage IV facilities would offer enormous gains in knowledge of the Universe beyond their dark-energy studies, at very little marginal cost.

4. A means of quantifying the increase in our understanding of dark energy from the suite of experiments should be developed.

- The figure of merit developed by the Task Force is a first effort in this direction.  It has proved very valuable in organizing and comparing alternative proposed programs to study dark energy.

Summary of DETF recommendations:

**I. We strongly recommend that there be an aggressive program to explore dark energy as fully as possible, since it challenges our understanding of fundamental physical laws and the nature of the cosmos.**

**II. We recommend that the dark energy program have multiple techniques at every stage, at least one of which is a probe sensitive to the growth of cosmological structure in the form of galaxies and clusters of galaxies.**

**III. We recommend that the dark energy program include a combination of techniques from one or more Stage III projects designed to achieve, in combination, at least a <u>factor of three</u> gain over Stage II in the DETF figure of merit, based on critical appraisals of likely statistical and systematic uncertainties.**

**IV. We recommend that the dark energy program include a combination of techniques from one or more Stage IV projects designed to achieve, in combination, at least a <u>factor of ten</u> gain over Stage II in the DETF figure of merit, based on critical appraisals of likely statistical and systematic uncertainties. Because JDEM, LST, and SKA all offer promising avenues to greatly improved understanding of dark energy, we recommend continued research and development investments to optimize the programs and to address remaining technical questions and systematic-error risks.**

**V. We recommend that high priority for near-term funding should be given as well to projects that will improve our understanding of the dominant systematic effects in dark energy measurements and, wherever possible, reduce them, even if they do not immediately increase the DETF figure of merit.**

**VI. We recommend that the community and the funding agencies develop a coherent program of experiments designed to meet the goals and criteria set out in these recommendations.**

# VI. A Dark Energy Primer

In General Relativity (GR), the growth of the Universe is described by a scale factor $a(t)$, defined so that at the present time $t_0$, $a(t_0) = 1$. The time evolution of the expansion in GR obeys

$$\frac{\ddot{a}}{a} = -\frac{4\pi G}{3}\left(\rho + 3P\right) + \frac{\Lambda}{3},$$

where $P$ and $\rho$ are the mean pressure and density of the contents of the Universe, and $\Lambda$ is the *cosmological constant* proposed and then discarded by Einstein. Remarkably, several lines of evidence (described below) confirm that at the present time, $\ddot{a} > 0$. This acceleration immediately implies that either

1. The Universe is dominated by some particle or field (*dark energy*) that has negative pressure, in particular $w = P/\rho < -1/3$; ***or***

2. There is in fact a non-zero cosmological constant; ***or***

3. The theoretical basis for this equation, GR or the standard cosmological model, is incorrect.

Any of these three explanations would require fundamental revision to the underpinning theories of physics. It is of great interest to determine which of these three explanations is correct.

**The Observable Consequences of Dark Energy**

Within the context of GR, a convenient expression of the equation for the expansion is

$$\left(\frac{\dot{a}}{a}\right)^2 \equiv H^2(a) = \frac{8\pi G_N \rho}{3} - \frac{k}{a^2} + \frac{\Lambda}{3},$$

where $k$ is the curvature. The value of $H$ today, $H_0$, is the Hubble constant, $72\pm8$ km s$^{-1}$ Mpc$^{-1}$. From these two equations it follows that

$$\dot{\rho} = -3H(\rho + P),$$

which holds separately for each contributor to the energy density. For non-relativistic matter, $P/\rho$ is of order $(v/c)^2$, and can be ignored, and the equation becomes

$$\dot{\rho}_m = \dot{a}\frac{d\rho_m}{da} = -3\frac{\dot{a}}{a}\rho_m$$

so $d\rho_m/da = -3(\rho_m/a)$ and $\rho_m = \rho_{m0}/a^3$, where $\rho_{m0}$ is the density of non-relativistic matter today. More generally, if $w = P/\rho$ is constant, then

$$\rho = \rho_0 a^{-3(1+w)}.$$

For non-relativistic matter, we define

$$\Omega_m = \frac{8\pi G_N \rho_{m0}}{3H_0^2},$$

and we define analogously $\Omega_r$ for the density of relativistic matter (and radiation), for which $P/\rho = 1/3$. To obtain an attractive equation we introduce

$$\Omega_k = -\frac{k}{H_0^2},$$

Now we can write

$$H^2(a) \equiv \left(\frac{\dot{a}}{a}\right)^2 = H_0^2 \left[\Omega_m a^{-3} + \Omega_r a^{-4} + \Omega_k a^{-2} + \Omega_X a^{-3(1+w)}\right],$$

The term $\Omega_X$ represents the cosmological constant if $w = -1$. Otherwise, it represents dark energy with constant $w$. This generalizes easily for non-constant $w$ with the replacement

$$a^{-3(1+w)} \rightarrow \exp\left(3\int_a^1 \frac{da'}{a'}\left[1 + w(a')\right]\right).$$

The quantity $\Omega_k$ describes the current curvature of the universe. For $\Omega_k < 0$, the Universe is closed and finite; for $\Omega_k > 0$ the Universe is open and potentially infinite; while for $\Omega_k = 0$ the geometry of the Universe is Euclidean (flat).

The cosmic microwave background radiation (CMB) gives very good constraints on the matter and radiation densities $\Omega_m H_0^2$ and $\Omega_r H_0^2$, so it appears one could determine the time history of the dark-energy density, modulo some uncertainty due to curvature, if one could accurately measure the expansion history $H(a)$. When a distant astronomical source is observed, it is straightforward to determine the scale factor $a$ at the time of emission of the light, since all photon wavelengths stretch during the expansion; this is quantified by the *redshift z*, with $(1+z) = a^{-1}$. The derivative $\dot{a}$ is more difficult, however, since time is not directly observable. Most cosmological observations instead quantify the *distance* to a given source at redshift $z$, which is closely related to the expansion history since a photon on a radial path must satisfy

$$ds^2 = dt^2 - a^2 \frac{dr^2}{1 - kr^2} = 0.$$

This implies that the distance to a source at redshift $z$, defined as $D(z)$, is given by

$$D(z) = \int_0^r \frac{dr'}{\sqrt{1-kr'^2}} = \int_t^{t_0} \frac{dt'}{a(t')} = \int_0^z \frac{dz'}{H(z')}.$$

This procedure also can be used to express the coordinate $r$ in terms of the redshift:

$$r(z) = |k|^{-1/2} S_k \left[ |k|^{1/2} \int_0^z \frac{dz'}{H(z')} \right] = |k|^{-1/2} S_k \left[ |k|^{1/2} D(z) \right],$$

where the function $S_k[x]$ is given by

$$S_k[x] = \begin{cases} \sin x & k > 0 \\ x & k = 0 \\ \sinh x & k < 0. \end{cases}$$

The coordinate $r(z)$ has several measurable consequences. This function, or closely related ones, determine: a) the apparent flux of an object of fixed luminosity (*standard-candle method*); b) the apparent angular size or redshift extent of an object of fixed linear size (*standard-ruler method*); or c) the apparent sky density of an object of known space density. The distance functions related to each of these observations are given in the table below. [Recall that $k = H_0^2 (\Omega_0 - 1)$.]

| measurable | Definition |
|---|---|
| proper distance | $D(z) = \int_0^z \dfrac{dz'}{H(z')} = \begin{cases} \|k\|^{-1/2} \sin^{-1} \left[ \|k\|^{1/2} r(z) \right] & k > 0 \\ r(z) & k = 0 \\ \|k\|^{-1/2} \sinh^{-1} \left[ \|k\|^{1/2} r(z) \right] & k < 0 \end{cases}$ |
| luminosity distance | $d_L(z) = r(z)(1+z)$ |
| angular diameter distance | $d_A(z) = r(z)/(1+z)$ |
| volume element | $dV = \dfrac{r^2(z)}{\sqrt{1-kr^2(z)}} \, dr \, d\Omega$ |

Dark energy enters through the dependence of $H(z)$ on dark energy. In turn, the dependence of the expansion rate on dark energy results in a dark-energy dependence to $r(z)$.

The existence of dark energy has a second observable consequence: it affects the growth of density perturbations. Quantum fluctuations in the early Universe create density fluctuations. These are measured in great detail as temperature fluctuations in the CMB at redshift $z = 1088$.

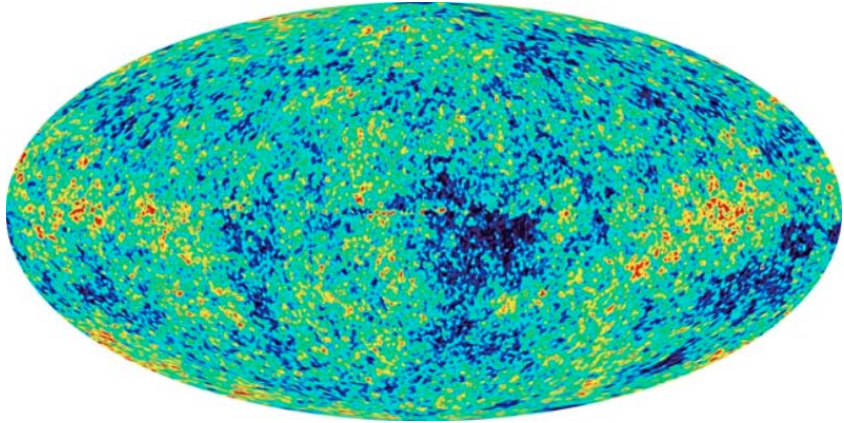

*Fig. VI-1: Fluctuations in the temperature of the early Universe, as measured by the WMAP experiment.*

In a static Universe, overdense regions will increase their density at an exponential rate, but in our expanding Universe there is a competition between the expansion and gravitational collapse. More rapid expansion – as induced by dark energy – retards the growth of structure. GR provides the following relation, in linear perturbation theory, between the *growth factor $g(z)$* and the expansion history of the Universe:

$$\ddot{g} + 2H\dot{g} = 4\pi G \rho_m g = \frac{3\Omega_m H_0^2}{2a^3} g \ .$$

Because the fluctuations at $z = 1088$ are accurately quantified by CMB measurements, **the amplitude of matter fluctuations provides an additional observable manifestation of dark energy via the growth-redshift relation $g(z)$.**

Within the context of GR, this differential equation provides a one-to-one relation between the two observable quantities $D(z)$ and $g(z)$. Inconsistency between these two quantities would indicate that GR is incorrect on the largest observable scales in the Universe (or that dark energy contributes to the growth of clustering in an unexpected manner). If both quantities can be measured, the veracity of this relation can be checked, **permitting a test of the underlying GR theory**.

Figure IV-2 illustrates the effect of dark energy on the distance-redshift and growth-redshift relations, highlighting the need for percent-level precision in these quantities if we are to constrain the dark-energy equation of state to about 0.1 accuracy.

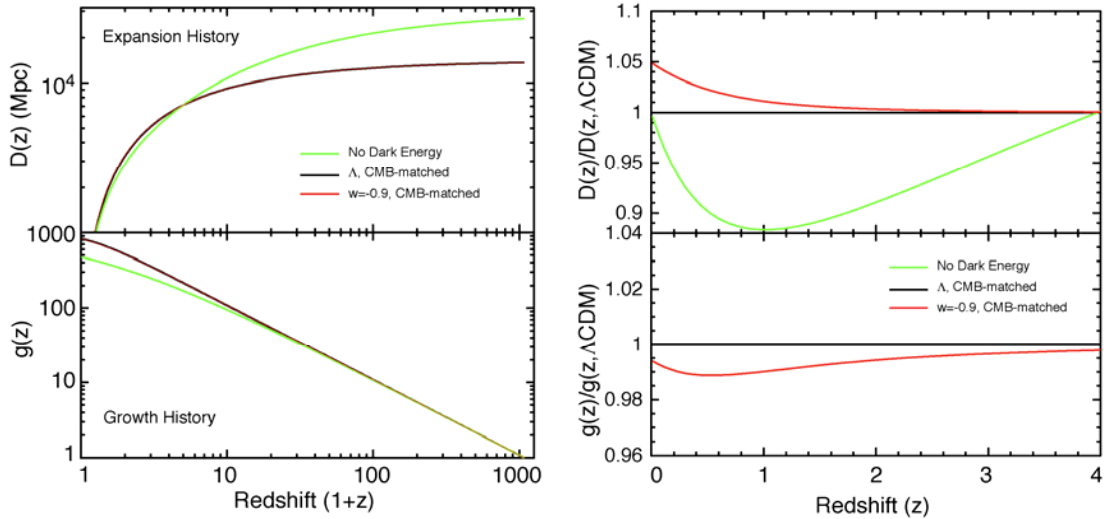

*Fig. VI-2: The primary observables for dark-energy – the distance-redshift relation $D(z)$ and the growth-redshift relation $g(z)$ – are plotted vs. redshift for three cosmological models. The green curve is an open-Universe model with no dark energy at all. The black curve is the "concordance" $\Lambda CDM$ model, which is flat and has a cosmological constant, i.e., $w = -1$. This model is consistent with all reliable present-day data. The red curve is a dark-energy model with $w = -0.9$, for which other parameters have been adjusted to match WMAP data. At left one sees that dark-energy models are easily distinguished from non-dark-energy models. At right, we plot the ratios of each model to the $\Lambda CDM$ model, and it is apparent that distinguishing the $w = -0.9$ model from $\Lambda CDM$ requires percent-level precision on the diagnostic quantities.*

**Four Astrophysical Approaches to Dark Energy Measurements**

## 1. Type Ia Supernovae

Type Ia supernovae are believed to be the explosive disintegrations of white-dwarf stars that accrete material to exceed the stability limit of 1.4 solar masses derived by Chandrasekhar. Because the masses of these objects are nearly all the same, their explosions are expected to serve as standard candles of known luminosity $L$, in which case the relation $f = L/4\pi d_L^2$ can be used to infer the luminosity distance $d_L$. Spectral lines in the supernova light may be used to identify the redshift, as can spectral features of the galaxy hosting the explosion.

Type Ia supernovae observed from the ground and the Hubble Space Telescope (HST) have been used successfully to deduce the acceleration of the Universe after $z = 1$, as illustrated in Fig. IV-2. In practice one finds that Type Ia supernovae are not homogeneous in luminosity. However, variations in luminosity appear to be correlated with other, distance-independent, features of the events, such as the rest-frame duration of the event or its spectral features. Thus Type Ia SNe are *standardizable*, to some yet-unknown degree of precision. Theoretical modeling of SN explosions is extremely difficult; it is not expected that this theory will ever deduce the absolute magnitude nor the standardization process to the accuracy required for dark-energy study. Hence the standardization process must be empirical, and its ultimate accuracy or evolution with cosmic time are very difficult to predict.

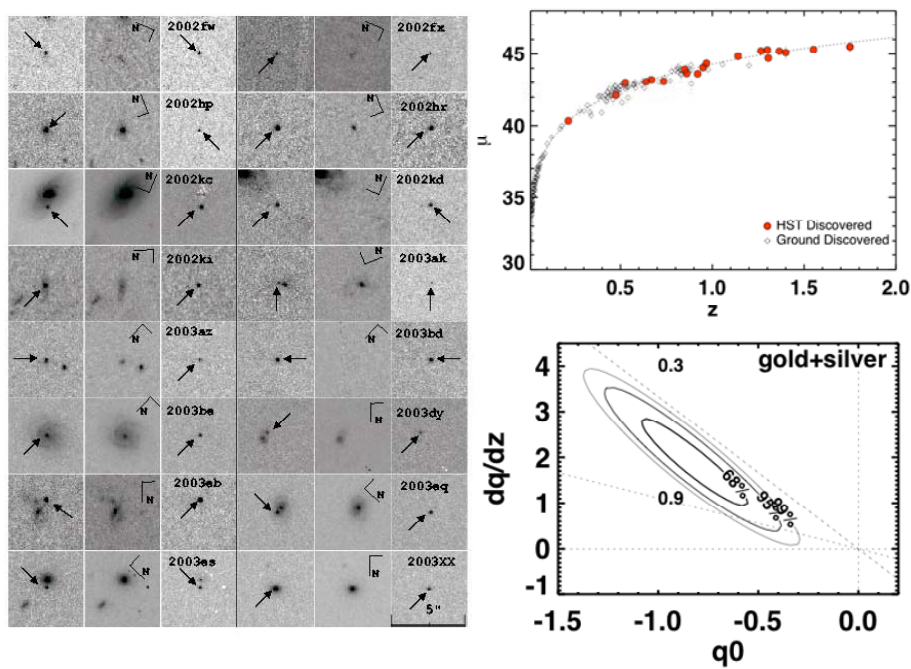

*Fig VI-3: Left: High-redshift supernovae observed from HST by Riess et al (2004). Right: Cosmological results from the GOODS SNe (Riess et al. 2004). Upper panel: distance ($\mu = 5 \log_{10} d_L + $ const.) vs. redshift; lower: constraints on present-day acceleration.*

Other standard(izable)-candle sources may be available in the future: other types of SNe, gamma-ray bursts, or gravity-wave sources. There is not yet evidence that any of them will exceed the precision of Type Ia SNe over the critical $0 < z < 2$ range in the coming decade.

## 2. Baryon Acoustic Oscillations

From the moment inflation ends, the Universe is filled with an ionized plasma. Pressure waves propagate in this baryon-photon fluid at the sound speed of $c_s \simeq c/\sqrt{3}$. From any initial density fluctuation, a expanding spherical perturbation propagates until the time, approximately 370,000 years after the Big Bang, when electrons and protons combine to form neutral hydrogen. At this moment the pressure waves cease to expand, and are frozen into the matter distribution. The total propagation distance $r_s$, is called the *sound horizon*, and the matter distribution is imprinted with this characteristic size. The physics of these *baryon acoustic oscillations* (BAO) is well understood, and their manifestation as wiggles in the CMB fluctuation spectrum is modeled to very high accuracy. The value of $r_s$ is found to be 148±3 Mpc, by the Wilkinson Microwave Anisotropy Probe (WMAP) 3-year data (Spergel *et al.* 2006). The sound horizon scale can thus serve as a standard ruler for distance measurements. Indeed their presence in the CMB allows the distance to $z = 1088$ to be determined to very high accuracy. If we consider that galaxies roughly trace the (dark) matter distribution, then a survey of the galaxy density field should reveal this characteristic scale.

The largest galaxy survey to date, the Sloan Digital Sky Survey, has yielded the first detection of the BAO signal outside of the CMB, as illustrated in Fig. IV.3. The identification of the horizon scale as a transverse angle determines the distance ratio $D(z)/r_s$ (modulo the curvature contribution), while its determination along the line of sight determines $H(z)r_s$.

The density survey to find the BAO feature can use galaxies as the target, in optical, near-IR, or 21-cm emission, or it may be possible to identify the BAO feature in the distribution of neutral hydrogen at redshifts $z > 5$.

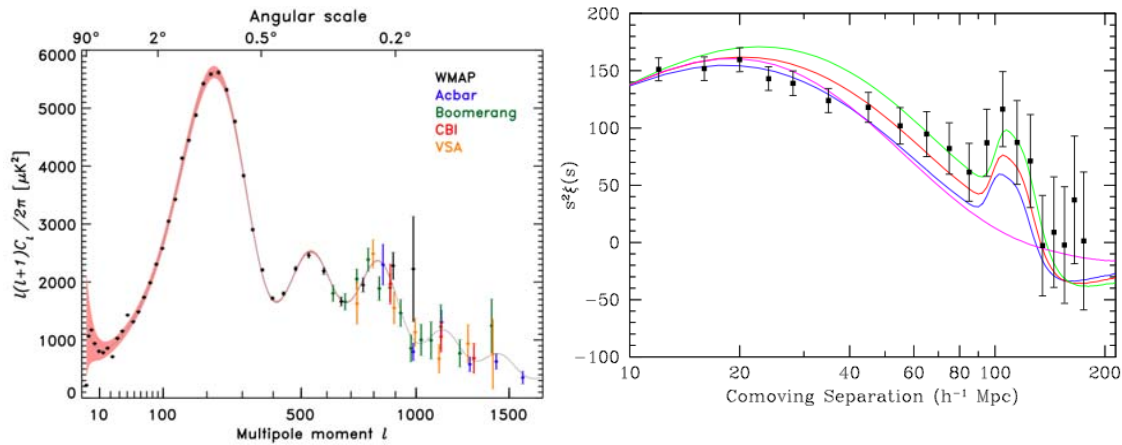

*Fig. VI-4: The baryon acoustic oscillations are seen as wiggles in the power spectrum of the CMB (left, Hinshaw et al. 2003), and have now been detected as a feature in the correlation function of nearby galaxies using the Sloan Digital Sky Survey (right, Eisenstein et al 2005).*

## 3. Galaxy Cluster Counting

Clusters of galaxies are the largest structures in the Universe to have undergone gravitational collapse, and they serve as markers for those locations which were endowed with the highest density fluctuations in the early Universe. Analytic prediction is possible for the *mass function dN / (dM dV)* of these rare events per unit comoving volume per unit cluster mass. Gravitational *N*-body modeling can produce even more precise predictions of the mass function. One can in principle measure the abundance of clusters on the sky, *dN / (dM dΩ dz)*. This is sensitive to dark energy in two ways: First, the comoving volume element depends on dark energy, so cluster counts depend upon the expansion history. Second, the mass function itself is sensitive to the amplitude of density fluctuations; in fact it is exponentially sensitive to the growth function *g*(*z*) at fixed mass *M*.

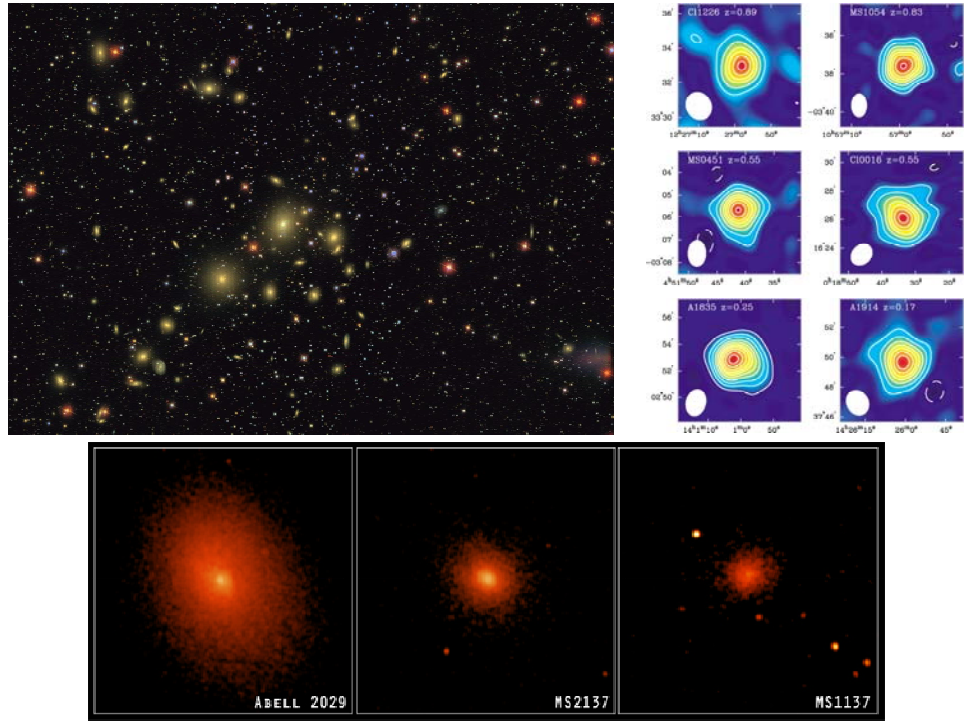

*Fig. VI-5: Galaxy clusters as viewed in three different spectral regimes: top left, an optical view showing the concentration of yellowish member galaxies (SDSS); top right, Sunyaev, Zel'dovich flux decrements at 30 GHz (Carlstrom, et al. 2001); bottom, x-ray emission (Chandra Science Center). These images are not at a common scale.*

Galaxy clusters can be and have been detected in several ways: originally, by the optical detection of their member galaxies; then by the x-ray emission from the hot electrons confined by the gravitational potential well; by the *Sunyaev-Zeldovich effect*, whereby these hot electrons up-scatter the CMB photons, leaving an apparent deficit of low-frequency CMB flux in their direction; and, most recently, by their weak gravitational lensing effect on background galaxy images (see below). The main obstacle to cluster counting is that none of the first three of these techniques measure mass directly Rather they measure some proxy quantity such as galaxy counts, x-ray flux and/or temperature, or the Sunyaev-Zeldovich decrement. The mass function is exponentially sensitive to errors in the calibration of this mass-vs-observable relationship, just as it is exponentially sensitive to the mass itself. These relations are harder to model than pure gravitational growth because they involve complex baryonic physics, *e.g.,* hydrodynamics and galaxy formation.

## 4. **Weak Gravitational Lensing**

Foreground mass concentrations deflect the photons from background sources on their way to Earthbound observers, causing us to see the background source at a position deflected from the "true" direction. The size of the deflection angle depends both on the mass of the foreground deflector and upon the ratios of distances between observer, lens, and source. Like cluster counting, gravitational lensing observations hence probe the dark energy via both the expansion history, $D(z)$, and the growth history of density fluctuations, $g(z)$.

The deflection angles are not observable in general, because we are not at liberty to remove the foreground lens structures to observe the unlensed position. In rare cases the deflection is strong enough to deflect two distinct ray bundles to the observer, who will then see two (or more) distinct images of the same source, and can deduce the deflection angles. But in the more common and general case of *weak lensing*, we can measure the gradient of the deflection angle because any anisotropy in this gradient makes circular source galaxies look slightly elliptical. On a typical line of sight in the Universe, this *shear* amounts to about a 2% stretch along the preferred axis. Since most galaxies are far from circular even in an unlensed view, it is not possible to deduce the lensing signal from a single background galaxy image. However when large numbers of galaxies are observed, the lensing signal can be discerned as a slight tendency for nearby galaxies to have aligned shapes (the intrinsic galaxy shapes need not behave in this manner). The signal-to-noise ratio for weak lensing can be very large if $10^8$-$10^9$ galaxy images are surveyed, as planned for future projects.

This *cosmic shear* effect was first detected in 2000, because of the large volumes of deep digital imaging that are necessary, and because the signal is very subtle and must be carefully distinguished from image distortions caused by the atmosphere and telescope optics. Levels of accuracy are advancing quickly but still far from those needed for the best possible dark-energy measurements. But weak-lensing data is very rich. The cosmic-shear patterns can be measured in many ways, especially if the source galaxies can be divided by redshift. There are power spectra, cross-spectra for every pair of source distances, cross-spectra between the shear patterns and the foreground galaxy distribution, and non-Gaussian statistics such as bispectra. In addition, the peaks in the shear field are a form of cluster counting. It is thus possible to diagnose and correct for many sources of systematic error (but not all) using internal comparisons of different weak-lensing statistics.

While weak lensing has only been detected on the images of background galaxies, it should also be possible to use the CMB itself as the background "wallpaper." Similarly, the 21-cm emission from neutral hydrogen at $z > 5$ may be a viable lensing source. The galaxies at $z < 5$ can be observed with 21-cm and near-IR detectors as well in visible light.

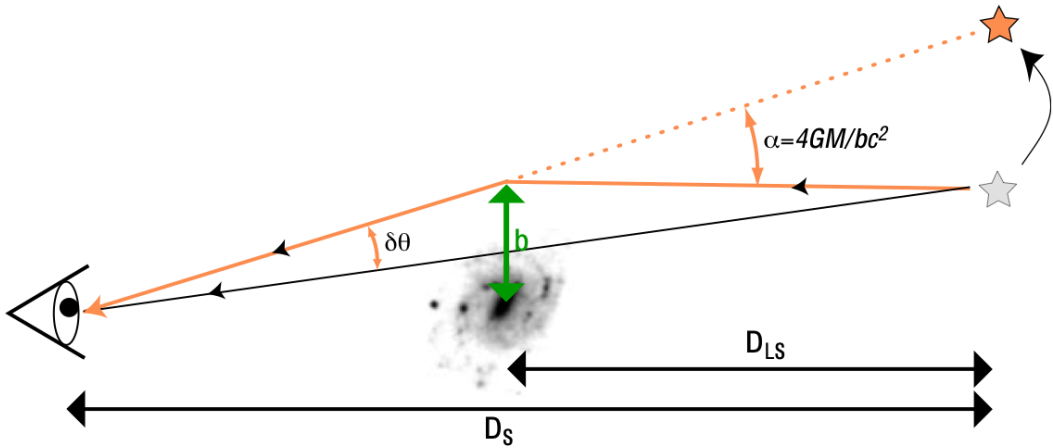

*Fig. VI-6: Schematic of gravitational lensing: the deflection angle apparent to the observer at left depends both upon the mass of the deflector and on the distance ratios between source, lens, and observer.*

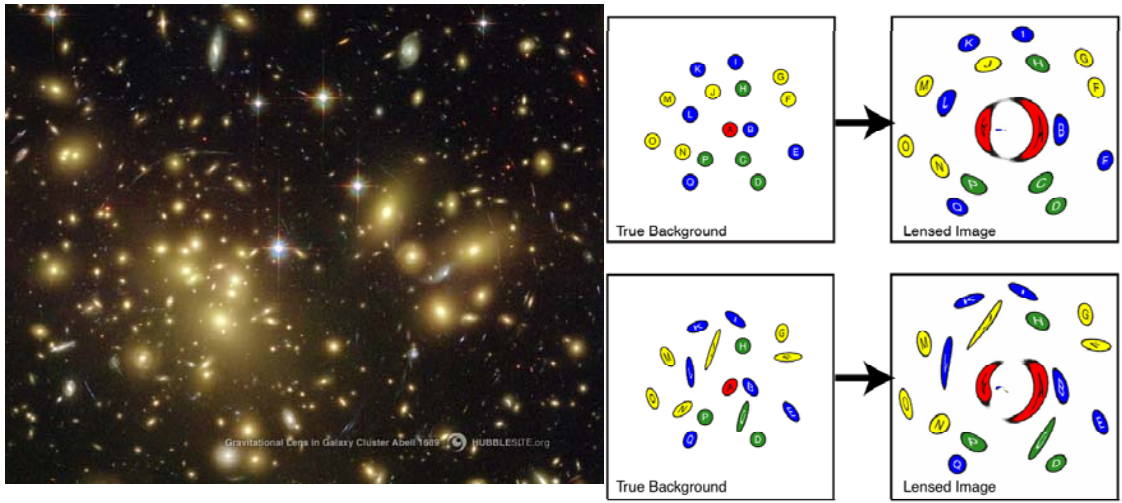

*Fig. VI-7: At left is an image of a galaxy cluster from the Hubble Space Telescope, exhibiting arc-like images of faint background galaxies that are characteristic of strong gravitational lensing. At left: the upper panels show a fictitious collection of circular background galaxies before (left) and after (right) lensing by a foreground mass concentration. While galaxy A, on axis, is grossly distorted into a ring, **all** the other galaxies undergo a slight shearing by the lens. On the lower panels, the galaxies have a variety of initial shapes, so the lensing shear pattern is less obvious, but would be detectable by statistical analysis.*

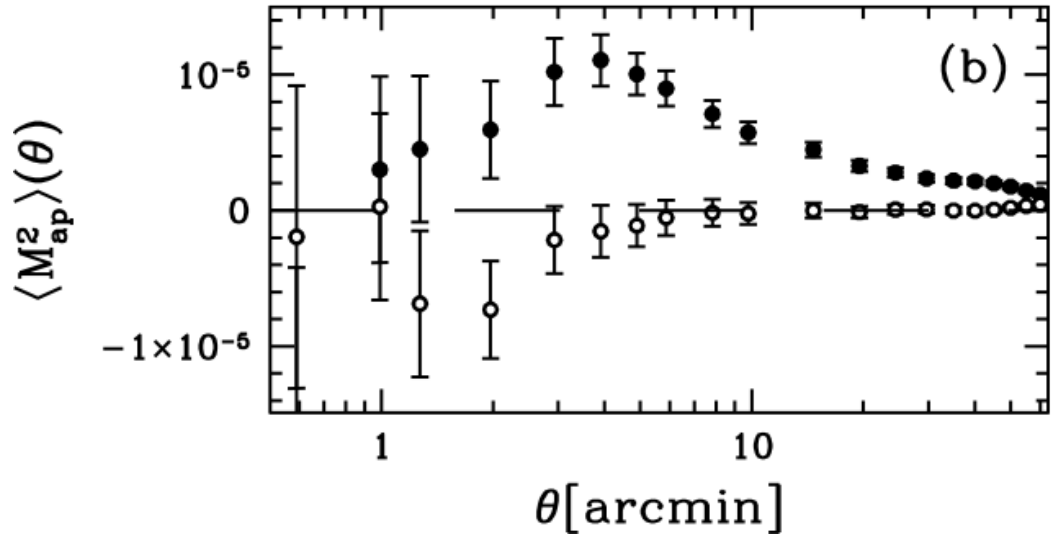

*Fig. VI-8: Measurements of the power of the cosmic-shear effect vs angular scale (Hoekstra et al. 2005). The filled circles represent the "E mode" of the shear pattern, while the open circles are the "B mode," which should be negligible in the absence of systematic errors.*

# VII. The DETF Fiducial Model and Figure of merit

We wish to predict how well future projects would do in constraining dark energy parameters.

The first step is to construct a cosmological model. With the choice of the equation of state parameterization $w(a) = w_0 + (1-a)w_a$, the dark energy cosmological parameters are $w_0$, $w_a$, and $\Omega_{DE}$. (In general, for any component '$i$', $\Omega_i$ is the present-day value of $\rho_i/\rho_C$, where the critical density is $\rho_C = 3H_0^2/8\pi G$, except that we define $\Omega_k = 1 - \Omega_{DE} - \Omega_M$.)

Including the dark energy parameters, the DETF cosmological model is described by eight cosmological parameters:
  A. $w_0$ : the present value of the dark energy equation of state parameter
  B. $w_a$ : the rate of change of the dark energy equation of state parameter
  C. $\Omega_{DE}$ : the present dark energy density
  D. $\Omega_M$ : the present matter density
  E. $\Omega_B$ : the present matter density in the form of baryons
  F. $H_0$ : the Hubble constant
  G. $\delta_\zeta = (k^3 P_\zeta / 2\pi^2)^{1/2}$, the rms primordial curvature fluctuation per $e$-fold evaluated at $k = 0.05$ Mpc$^{-1}$.
  H. $n_S$ : the spectral index of cosmological perturbations.

We do not assume a flat-space prior; *i.e.,* we do not set to zero the curvature contribution to $\Omega$ ($\Omega_k = 1 - \Omega_{DE} - \Omega_M$). While current data are consistent with zero curvature, and most inflation models predict $|\Omega_k| \sim 10^{-5}$, this remains a theoretical prejudice. Given that the acceleration phenomenon was unanticipated by theory, it seems prudent to rely upon observational constraints for curvature rather than accept the theoretical prejudice.

With regard to cosmological perturbations, we assume a pure power-law spectral index, no massive neutrinos, and pure adiabatic perturbations. Allowing for such complications (or others such as running of the spectral index) would weaken the derived dark-energy constraints for some techniques. In general, such effects are minor, and more importantly they tend to have very little impact on the *relative* merit of dark-energy constraints from different experiments.

CMB temperature and polarization data provide constraints on the cosmological parameters, and also provide the distance to last scattering. We model the data anticipated from the Planck satellite mission as detailed in the Technical Appendix, and take these CMB constraints as prior information for any dark-energy experiment.

We also assume as a prior the result on the Hubble constant from the Hubble Space Telescope Key Project: $H_0 = 72 \pm 8$ km s$^{-1}$ Mpc$^{-1}$ [Freedman, et al. (2001)].

The dark-energy parameters and cosmological parameters for the DETF fiducial model were chosen to be consistent with existing observations, including the first-year WMAP results.

1. $w_0 = -1.0$
2. $w_a = 0.0$
3. $\Omega_{DE} = 0.73$
4. $\Omega_M = 0.27$
5. $\Omega_B = 0.046$
6. $H_0 = 72 \text{ km s}^{-1} \text{ Mpc}^{-1}$
7. $\delta_\zeta = 5.07 \times 10^{-5}$ at $\tau = 0.17$.
8. $n_S = 1.0$

The next step is to model the quality and quantity of the data expected for particular experimental implementations of the four dark energy techniques. Each data model incorporates information on anticipated statistical and systematic errors. In the Section IX and in the Technical Appendix we give details about the DETF data models.

For the four techniques we examine in detail (BAO, CL, SN, and WL), we construct data models describing the evolution of a Dark Energy Program in various stages:

A. Stage I represents what is now known.
B. Stage II represents the anticipated state of knowledge upon completion of ongoing projects that are relevant to dark-energy.
C. Stage III comprises near-term, medium-cost, currently proposed projects.
D. Stage IV comprises a Large Survey Telescope (LST), and/or the Square Kilometer Array (SKA), and/or a Joint Dark Energy (Space) Mission (JDEM).

We use Fisher-matrix techniques (described in the Technical Appendix) to predict how well an individual model experiment would be able to restrict the dark energy parameters $w_0$, $w_a$, and $\Omega_{DE}$. This information can be expressed in terms of the standard deviations $\sigma(w_0)$, $\sigma(w_a)$, and $\sigma(\Omega_{DE})$. Since in some sense theoretical predictions for $\Omega_{DE}$ are off by 120 orders of magnitude, the DETF has not placed high priority on precision measurements of $\Omega_{DE}$. Of more relevance is the precision in $w_0$ and $w_a$. The information may be presented in terms of a diagram in the $w_a - w_0$ plane with a contour enclosing some confidence level (C.L.) after marginalization over the other six cosmological parameters and any other nuisance parameters specific to the experiment. An example is given below.

**All diagrams in this report will show contours enclosing 95% C.L., *i.e.*, $\Delta\chi^2 = 6.17$ for our assumption of Gaussian uncertainties in two dimensions.**

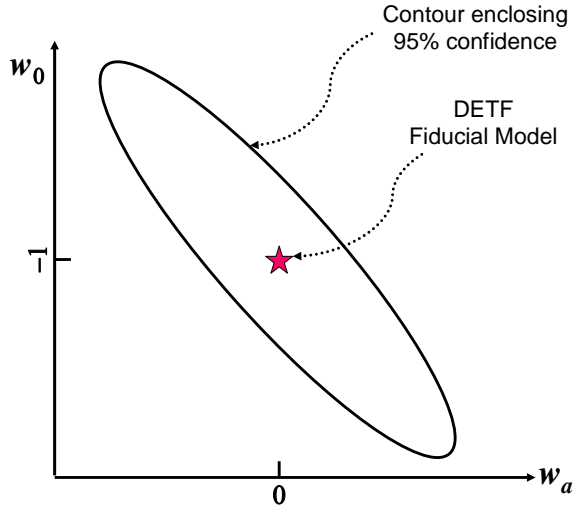

*The DETF figure of merit is defined as the reciprocal of the area of the error ellipse in the $w_0$–$w_a$ plane that encloses the 95% C.L. contour. (We show in the Technical Appendix that the area enclosed in the $w_0$–$w_a$ plane is the same as the area enclosed in the $w_p$–$w_a$ plane.)*

Note that if dark-energy uncertainties are dominated by a noise source that scales as $Q^{-0.5}$ for some quantity $Q$, such as survey area or source counts, then the figure of merit will scale as $Q$.

Recall that a goal of a dark energy program is to test whether dark energy arises from a simple cosmological constant, ($w_0 = -1$, $w_a = 0$). A given data model may do a better job excluding $w_0 = -1$ and $w_a = 0$ than is apparent from simply quoting $\sigma(w_0)$ and $\sigma(w_a)$. This is because the effect of dark energy is generally not best constrained at the present epoch ($z = 0$; $a = 1$). For each data model the constraint on on $w(a) = w_0 + (1-a)w_a$ varies with $a$. However there is some pivot value of $a$, denoted as $a_p$, where the uncertainty in $w(a)$ is minimized for a given data model. The idea is illustrated in the figure below.

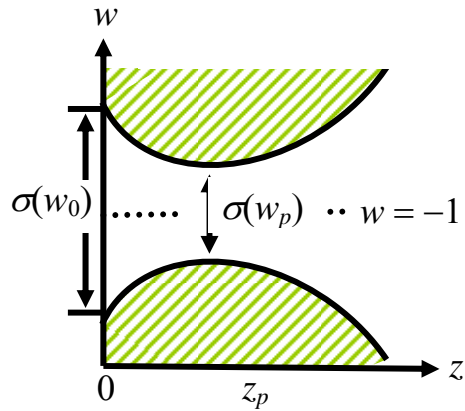

Each data model results in values for $\langle \delta w_0^2 \rangle = [\sigma(w_0)]^2$, $\langle \delta w_a^2 \rangle = [\sigma(w_a)]^2$, and the correlation $\langle \delta w_a \delta w_0 \rangle$, which determine the error ellipse. With $w_p = w_0 + (1-a_p)w_a$, the uncertainty in $w_p$ is least when $1 - a_p = -\langle \delta w_0 \delta w_a \rangle / \langle \delta w_a{}^2 \rangle$. We demonstrate in the Technical Appendix that:

A. The errors on $w_p$ and $w_a$ are uncorrelated, *i.e.,* the error ellipse in the $w_p - w_a$ plane is not tilted;

B. The area of the error ellipse in the $w_p - w_a$ plane is the same as that in the $w_0 - w_a$ plane, so the DETF figure of merit is proportional to $[\sigma(w_p) \times \sigma(w_a)]^{-1}$;

C. The uncertainty $\sigma(w_p)$ is the same as the uncertainty that one would have in $w_0$ if the equation of state parameter $w$ that was assumed constant in time.

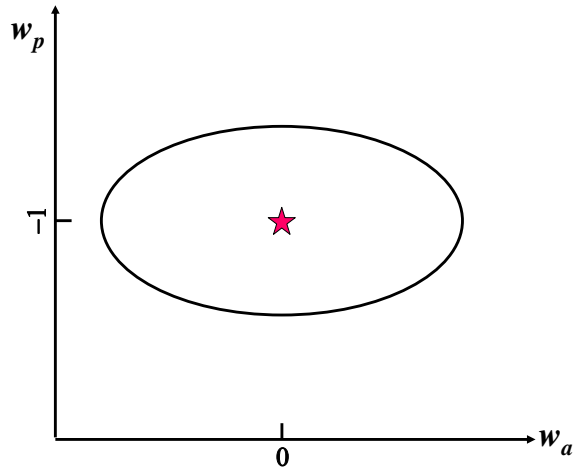

**The DETF figure of merit, which is defined to be the reciprocal of the area in the $w_0 - w_a$ plane that encloses the 95% C.L. region, is also proportional to $[\sigma(w_p) \times \sigma(w_a)]^{-1}$.**

For each data model the results will be presented in tabular form and in the form of a figure in the $w_p - \Omega_{DE}$ plane with 95% C.L. contours of what our task force experts feel will be reasonable optimistic and pessimistic estimates including systematic errors. Note that the plots are *not* in the $w_0 - w_a$ plane where the figure of merit is defined. Because $w_0$ and $w_a$ are uncorrelated, the ellipses would not be more informative than the tabulated data. We plot the $w_p - \Omega_{DE}$ contours so that one can perhaps see how different experiments break degeneracies with this additional parameter.

For each model there will be a table of possible origins of systematic errors and how well a project has to perform to be within the systematic errors of the technique.

An example is given here for the data model CL-IIIp:

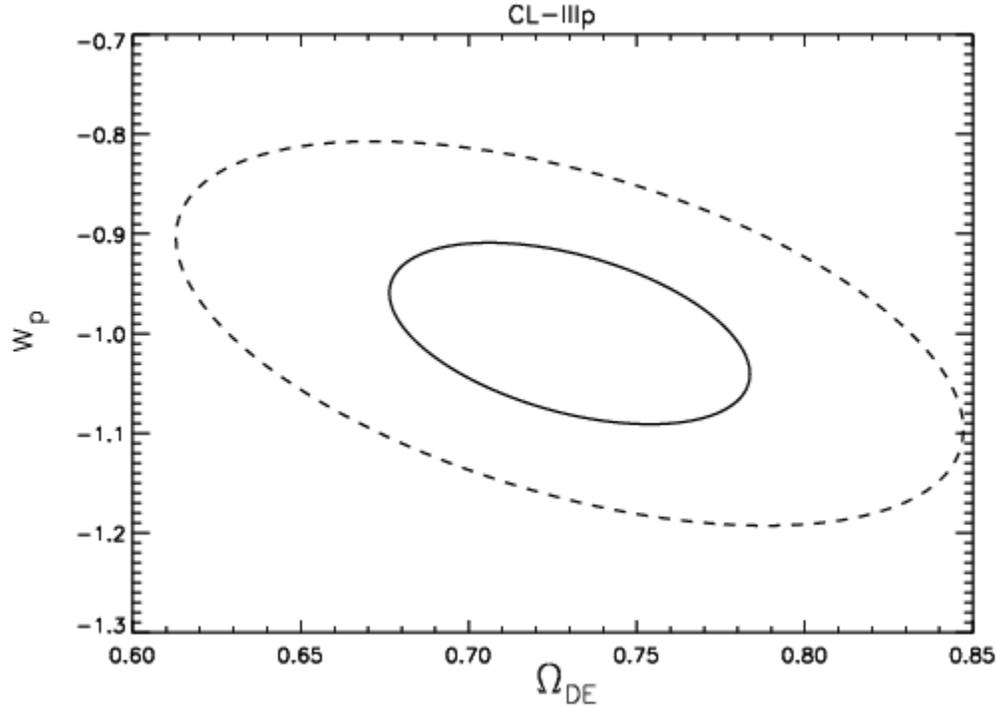

**Dashed contours represent pessimistic projections and solid contours represent optimistic projections.**

| MODEL | $\sigma(w_0)$ | $\sigma(w_a)$ | $\sigma(\Omega_{DE})$ | $a_p$ | $\sigma(w_p)$ | $[\sigma(w_a) \times \sigma(w_p)]^{-1}$ |
|---|---|---|---|---|---|---|
| CL-IIIp-o | 0.256 | 0.774 | 0.022 | 0.672 | 0.037 | 35.21 |
| CL-IIIp-p | 0.698 | 2.106 | 0.047 | 0.670 | 0.078 | 6.11 |

Data models are denoted by
TECHNIQUE-STAGE+QUALIFIER-OPTIMISTIC/ PESSIMISTIC.

| TECHNIQUE | STAGE | | QUALIFIER | OPTIMISTIC/PESSIMISTIC | |
|---|---|---|---|---|---|
| BAO | I | s | spectroscopic survey | o | optimistic |
| CL | II | p | photometric survey | p | pessimistic |
| SN | III | LST | Large Survey Telescope | | |
| WL | IV | SKA | Square Kilometer Array | | |
| | | S | Space | | |

**For each data model we present the assumptions regarding statistical and systematic uncertainties. While the statistical performance is reasonably straightforward, the key is systematic errors. Considerable effort and thought went into our projections. It is absolutely crucial that any proposed project justify its systematic error budget.**

# VIII. Staging Stage IV: Ground and Space Options

Stage IV of the dark-energy program will aim for full exploitation of the available measurement techniques. In this Section we summarize the strengths and weaknesses of the four most prominent measurement techniques and compare the three types of observational platforms that have been proposed (space mission, ground-based Large Survey Telescope, and Square Kilometer Array for radio observations). Each platform has unique advantages: as a result, none of these three platforms can at present be judged as redundant even if another one or even two were to be built. There are very strong motivations, from dark-energy science and more general astrophysics, for continuing development of all three projects. The relative benefits, risks, and costs of these projects for dark-energy science should be much better known on a time scale of a few years, if their development and supporting research on systematic errors are pursued aggressively.

**Analysis of the four techniques:**

- **Baryon Acoustic Oscillations (BAO) [*Dark-energy Observables: D(z), H(z)*]**
  - ○ *Strengths:* This is the method least affected by systematic uncertainties, and for which we have the most reliable forecasts of resources required to accomplish a survey of chosen accuracy. This method uses a standard ruler understood from first principles and calibrated with CMB observations. The BAO technique can constrain $D(z)$ (from oscillations viewed transversely) and $H(z)$ (from oscillations viewed radially) well at high $z$, which complements other techniques. If the dark-energy approximates a cosmological constant, then it is unimportant at high $z$, so high-redshift measures are useful for controlling curvature and testing the $\Lambda$CDM model independent of dark energy. If dark energy is more prominent at high redshifts than in the $\Lambda$CDM model, then high-$z$ measures of $D(z)$ and $H(z)$ become useful for dark-energy constraints.
  - ○ *Weaknesses:* This method is the one with the least statistical power to detect departures from the fiducial $\Lambda$CDM model within the $(w_0, w_a)$ parameterization, since the most precise measurements are made at $z > 1$, where dark energy is relatively unimportant if dark energy approximates a cosmological constant. Relying on photometry in place of spectroscopy for redshift determination probably sacrifices the ability to probe $H(z)$ directly and reduces the signal used to determine $D(z)$. If $z$ is determined photometrically, errors in the redshift distributions must be controlled very well in order to avoid significant biases in cosmological parameter estimates.
  - ○ *Potential Advantages of LST:* A survey that foregoes spectroscopy can largely compensate for the increased statistical errors on $D(z)$ by covering very large amounts of sky. Obtaining high galaxy number densities, as is possible with very deep imaging, means that one can retain only the galaxies with the very best photometric redshifts, discarding the rest without significantly increasing the statistical errors in $D(z)$.

- o ***Potential Advantages of SKA:*** 21-cm detection of galaxies yields high-precision redshifts without additional effort. An SKA with wide-field capabilities can conduct such a spectroscopic survey over the full hemisphere.
- o ***Potential Advantages of Space Mission:*** It is likely that low-background NIR spectra can obtain redshifts more quickly than ground-based surveys, over much of the interesting redshift range.
- o ***Steps to Sharpen Forecasts:*** Uncertainty in the effect of nonlinear processes on the galaxy power spectrum can be reduced with further theoretical and numerical studies rather than the execution of a precursor survey. Further development of the photometric-redshift technique is required just as in the case of weak lensing. The redshift limit of a SKA BAO survey depends upon the evolution of neutral hydrogen content of galaxies, which is poorly determined at present.

- **Galaxy Cluster Counting (CL) [*Dark-energy Observables:* $D^2(z)/H(z)$ and $g(z)$]**
  - o ***Strengths:*** Galaxy-cluster abundances are sensitive to both the expansion and growth histories of the Universe, in this case with extremely strong dependence on the growth factor. There are multiple approaches to cluster detection: the Sunyaev-Zeldovich (SZ) effect, x-ray emission, lensing shear, and of course optical detection of the cluster galaxies. A large SZ cluster survey (SPT) is already funded, and is the only funded project in our Stage III class.
  - o ***Weaknesses:*** While *N*-body simulations will be able to predict the abundance of clusters vs. mass and vs. lensing shear to high accuracy, the prediction of SZ, x-ray, or galaxy counts is subject to substantial uncertainties in the baryonic physics. Dark-energy constraints are very sensitive to errors in these "mass-observable" relations, which are likely to dominate the error budget. This method is the one for which our forecasts are least reliable, due to this large astrophysical systematic effect.
  - o ***Potential Advantages of LST:*** LST can detect galaxy clusters via the effect of their mass on shear patterns and also via the overdensities of the cluster galaxies themselves. Deep weak-lensing observations would play a key role for calibrating the mass-observable relation for optical (LST) observables as well as SZ and x-ray observables of spatially overlapping SZ or x-ray surveys.
  - o ***Potential Advantages of Space Mission:*** An x-ray cluster survey, of course, requires a space mission. With an optical/NIR-imaging space mission, lensing-selected cluster surveys benefit from in the same way as WL surveys do, by offering lower noise levels for WL mapping due to higher density of resolved background galaxies. We subsume consideration of lensing-selected clusters into our WL category because any cosmic-shear survey is also a cluster survey. A similar statement can be made for optically-selected galaxy clusters.
  - o ***Potential Advantages of SKA:*** None recognized: cluster galaxies tend to be deficient in neutral hydrogen, so cluster detection is not a strength of SKA.
  - o ***Steps to Sharpen Forecasts:*** "Self-calibration" methods can potentially recover much of the information lost to the mass-observable uncertainties, but their efficacy depends critically on the complexity/diversity of cluster baryon evolution. A better understanding of cluster baryonic physics will likely result from the SZ surveys about to commence. Weak-lensing observations of the detected clusters in these surveys may help as well; more generally, intercomparison of all four kinds of observables could constrain many of the uncertain parameters in the mass-observable relations.

- **Supernovae (SN)** [*Dark-energy Observables:* $D(z)$]
  - ○ *Strengths:* The most established method and the one that currently contributes the most to the constraint of dark energy. If Type Ia SN luminosities were exactly standard(izable) over the full redshift range $0 < z < 2$, then the statistical precision of SN method would ultimately be limited only by the accuracy to which we can establish the astronomical flux scale across visible/NIR spectrum.
  - ○ *Weaknesses:* Changes in the population of Type Ia events and foreground extinction over time will bias dark energy parameters unless they can be identified by signatures in the colors/spectra/light curves of individual events. Estimates of the systematic errors that will affect given surveys are very difficult in the absence of a quantitative understanding of the diversity of SN events and their foreground extinction, and the ways in which this diversity is manifested in the spectral/temporal observables of each event.
  - ○ *Potential Advantages of LST:* High throughput enables discovery of SNe at very high rate (tens of thousands per year) with densely sampled light curves with high signal-to-noise ratios in optical bands at $z < 1$. These large numbers of high signal-to-noise events would be useful in the search for further parameters to improve supernovae as standard candles and control evolutionary effects.
  - ○ *Potential Advantages of Space Mission:* NIR coverage offers light curves less affected by extinction at low $z$, and rest-frame-visible light curves and spectra for $z > 1$. We expect unified, stable photometric calibration across the visible/NIR to be better above the atmosphere as well. In the long term, JWST observations of SNe at $z > 2$ may constrain evolution of SNe and extinction.
  - ○ *Potential Advantages of SKA:* None – radio detection/follow-up has not been proposed as a principal observational tool for SNe.
  - ○ *Steps to Sharpen Forecasts:* A large low-$z$ SN Ia survey (many hundreds of events) could quantify the diversity and key observational signatures of SNe variation, well before high-$z$ surveys are conducted. Forecasting the quality of photometric redshifts requires better understanding of the variety of spectra in the rest-frame wavelength range probed by optical observations of SNe in the $0 < z < 1$ range. High-$z$ surveys should ultimately be designed to measure all the observables needed to diagnose and correct for this diversity.

- **Weak Gravitational Lensing (WL)** [*Dark-energy Observables:* $D(z)$ and $g(z)$]
  - ○ *Strengths:* The method with the greatest potential for constraining dark energy. The multitude of WL statistics (power spectra, cross-spectra, bispectra, *etc.*) allows internal tests for, and correction of, many potential systematic errors. Both expansion and growth history may be determined from WL data.  The theory of WL is still developing but there appear no fundamental barriers to doing all the calculations needed to exploit the data. WL surveys produce shear-selected galaxy-cluster counts and photo-$z$ BAO data at no additional cost.

o  **Weaknesses:** WL is likely to be limited by systematic errors arising from incomplete knowledge of the error distributions of photometric redshifts (except for SKA survey). While NIR data are known to be of great utility in producing reliable photo-$z$'s, their ultimate impact on photo-$z$'s and WL dark-energy constraints cannot be ascertained without a quantitative understanding of the diversity of galaxy spectra at modest redshifts. The methodology of WL is progressing rapidly but not yet mature: it is not yet demonstrated that one can measure galaxy shapes to the statistical limits, especially from the ground.

o  **Potential Advantages of LST:** A ground-based telescope can be built with a large collecting area and large field-of-view that would be prohibitively expensive in space. This higher-throughput instrument could rapidly map a full hemisphere of sky, reducing statistical errors, and enabling repeated observations with varying observational parameters to evaluate and control systematic errors.

o  **Potential Advantages of Space Mission:** Deep wide-field NIR imaging would improve photo-$z$ accuracy and reliability, and extend the galaxy sample to higher redshifts.  Higher angular resolution triples the number density of galaxies with measurable shapes.  Absence of thermal, gravity-loading, and atmospheric variations allows improved correction for instrumental effects on galaxy shapes. Observations above the atmosphere may permit more accurate photometric calibrations, improving photo-$z$'s.

o  **Potential Advantages of SKA:** 21-cm observations yield precise redshift information for every detected galaxy. An SKA with wide-field capabilities can conduct such a survey over the full hemisphere.  Radio interferometric observations measure galaxy shapes directly in Fourier space, making it much simpler to understand and correct for instrumental effects – if the array contains sufficiently long baselines to resolve the source galaxies' neutral-hydrogen emission.  Atmospheric effects on radio imaging are much less severe than in the visible/NIR.

o  **Steps to Sharpen Forecasts:** Since shape measurement is an instrumental and data-processing problem, rather than an astrophysical unknown, it can be improved and evaluated with further simulation and testing.  Experiments with new-generation telescopes with improved monitoring and control of wavefront quality can be used to better understand the limitations of ground-based shape measurements. Improvements in forecasts of the resources needed for photo-$z$ training sets are possible by quantitative study of the diversity of galaxy spectra in the SDSS, for example, but new, high-quality, NIR imaging, deep spectroscopic surveys (or deep imaging in tens of narrow bands) will be needed to fully understand the limitations of photo-$z$'s at $z > 1$. Deep 21-cm observations to ascertain the HI size and flux distribution of moderate-redshift galaxies would reduce the uncertainty in forecasts for a SKA survey.

**Advantages of an optical ground-based dark-energy experiment (LST):**

1. Weak lensing is potentially the most powerful probe of dark energy. An experiment that can measure weak lensing of source galaxies to high redshifts over half of the sky, while designing for the reduction and control of systematic errors, is thus a very attractive proposition. The combination of high sky coverage and depth is enabled by the product of large collecting area with large field-of-view ("throughput"); a ground-based telescope can offer substantially larger throughput. The ultimate limit would be set by the extent to which the systematics can be controlled. All WL statistics have greater constraining power with greater sky coverage. In particular, the WL three-point statistics (bispectrum) and peak-counting statistics (cluster counts), which have not been incorporated into the DETF forecasts for WL, would gain strength with increased survey area or volume.

2. A well-studied design exists for combining photo-$z$ WL and BAO in an LST mission. Combination with SNe and clusters may also be feasible. Indeed, a WL survey might have as a byproduct a cluster catalog of substantial statistical power. LST thus appears capable of providing multiple probes of DE, including sets that can measure expansion and growth histories independently.

3. If an LST does not improve over the nominal systematics levels assumed for our Stage III experiments, then its gain in our DE figure of merit is modest: a factor of 2.5 for WL+BAO relative to knowledge at the end of Stage II, a factor of 3.4 for WL+BAO+SN. If we anticipate lower systematic errors for all techniques with an LST, then the figure of merit of an LST for WL+BAO is 11 times better than the nominal Stage II. A triple-method LST, WL+BAO+SN, would have 17 times higher figure of merit than Stage II nominal results.

4. The huge numbers of SNe Ia (tens of thousands per year) with high signal-to-noise well-sampled light curves in 5 optical bands at $z < 1$ would be useful for better understanding the demography of SNe in ways that are complementary to a JDEM SN survey, which would have much smaller numbers of supernovae, though much richer data on each one.

5. An LST mission would make gains in many dimensions of observational phase space, providing great opportunities for science beyond dark energy and the potential to make completely unanticipated discoveries.

6. Performance of an LST mission depends critically on control of systematic errors in both shape measurement and photometry. Potential LSTs should be judged by their ability to control systematic errors.

7. We expect significant progress in understanding the limitations of photometric redshifts and ground-based shape measurements. A clearer picture will emerge over the next few years, especially if the "steps to sharpen forecasts" are pursued vigorously in the interim.

**Advantages of a space-based dark-energy experiment (JDEM):**

1. A single space mission appears capable of providing multiple probes of DE, including sets that can measure expansion and growth histories independently. For example, a well-studied design exists for combining SN+WL on JDEM-class missions, with significant optically-selected and lensing-selected cluster catalogs as a potential byproduct. Missions have also been described that exploit BAO simultaneously with other techniques. None of the DETF White Papers describe missions that would simultaneously exploit three techniques to the levels assumed in the DETF data models, but the JDEM Concept Studies will better explore the breadth achievable in a single JDEM mission.

2. The primary virtue of a space mission is that it offers opportunities to reduce significantly the systematic uncertainties associated with all methods (save BAO, for which a statistical advantage may exist in space). As a result, there is less down-side risk in a space mission than in a purely ground-based program. With our pessimistic projections, a combining WL and SN would provide an improvement of a factor of 7.7 over the Stage II results. Adding BAO, again with pessimistic systematics would bring this to 8.8. With our optimistic projection of systematics, WL+SN provides a factor of 13 increase over Stage II, while adding BAO would bring this to 17.

3. JDEM may be, in this sense, the lower-risk step into Stage IV, because space-based surveys are capable of collecting richer data sets that are more likely to include the key pieces of information needed to ameliorate key systematic errors.

4. A JDEM with NIR capabilities could make use of SNe at higher redshifts than is possible with ground-based visible or radio detection. NIR imaging would also lead to a higher median redshift for WL source galaxies with good photometric redshifts, which strengthens dark-energy constraints.

5. The choice of methods to include on JDEM should be weighted toward those that can best improve systematic errors over their ground-based alternatives, those that can be most effectively combined with ground-based Stage III and IV results, and on those that can be combined into a single mission within the expected JDEM budget.

6. In this regard an x-ray cluster survey stands alone since the other JDEM approaches use widefield optical/NIR imaging/spectroscopy. It is also possible that a visible/NIR WL survey would have as a byproduct a cluster catalog of substantial statistical power.

7. While executing both JDEM and LST would certainly improve upon the constraints obtained from one of the two alone, the degree of improvement is difficult to forecast at this time due to the uncertainties in the systematic-error levels of both.

8. Most proposed forms of JDEM observatory would greatly expand our capabilities for astronomical investigations beyond dark energy, for example by providing unique high-throughput, high-resolution, visible/NIR imaging capabilities.

9. We expect the Concept Studies that will soon be funded by NASA will serve to refine estimates of both the cost and the benefit (*i.e.,* systematic-error reduction) of a space mission. A clearer picture will emerge at this time, especially if the "steps to sharpen forecasts" are pursued vigorously in the interim.

**Advantages of a radio-interferometer dark-energy experiment (SKA):**

1. An interferometric array with sufficiently large collecting area, baselines, field of view, and correlator throughput can efficiently survey the 21-cm-emitting galaxy population over half of the entire sky.

2. The combination of high-precision redshift information with stable interferometric imaging can produce BAO and WL data that is unsurpassed in statistical and systematic quality, over the volume that is accessible to the SKA.

3. The limiting redshift and galaxy density of a SKA survey are quite uncertain, due to our ignorance of the 21-cm emission properties of moderate-redshift galaxies.

4. Current plans for the SKA envision an array capable of operation to much shorter wavelengths than 21 cm. This higher-frequency capability may produce dark-energy science through other investigations, for instance $H_0$ constraints from extragalactic masers, or weak-lensing observations of continuum-detected sources.

5. The SKA would be a large multi-purpose facility, with dark-energy science as just on of many primary goals. Studies of its design and cost are underway, and continued attention to its capabilities for dark-energy science is well justified.

# IX. DETF Technique Performance Projections

The DETF has modeled the dark-energy constraints that will become available from future experiments using the four predominant techniques. This section presents brief descriptions of the DETF models for statistical and systematic uncertainties in each technique; the Technical Appendix details the calculations. Since in most cases the theoretical analyses and the state of experimental art of these techniques are still evolving, the DETF methodologies and estimates necessarily represent a snapshot of our knowledge. We presume that more accurate forecasts will be developed; in particular, ***it should be incumbent on proposers of Stage III and IV projects to derive bounds on systematic and statistical errors for their projects to a level of detail that well exceeds the rough estimates made by this Task Force.***

For each of the four techniques investigated, we delineate the parameters that we use to characterize the performance of future experiments. We then tabulate the parameters that we have taken to represent optimistic and pessimistic estimates of the performance of experiments at Stages II, III, and IV. These postulated future experiments are generic scenarios, neither required to model nor limited to the specific projects that have been proposed or described in the White Papers submitted to the DETF.

# Baryon Acoustic Oscillations Data Models

Before the universe had cooled sufficiently for neutral atoms to persist, the plasma of electrons, protons, other light nuclei, and photons was capable of propagating sound waves. Each density fluctuation initially created in the plasma and dark matter distribution was the source of a wavefront which expanded until the neutralization of the plasma at redshift $z \sim 1000$. The pattern of initial perturbations and expanding wavefronts is seen in the CMB, and is ultimately imprinted on the matter distribution. The primary manifestation of these *baryon acoustic oscillations (BAO)* is a feature at the "sound horizon" length $r_s$, which is the distance traveled by the acoustic waves by the time of plasma recombination. The sound horizon is known from CMB measurements, thus providing a standard cosmic meter stick.

The meter stick can be measured both in an orientation transverse to the line-of-sight and oriented along the line-of-sight. The sound horizon becomes apparent in the two-point correlations between galaxies. In the transverse orientation, the angle subtended by the sound horizon feature gives a measurement of $d_A(z)/r_s$, *i.e.* the angular-diameter distance to the redshift $z$ at which it is observed. Measurement of the radial scale of the sound-horizon feature gives $H(z)r_s$.

An advantage of BAO is that it does not require precision measurements of galaxy magnitudes, though if photo-$z$'s are used then precision in galaxy colors is important. In contrast to weak lensing, BAO does not require that galaxy images be resolved; only their three-dimensional positions need be determined.

The statistical power of a BAO experiment to measure $d_A(z)$, $H(z)$, and hence dark energy, depends on the volume of sky surveyed, the range of $z$, and the precision with which $z$ is measured. The survey volume determines the level of sample variance in the power spectrum or correlation function that is used to identify the sound-horizon scale. If too few galaxies are used, shot noise will dominate the sample variance. We will assume that future BAO surveys will be designed for a density $n$ of surveyed galaxies to yield $nP \gtrsim 3$, so that shot noise is a minor contributor.

Measurements of BAO fall into two classes: those using spectroscopic or other high-accuracy measurements of $z$; and those that determine $z$ photometrically. Results of both sorts are now available from SDSS (Eisenstein *et al.* 2005; Padmanabhan *et al.*, 2005), demonstrating the feasibility of the techniques. The spectroscopic and photometric approaches have contrasting strengths. The large surveys using photo-$z$'s would have little or no information in the radial direction. Spectroscopy of large numbers of distant galaxies is, however, much more expensive than imaging them, so at fixed cost a photo-$z$ survey can survey a larger volume than a spectroscopic survey. A SKA 21-cm survey would have spectroscopic redshifts inherently over its full survey volume.

To compute the expected statistical uncertainties of BAO experiments, we first posit the solid angle, redshift coverage, and redshift precision to be expected. The resultant accuracy on the determination of $d_A(z)/r_s$ and $H(z)r_s$ are taken from the estimates of Blake et al. (2005), as described in greater detail in the Technical Appendix.

Systematic errors in the spectroscopic BAO method are more likely to arise from the underlying theory than from the measurement process. The acoustic-oscillation information at higher wave numbers is degraded by non-linear effects in the growth of structure. Theoretical modeling will be limited by our understanding of "bias," the difference between the distribution of galaxies (the measured quantity) and the distribution of matter (the predicted quantity). Bias and non-linearities in the *velocities* of galaxies will degrade the accuracy of sound-horizon determinations in the radial (redshift) direction.

Our pessimistic model for the cumulative effects of errors in the theory for non-linearities and bias assumes independent 1% uncertainties in each $d_A(z)/r_s$ and $H(z)r_s$ in each bin of width 0.5 in $z$. These systematic errors are added in quadrature to the statistical errors. An optimistic view is that future theoretical developments will reduce these uncertainties well below future statistical errors.

There are additional issues to be addressed for photo-$z$ BAO surveys. In practice, a large number of galaxies will be studied both photometrically and spectroscopically and this sample will be used to calibrate the photo-$z$'s. This process will have both statistical and systematic errors. The dispersion $\sigma_F$ in difference between the true $z$ and that inferred photometrically (more precisely, $\sigma_F{}^2 = \text{Var}(z\text{-}z_{\text{phot}})/(1+z)^2$) is a parameter in the statistical-error model of Blake *et al.*, as it degrades the line-of-sight resolution of the acoustic scale. The simplest systematic error is an overall bias in the photometric redshift scale. We presume the bias is bounded by a spectroscopic survey of $N = 1000$ galaxies per redshift bin, so that the bias has a prior uncertainty of $\sigma_F (1+z) / \sqrt{N}$.

In addition to a simple bias, catastrophic errors can occur when photometry gives an ambiguous redshift as a consequence of the diversity of galaxy types. Our models specify both the dispersion and the bias as a function of $z$, but do not explicitly include the possibility of catastrophic error. More generally, we do not look at any non-Gaussian distribution in the discrepancy between the true and measured $z$. We note that BAO surveys require redshift information for only a small fraction of the galaxies in a given volume in order to maintain $nP \gtrsim 3$. Hence a photo-$z$ survey can choose to use only those galaxies with the most reliable redshifts, making the calibration task easier than for weak-lensing surveys.

The specific experimental configurations analyzed by the DETF are as follows:

We consider no **Stage II** BAO experiment.

The **Stage III spectroscopic** experiment would cover 2000 square degrees with $0.5 < z < 1.3$, plus 300 square degrees with $2.3 < z < 3.3$. The interval between $z$ of 1.3 and 2.3 is less amenable to an efficient terrestrial spectroscopic redshift survey. This experiment would obtain $10^7$ spectra. While this ambitious program is included in Stage III, we would not expect it to be completed for a long time, perhaps by 2016.

The **Stage III photometric** BAO experiment would cover 4000 square degrees, with $z$ from 0.5 to 1.4. We take photo-$z$ dispersion $\sigma_F$ to be 0.01 in the optimistic alternative and 0.05 in the pessimistic alternative. The RMS bias per redshift bin is taken to be $\sigma_F (1 + z) / \sqrt{N}$, with $N = 1000$.

For the **Stage IV ground-based (LST)** experiment, we consider a large photo-$z$ survey, covering 20,000 square degrees, with $0.2 < z < 3.5$. The photo-$z$ capabilities are again the optimistic and pessimistic scenarios used for the Stage III photo-$z$ BAO experiment.

The **Stage IV SKA** measurement of BAO, which by its very nature has spectroscopically measured $z$'s, assumes coverage of 20,000 square degrees and $0.01 < z < 1.5$. The redshift range corresponds to the median of three possible models given by Abdalla & Rawlings (2005) for the evolution of the 21-cm luminosity function.

For the **Stage IV space-based** BAO experiment, we postulate spectroscopic coverage of 10,000 square degrees over $0.5 < z < 2$.

| Data Model | $f_{sky}$ | $z$ range | $\sigma_F$ | $N$ (spectra per $z$ bin) |
|---|---|---|---|---|
| BAO-IIIs | 0.05 | 0.5–1.3 | … | … |
|  | 0.0075 | 2.3–3.3 |  |  |
| BAO-IIIp-o | 0.1 | 0.5–1.4 | 0.01 | 1000 |
| BAO-IIIp-p | 0.1 | 0.5–1.4 | 0.05 | 1000 |
| BAO-IVLST-o | 0.5 | 0.2–3.5 | 0.01 | 1000 |
| BAO-IVLST-p | 0.5 | 0.2–3.5 | 0.05 | 1000 |
| BAO-IVSKA | 0.5 | 0.01–1.5 | … | … |
| BAO-IVS-p | 0.25 | 0.5–2.0 | … | … |

# Baryon Acoustic Oscillations 95% C.L. Contours

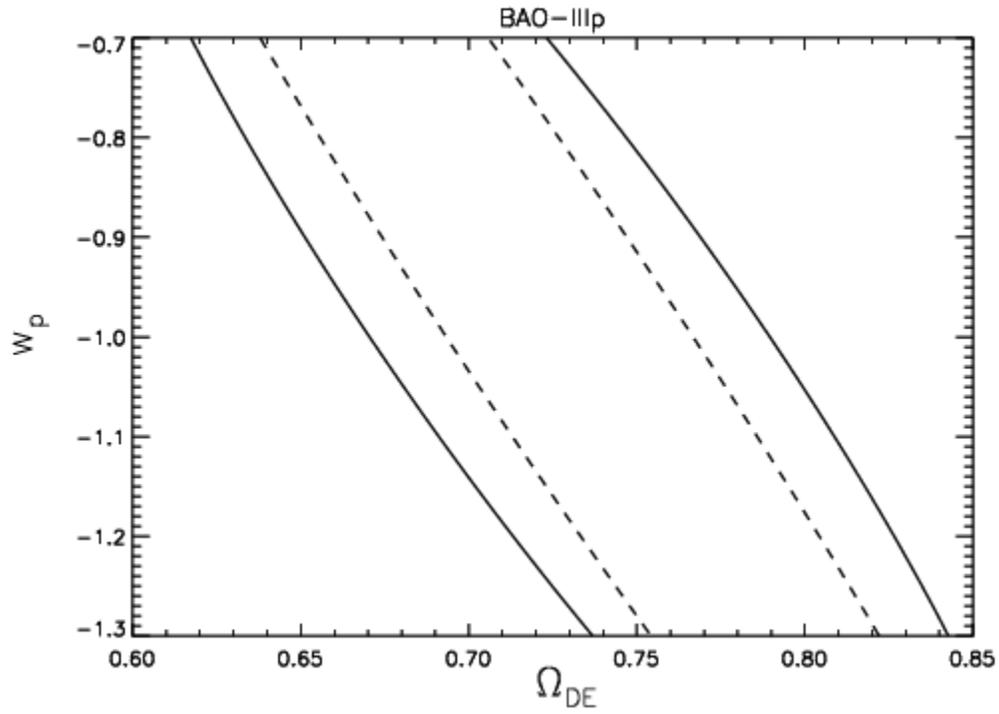

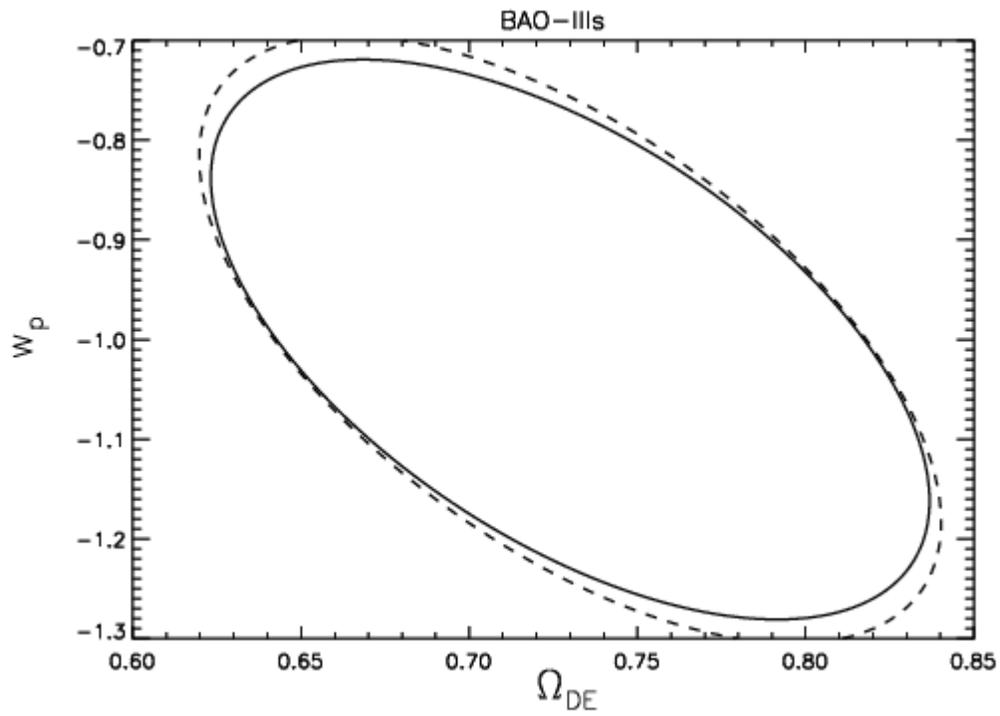

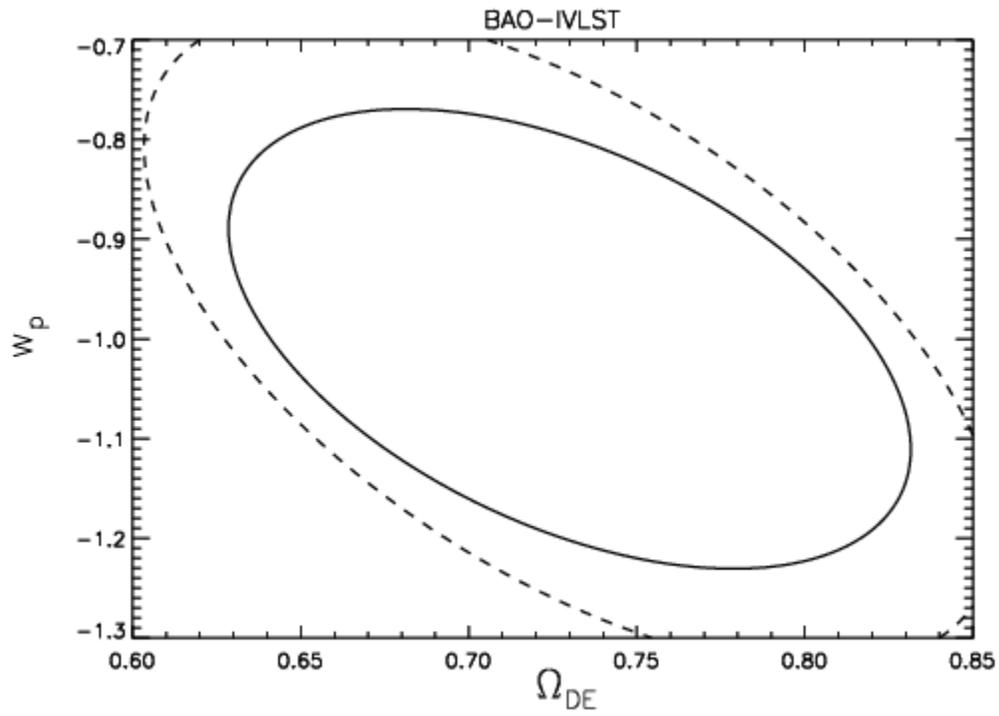

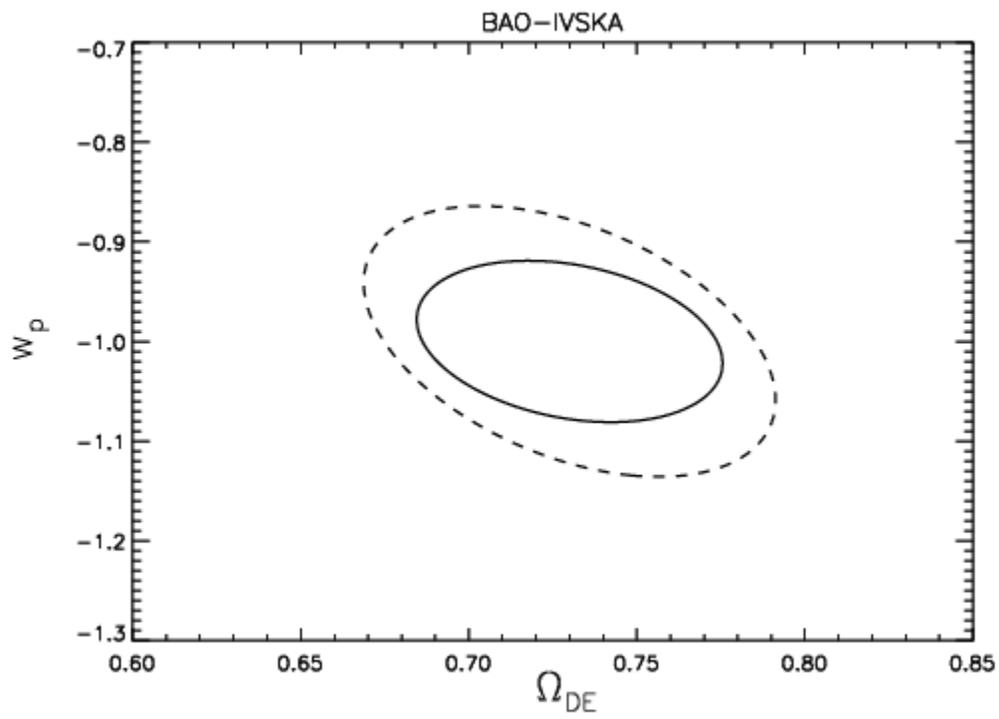

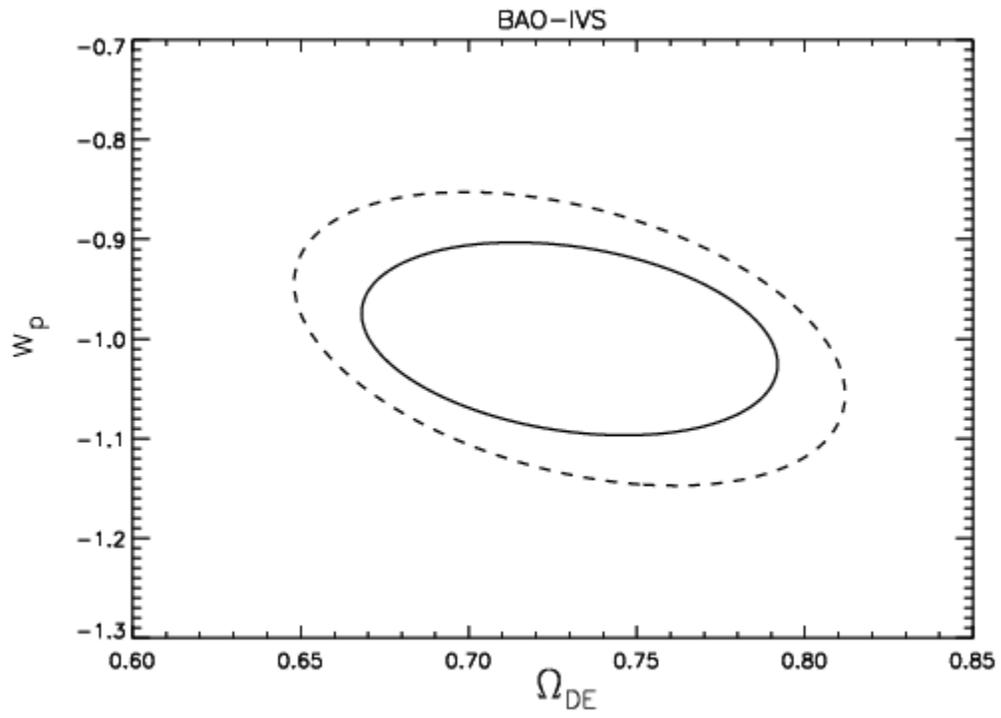

# Cluster Data Models

The number of rare clusters of galaxies as a function of their mass $M$ is exponentially sensitive to the linear density field, and hence its rate of growth, as well as linearly sensitive to the volume element. Thus, the CL method is sensitive to dark energy both through $dV/d\Omega\,dz \propto D^2(z)\,/H(z)$ and through the growth rate of structure.

The statistical errors in a cluster-counting experiment are specified by the survey redshift range, solid angle, and mass threshold. The latter two may be functions of redshift, and one may specify the total number of detected clusters in lieu of one of these parameters.

We will assume that cosmological $N$-body simulations will in the future calibrate the number density or mass function to the required accuracy (see below). The main challenge for using cluster counts for dark energy tests is that the mass of a cluster is not directly observable. A selection that is based on a threshold in an observable property of clusters – *e.g.,* Sunyaev-Zel'dovich flux decrement, x-ray temperature and surface brightness, optical richness, or lensing shear – will carry a corresponding selection function in the mass domain. On the other hand it is this richness in the available observables of a cluster that provides the opportunity to calibrate the selection empirically and the checks against systematic errors in the modeling. Furthermore, a cluster survey yields more dark-energy observables than the cluster abundance alone. For example, the spatial clustering of clusters also contains BAO information.

Uncertainties in the mass selection propagate from uncertainties in the mean relationship between the cluster observable and mass as well as uncertainties in the scatter, and, more generally, the mass-observable distribution. Instrumental effects and contamination from point sources can further distort the selection, especially near threshold. Given the steepness of the mass function near the detection threshold, characterizing the mass selection function is the main obstacle to extracting dark energy information from cluster counts.

Our approach to forecasting the performance of cluster dark-energy surveys correspondingly places the emphasis on determining the level of control on the mass selection required to reach a given a given level of dark energy performance. For illustrative purposes we examine two control functions: the mean of the mass-observable relation and its variance. We allow these functions to evolve in redshift arbitrarily by representing them as two independent nuisance parameters per redshift interval of $\Delta z = 0.1$ (see the Technical Appendix for details).

Our projections for optimistic and pessimistic levels of control on these parameters are currently highly uncertain and *it is critical that proposed surveys demonstrate their expected control of the mass-observable relation.* The pessimistic target reflects determinations of the control parameters that are likely to be available internally through "self-calibration" in any one given survey without multi-wavelength follow up. Self-calibration techniques, still in development, use the shape of the mass function, and the spatial clustering of clusters–both of which can be accurately calibrated as functions of mass using *N*-body simulations–to constrain the mass-observable relation and cosmological parameters simultaneously. Our approach to self calibration is to consider it as one of many possible priors on the selection function. The optimistic target reflects determinations that may be available through multiple mass determinations, *e.g.,* with high resolution x-ray follow up that can determine the mass and physical properties of the cluster gas, weak-lensing cluster-shear correlation functions for clusters near threshold, or advances in hydrodynamic simulations. This optimistic target is highly uncertain. There are projections in the literature that both exceed and fall short of this target by factors of a few.

Our control parameters are illustrative but not exhaustive. For example, non-Gaussian tails in the mass-observable relation and point-source contamination must be sufficiently controlled such that the abundant low-mass clusters do not cause uncertainties for the high-mass counts. For example (see the Technical Appendix) for typical surveys we consider, uncertainties in the mass selection at 1/3 of the mean threshold must be less 1% in order to measure $w_p$ to 10%. Likewise, the mass function itself must be calibrated with *N*-body simulations of volumes at least as large as the planned survey volume. Accurate cluster redshifts must be determined from follow-up observations, *e.g.,* optical photometric redshifts. We have not included uncertainties from photometric redshifts since the requirement per galaxy is less stringent here than in the other dark-energy techniques: red cluster galaxies are well suited to the photo-*z* technique, and random variance is reduced by the number of red galaxies available in each cluster.

Given our focus on systematic control, we do not attempt to model in detail the statistical errors of any one proposed cluster detection method or survey. Instead we describe a generic cluster survey that detects clusters to some mean mass threshold that is independent of cosmology and redshift. *A more detailed calculation of a specific survey may therefore achieve statistical errors that are moderately better or worse than our fiducial projections* for a given total number of detected clusters, but we expect the requirements for control over systematic errors to scale appropriately.

We note also that a cluster survey based on detection by gravitational shear will face substantially different systematic errors than x-ray, SZ, or optical cluster surveys. In this case the mass-observable relation is well defined, but it is the projected mass rather than the virial mass which is detected, and the impact of this difference on dark-energy constraints is not yet fully understood.

For **Stage II**, we model a 200 square-degree survey to a mean threshold of $10^{14} h^{-1} M_{\odot}$ with a total of approximately 4,000 to 5,000 clusters. For the projection we take the mean and variance of the mass-observable relation to be determined to 27% in redshift bins of $\Delta z = 0.1$. This choice corresponds to degradation in errors in $w_p$ of a factor of $N$ = 3 from the purely statistical uncertainty.

For **Stage III**, we model a 4,000 square-degree survey to a mean threshold of $10^{14.2} h^{-1} M_{\odot}$ with a total of about 30,000 clusters. For the pessimistic projection we take the control parameters to be determined to 14%, or $N$ = 3. For the optimistic projection we take the control parameters to be determined to 2%, corresponding to $N = 1.4$, or equal statistical and systematic errors.

For **Stage IV**, we model a 20,000 square-degree survey to a mean threshold of $10^{14.4} h^{-1} M_\odot$ with a total of about 30,000 clusters. For the pessimistic projection we take the control parameters to be determined to 11%, or $N = 3$. For the optimistic projection we take the control parameters to be determined to 1.6%, corresponding to $N = 1.4$, or equal statistical and systematic errors.

Note that Stage IV is conservatively intended to represent a survey that sacrifices depth for detailed measurements and control over systematic errors. Although we have kept the optimistic and pessimistic degradation factors constant between Stages III and IV to reflect the range of possibilities, we expect that a Stage IV survey will achieve a level that is closer to the optimistic projection. Moreover, if systematic errors in Stage III are demonstrably under control, then this Stage IV projection does not reflect an ultimate limitation since the statistical errors can be further reduced by lowering the mass threshold. Current projections in the literature employ up to 3 times the numbers assumed here (see the Technical Appendix for the scaling of errors with cluster numbers). Likewise, although we have kept the optimistic and pessimistic degradation factors constant between Stages III and IV to reflect the range of possibilities we expect that a Stage IV survey will achieve a level that is closer to the optimistic projection.

| Data Model | $f_{sky}$ | Mass threshold ($h^{-1} M_\odot$) | Cluster count | RMS error in mean/variance of mass, per $z$ bin | Degradation factor for $\sigma(w_p)$ |
|---|---|---|---|---|---|
| CL-II | 0.005 | $10^{14}$ | 4000 | 27% | 3 |
| CL-III-p | 0.1 | $10^{14.2}$ | 30,000 | 14% | 3 |
| CL-III-o | 0.1 | $10^{14.2}$ | 30,000 | 2% | 1.4 |
| CL-IV-p | 0.5 | $10^{14.4}$ | 30,000 | 11% | 3 |
| CL-IV-o | 0.5 | $10^{14.4}$ | 30,000 | 1.6% | 1.4 |

# Cluster 95% C.L. Contours

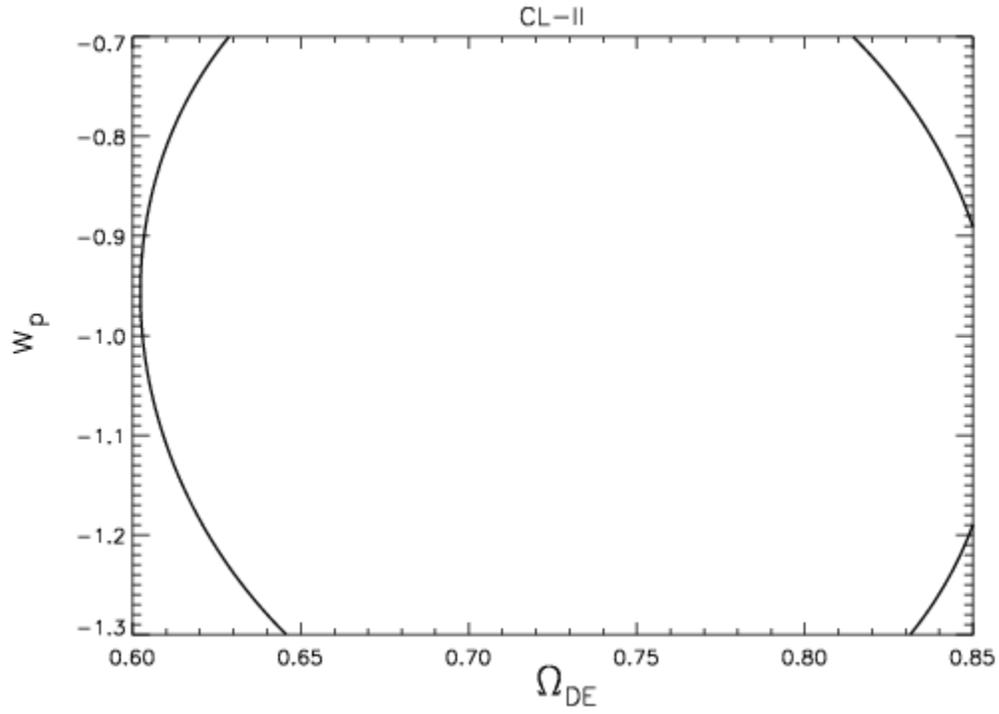

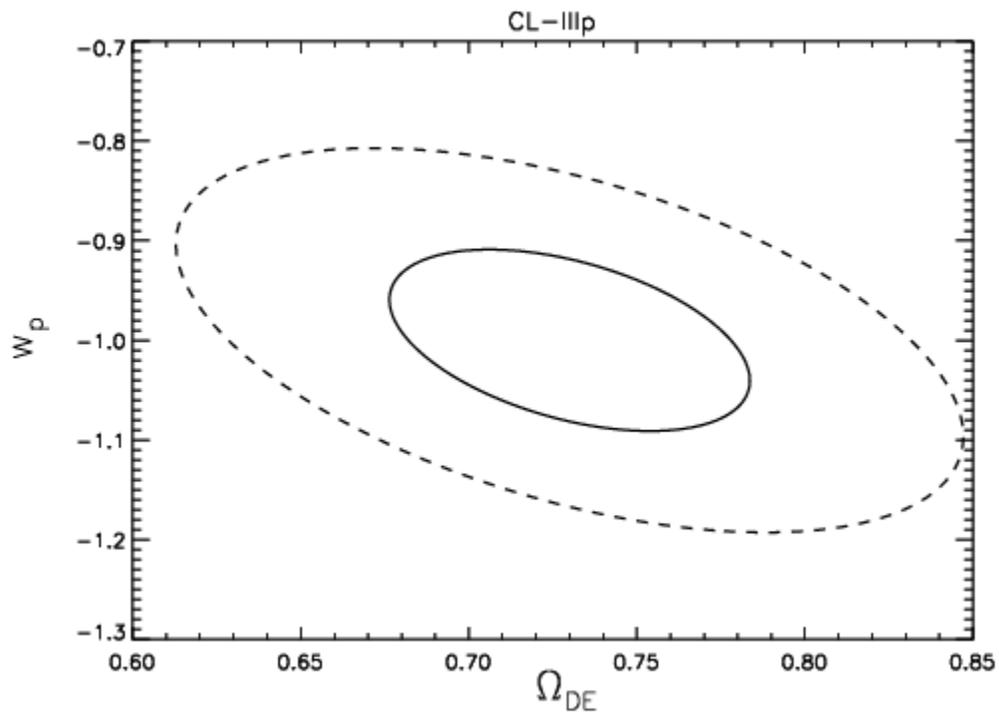

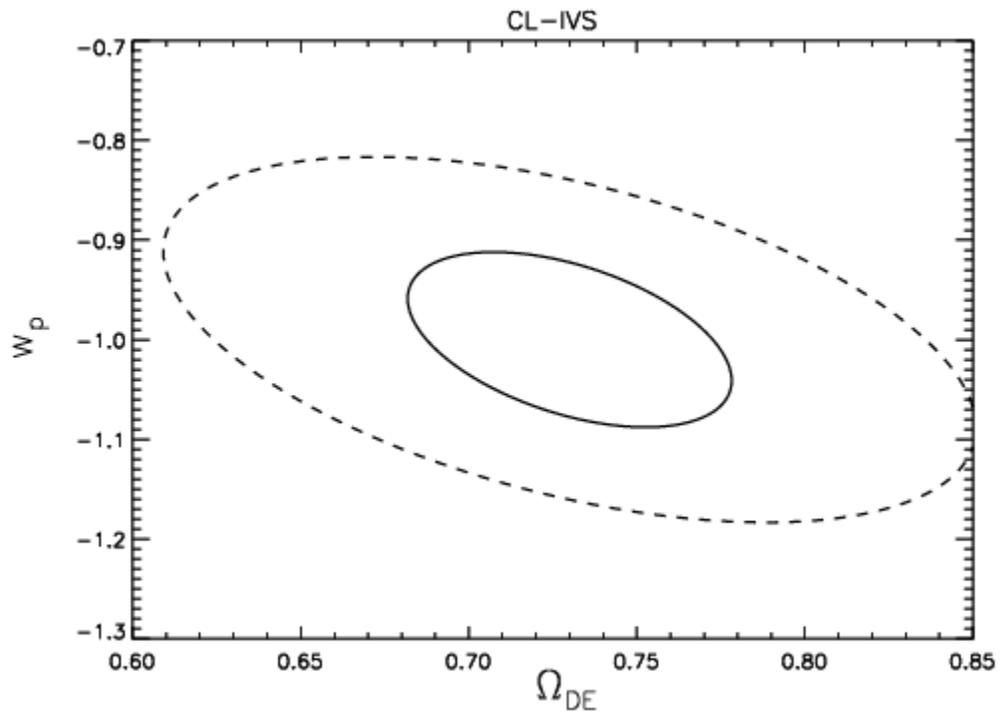

# Supernovae Data Models

To date, Supernovae Type Ia (SNIa) provide the most direct indication of the accelerating expansion of the universe. To a first approximation, these supernovae all have the same intrinsic luminosity. Thus measuring both their redshift and their apparent peak flux gives a direct measurement of their distance (more properly, their luminosity distance) as a function of redshift $z$. In practice, the peak luminosities of SNIa are not identical, but the variations do strongly correlate with the rate at which the supernovae decline in brightness. The observed rates of decline yield corrections to the peak brightnesses, reducing the dispersion in SNIa fluxes at fixed redshift. Measures of the SNe colors can similarly reduce dispersion due to variable dust extinction in the host galaxies. Obviously, calibration of the flux is crucial to precision measurements. In addition, each supernova must be studied carefully enough to determine whether it is truly of type Ia.

For a SNIa survey with spectroscopic followup, the statistical uncertainties in $D(z)$, and hence in dark energy, are determined by the number of observed SNe, their redshift distribution, and the standard deviation $\sigma_D$ in the SN absolute magnitude after all corrections for decline rate and extinction. Undoubtedly there will be further luminosity indicators discovered in the future as well. To the "intrinsic" dispersion $\sigma_D$ we must also add measurement error, which will be denoted as $\sigma_m$ when it is treated as a distinct quantity.

In addition to surveys covering redshift out to one or beyond, there is a need to anchor the Hubble diagram at the lower end. For this, we postulate a sample of 500 supernovae at low $z$ but still in the Hubble flow. The low-$z$ events are important because the (standardized) absolute magnitude of the SNIa is not known *a priori*.

An alternative to spectroscopic followup is to rely on photometrically determined $z$'s. Without the need for spectroscopy, it is possible to collect very much larger samples. However, the cost is loss of resolution in $z$, the loss of some detail that would help reduce dispersion, and possible contamination with supernovae of other types. For scenarios of SNIa surveys using photo-$z$'s, their dispersion is taken as $\sigma_z = \sigma_F (1+z)$, where $\sigma_F = 0.01$ (0.05) in optimistic (pessimistic) projections.

Systematic errors in SNIa measures of $D(z)$ will arise from two dominant sources, for which neither the functional form nor the amplitude are straightforward to forecast. First, wavelength-dependent errors in the astronomical flux scale propagate into $D(z)$ as the observed wavelength of the SNe redshifts through the visible and NIR spectrum. Second, any shift with redshift in the properties of the SNe or their host extinction propagates into $D(z)$, to the extent that it is not recognized and corrected through the use of decline rates or other observables. If, for example, some portion of the 0.10-0.15 mag "random" scatter in SNIa intrinsic luminosities is attributable to some physical variable that is systematically different in high-redshift events, then dark-energy results are biased unless this variable can be identified and measured in both nearby and distant events.

Given the ill-constrained nature of systematic errors in SNIa surveys, we adopt a simple generic model for them. We first posit an unknown offset $\Delta m$ in photometric calibration between the nearby sample and the distant sample, and then place a Gaussian prior of dispersion 0.01 mag on this offset. The possibility that there is an undiagnosed $z$-dependent "evolution" in the brightness of supernovae or extinction properties is simulated by adding to the expression for the observed magnitude a term $az + bz^2$, where the parameters $a$ and $b$ are drawn from independent Gaussian distributions with standard deviations $\sigma_a = \sigma_b = 0.01 / \sqrt{2}$ $(0.03 / \sqrt{2})$ in optimistic (pessimistic) scenarios.

In addition, we suppose that the calibration of the photo-$z$'s would be done in bins of $z$, with width $\Delta z = 0.1$. Each bin is susceptible to a bias, for which we take the prior $\sigma_z / \sqrt{N}$, with $N = 100$.

For **Stage II**, we take parameters representative of the Supernova Legacy Survey (SNLS) or the ESSENCE survey. The redshifts are determined spectroscopically for 700 supernovae, with $0.1 < z < 1.0$. These are supplemented by the postulated 500 nearby supernovae. The dispersion in observed magnitude is the sum in quadrature of a fixed $\sigma_D = 0.15$ and a second piece $\sigma_m$, which is 0.02 up to $z = 0.4$ but then increases until it reaches 0.3 at $z = 1$.

For **Stage III**, we consider both spectroscopic and photometric surveys. For the spectroscopic survey we simply scale up the SNLS program to 2000 supernovae and suppose that systematics can be reduced so that $\sigma_D = 0.12$, while the $z$-dependent measurement noise $\sigma_m$, is unchanged. The model for the photometric survey uses the same number of supernovae and the same distribution in $z$, but adds uncertainties specified by $\sigma_F = 0.01$ or 0.05.

For the **Space-based Stage IV** program, the $z$'s of the supernovae are assumed to be measured spectroscopically. The 2000 supernova are distributed over the range $z = 0.1$ to $z = 1.7$. $\sigma_D$ is reduced to 0.10 independent of $z$ with $\sigma_m = 0$.

For the **Ground-based Stage IV** program, we postulate 300,000 photometrically measured supernovae. The same dispersion, $\sigma_D = 0.10$, without a $z$-dependence, is used, with photo-$z$ errors represented by the choices $\sigma_F = 0.01$ or 0.05.

# Supernova 95% C.L. Contours

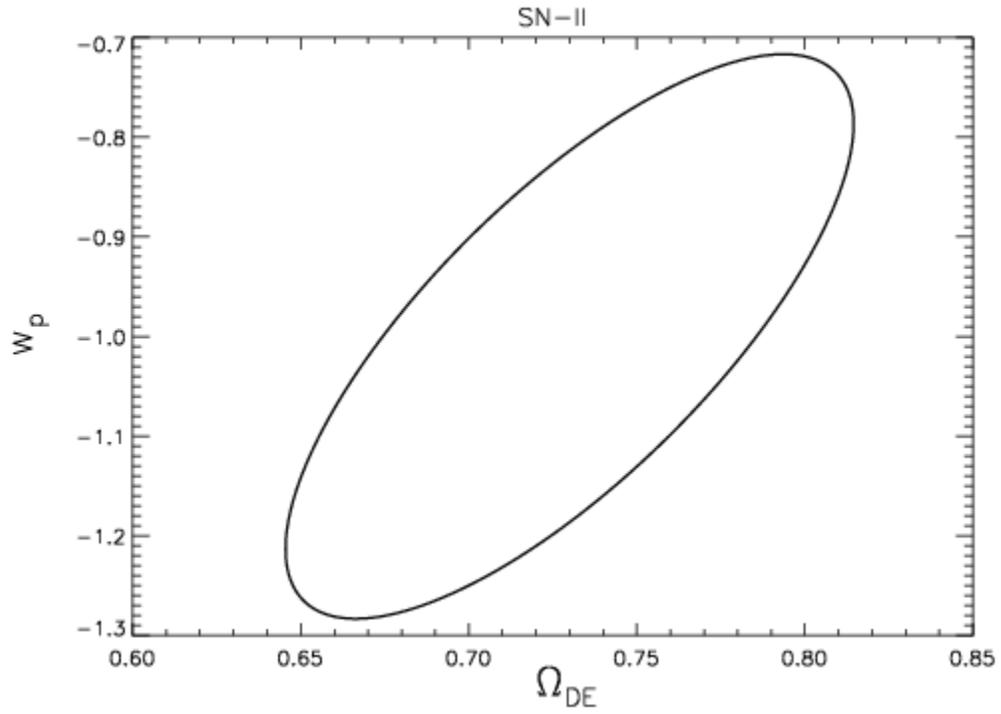

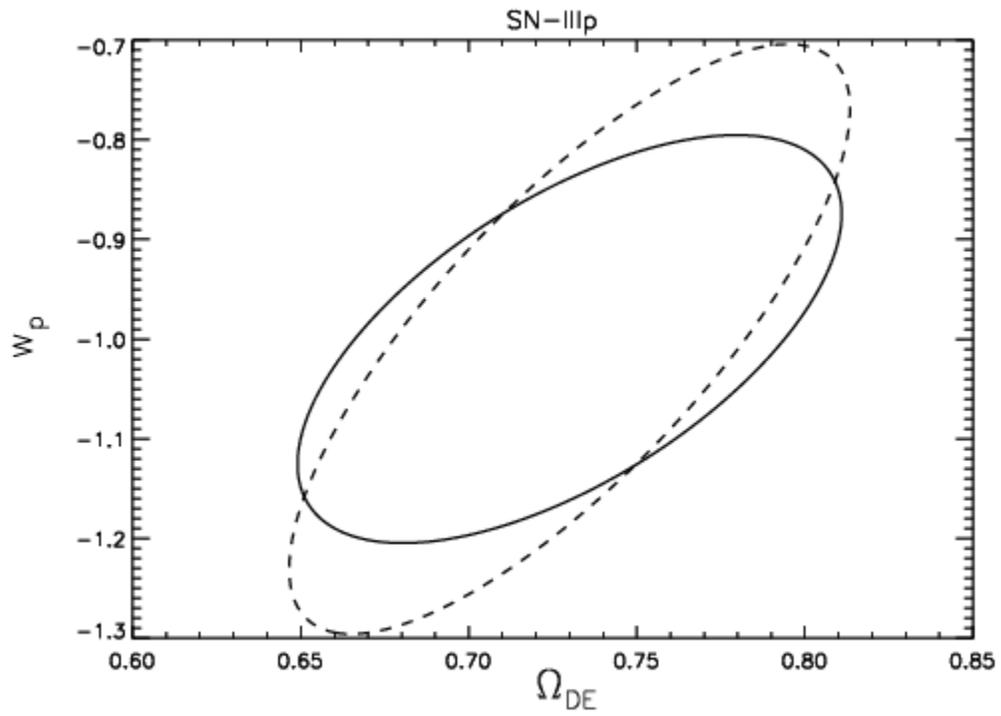

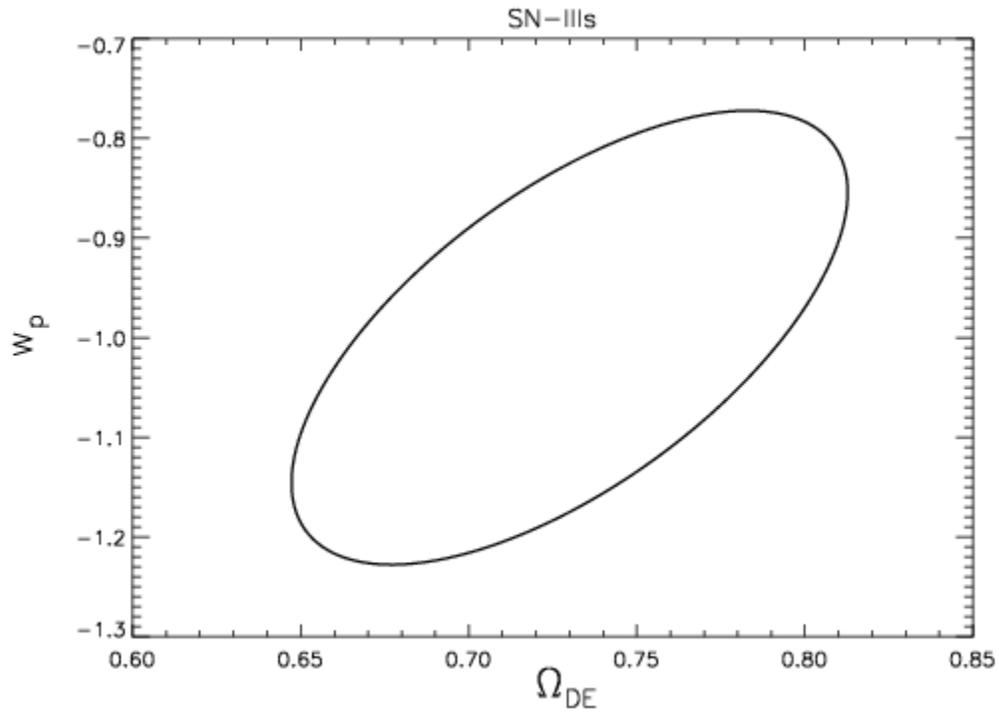

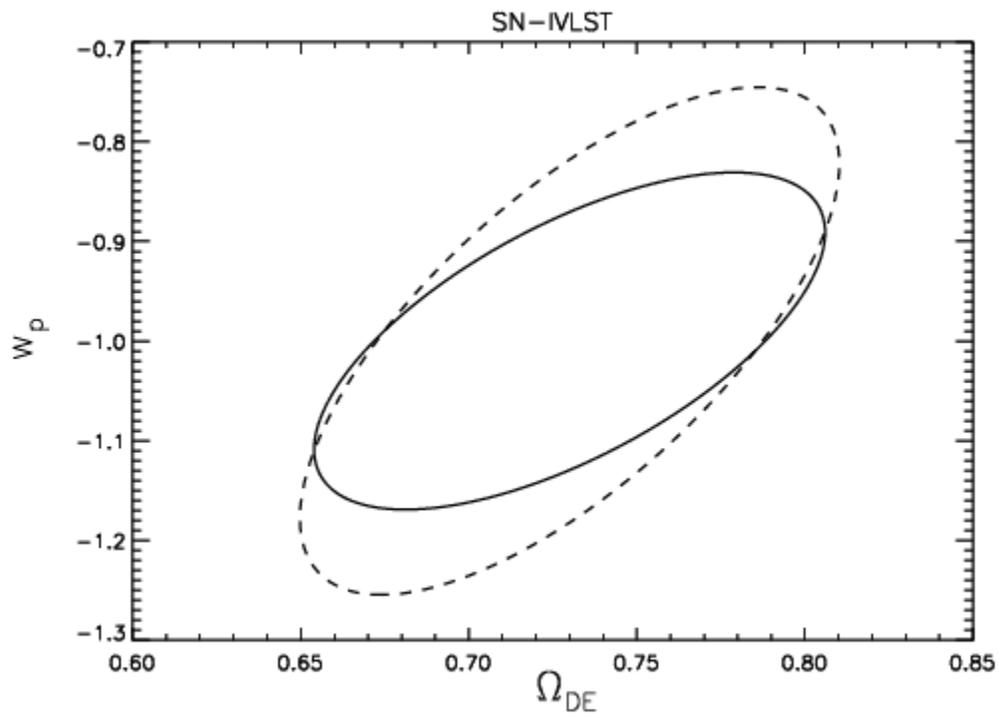

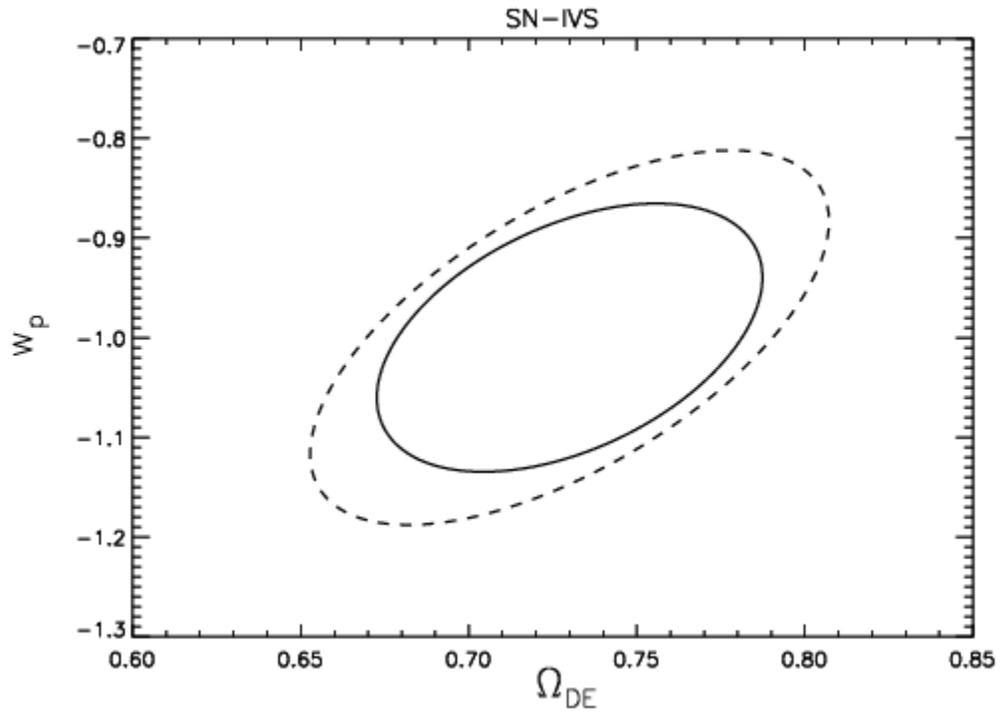

# Weak Lensing Data Models

The distortion pattern imparted on the images of distant light sources is sensitive to distance ratios between observer, lens, and source redshifts, as well as to the growth history of lensing mass structures between observer and source epochs. Schemes for exploiting the lensing effects to extract dark-energy information are diverse in several respects:

- The source of background photons: in present investigations, these are galaxies, in the range $0 < z < 5$ where they can be detected in great abundance. Future observations should also allow detection of lensing effects on CMB photons originating at $z \sim 1100$, and also photons emitted at 21 cm by neutral hydrogen in the "reionization epoch" of $6 < z < 20$.

- The wavelength of detection: galaxy shapes can be measured in large numbers either by optical/NIR detection, or by interferometric observations in the radio. CMB photons must of course be observed in the radio/mm regime, and 21-cm emission from reionization is redshifted well into the radio regime.

- The source and extent of redshift knowledge for sources (and lenses): at the very least, the distribution $dn/dz$ of sources must be known to high accuracy to extract dark-energy constraints. Much more powerful are surveys with redshift information on a source-by-source basis, enabling "tomography" by examination of the dependence of lens effects on source redshift. Low-precision photometric redshifts are available for galaxies from broadband visible/NIR imaging. Much higher precision and accuracy are available for the CMB, and for galaxies or reionization-era observations in the 21-cm line of hydrogen.

- The signature of lensing to be detected: the shear of source shapes is presently the dominant signature, but it may be possible to make use of lensing magnification as well.

- The statistics formed from the lensing signature(s):
  - The power spectrum (PS) of the lensing signal (including cross-spectra between different source-redshift bins) is the primary observable for Gaussian lensing fields.
  - The cross-correlation (CC) of the lensing signal with identified foreground structure provides additional information, even in the Gaussian regime.
  - The bispectrum (BS) or other higher-order correlators of the lensing signal carry significant information on scales where the mass distribution has become non-linear and hence non-Gaussian.
  - Cluster counts (CL) can be produced by counting peaks in the density inferred from the lensing signals.

The development of WL techniques is ongoing, and there is as yet no means for estimating the total amount of dark-energy information available from WL data. We restrict ourselves to the family of two-point statistics (PS and CC); additional WL constraints will undoubtedly be available from the non-Gaussian BS and CL data, but the theoretical tools for combining their information with the 2-point statistics are not yet in hand. We also consider only shear measurement, not magnification, because the experimental difficulties of the latter have not yet been surmounted, and the resultant dark-energy constraints would be improved by less than about $\sqrt{2}$. Only scenarios including tomographic information are constructed, since such experiments would have much stronger dark energy constraints and/or immunity to small errors in knowledge of the redshift distribution.

All WL case studies presume that lensing is detected by the shear of typical galaxy images. We examine cases in which these shapes are to be measured by broadband visible/NIR imaging, with photometric redshift information; or by detection of 21-cm emission by a Square Kilometer Array with the continental-scale baselines required to resolve galaxies at $z > 1$.

The statistical power of such surveys is determined by these parameters:

- The fraction $f_{sky}$ of the full sky which is surveyed.
- The noise density of the shear measurement $\sigma_\gamma^2 / n_{eff}$, where $\sigma_\gamma$ is the uncertainty in each component of the shear $\gamma$ per perfectly-measured galaxy, and $n_{eff}$ is the sky density of perfectly-measured galaxies which would yield the same shear noise as the (imperfectly) measured ensemble of galaxies.
- The redshift distribution of source galaxies $dn/dz$. For the visible/NIR surveys, $dn/dz$ is parameterized by the median source redshift $z_{med}$ and we assume the form $dn/dz \propto (z/z_0)^2 \exp\left[-(z/z_0)^{1.5}\right]$. Values of $\sigma_\gamma$, $n_{eff}$, and $z_0 = z_{med}/\sqrt{2}$ are estimated from current results, analysis of deep HST imaging, and extrapolation of redshift survey data to fainter magnitudes. For the SKA survey scenario we adopt the range of redshift distributions specified by models A, B, and C of Abdalla & Rawlings (2005).
- The fiducial correlation coefficient $r$ between the galaxy distribution and the dark-matter distribution, which is relevant to the CC method (*cf.*, Bernstein 2006). We assume $r = 0.5$, which is in the range suggested by halo-model calculations (Hu & Jain 2004).

Systematic errors will certainly be important in determining the dark-energy power of WL surveys. We include the following systematic errors in our projections:

- The theoretical power spectrum $P(k,z)$ of dark matter will be calculated from future *N*-body simulations, but baryonic physics will render these predictions inexact. This "theory systematic" ultimately limits the utility of the PS statistic. We presume that $P(k,z)$ in each $(k,z)$ bin will be uncertain by $f_P = 0.5$ of the difference between baryonic and no-baryon power spectra as estimated by Zhan & Knox (2004); see also Jing *et al.* (2006). Bins are 0.5 wide in $\log_{10}(k)$, and 0.15 wide in $\ln(1+z)$.
- The intrinsic correlations of galaxy shapes with each other and with local density (Hirata & Seljak 2004) are left as free parameters in each $(k,z)$ bin.
- The shear measurement is assumed to be miscalibrated by a factor $(1+f_{cal})$ that varies independently for each redshift bin. For a given scenario, a Gaussian prior of width $\sigma(f_{cal})$ is placed on the calibration factor of each redshift bin. The pessimistic scenarios assume that this calibration uncertainty is no better than that demonstrated by the best currently available methods in the STeP tests (Heymans *et al.* 2006).

- The bias and correlation coefficients between galaxies and dark matter are presumed to take different, unknown values in each $(k,z)$ bin. This again is relevant to the CC statistic.

- For the photo-$z$ surveys, the dominant systematic errors are uncertainties in the relation between photometric redshift and true redshift, $p(z|z_{phot})$. For WL purposes, it is not necessary that this distribution be very narrow, but rather that it be very well known (Ma, Huterer, & Hu 2006). We consider only the simplest possible error in this distribution, namely that the mean $z$ be biased with respect to $z_{phot}$ for galaxies in some redshift bin. We quantify this by $\sigma_{\ln(1+z)}$, the RMS bias in $\ln(1+z)$ for each bin of width 0.15 in $\ln(1+z)$. Pessimistic scenarios presume that this does not improve beyond the performance attained for photo-$z$'s of luminous red galaxies in the Sloan Digital Sky Survey (Padmanabhan *et al.* 2005).

A full formulation of the WL forecasting methodology is given in the Technical Appendix. The Table below gives the values of all parameters assumed for each WL scenario studied by this Task Force. A full exploration of the response of WL dark-energy constraints to choices of parameters is well beyond the scope of this document, but we note a few points.

First, WL data offers a rich variety of statistics, with cross-spectra between every pair of source-redshift bins, cross-correlations between every source and lens redshift bin, and a range of angular scales for every such 2-point statistic. These respond to dark energy and to systematic errors in different ways, making it possible to distinguish dark-energy signals from systematic errors at mild penalties in statistical accuracy. The exception is the photo-$z$ biases (*cf.* Huterer *et al.* 2006). Redshift errors lead us to make very precise measures of distance and growth, but assign them to the incorrect redshift and hence mis-characterize the dark energy. The SKA survey, with spectroscopic redshifts for each galaxy, is immune to this error; however the source distribution is currently poorly known.

Our neglect of BS and CL statistics, CMB or reionization-epoch lensing information, and magnification information renders our forecasts conservative. On the other hand our treatment of photometric-redshift errors is very simplistic, *e.g.,* Ma, Huterer, & Hu (2006) demonstrate that the variance as well as bias of the photo-$z$ estimator must be known to high accuracy in order to avoid degradation of dark-energy constraints.

| Data Model | $I_{sky}$ | $\sigma_\gamma$ | $n_{eff}$ (arcmin$^{-2}$) | $z_{med}$ | $\sigma(f_{cal})$ | $\sigma_{\ln(1+z)}$ |
|---|---|---|---|---|---|---|
| WL-II | 0.0042 | 0.25 | 15 | 1.0 | 0.02 | 0.02 |
| WL-IIIp-p | 0.1 | 0.25 | 15 | 1.0 | 0.01 | 0.01 |
| WL-IIIp-o | 0.1 | 0.25 | 15 | 1.0 | 0.01 | 0.003 |
| WL-IVLST-p | 0.5 | 0.25 | 30 | 1.0 | 0.01 | 0.01 |
| WL-IVLST-o | 0.5 | 0.25 | 40 | 1.2 | 0.001 | 0.001 |
| WL-IVSKA-p | 0.5 | 0.3 | (b) | (b) | 0.0001 | -0- |
| WL-IVSKA-o | 0.5 | 0.3 | (a) | (a) | 0.0001 | -0- |
| WL-IVS-p | 0.1 | 0.3 | 100 | 1.5 | 0.003 | 0.003 |
| WL-IVS-o | 0.1 | 0.3 | 100 | 1.5 | 0.001 | 0.001 |

(a) *dn/dz* from Abdalla & Rawlings Model A for 21-cm luminosity evolution.
(b) *dn/dz* from Abdalla & Rawlings Model B for 21-cm luminosity evolution.

# Weak Lensing 95% C.L. Contours

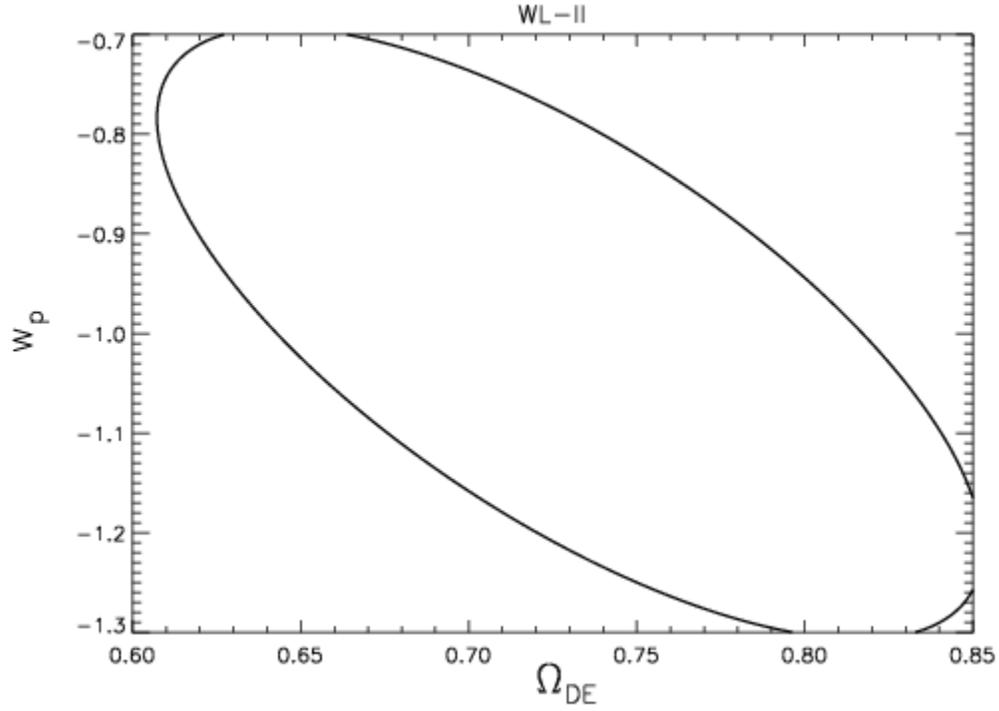

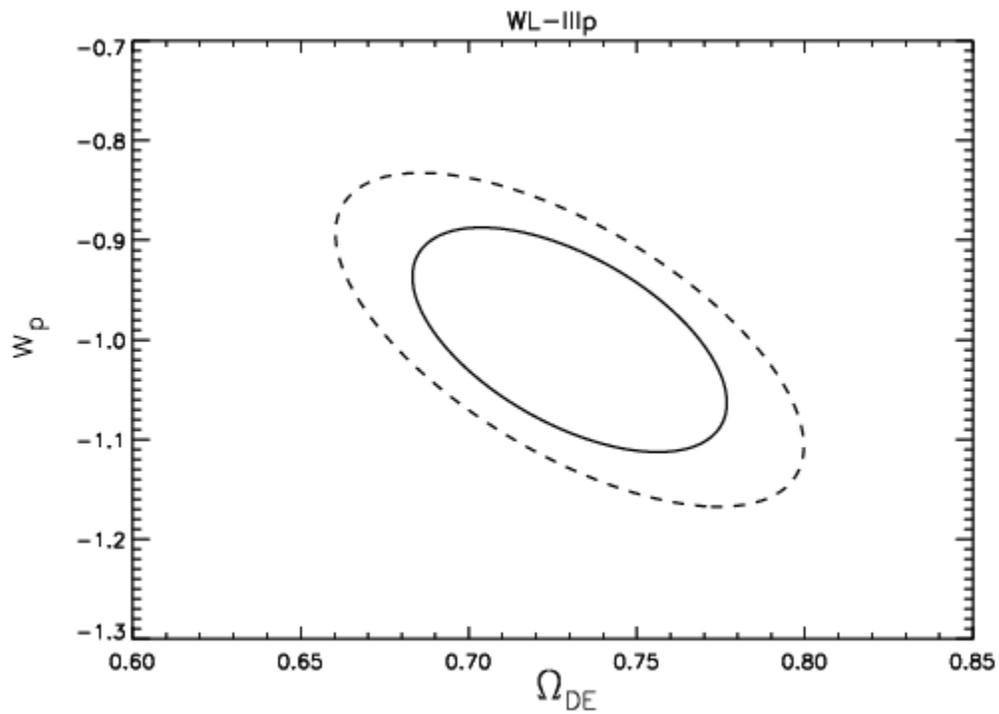

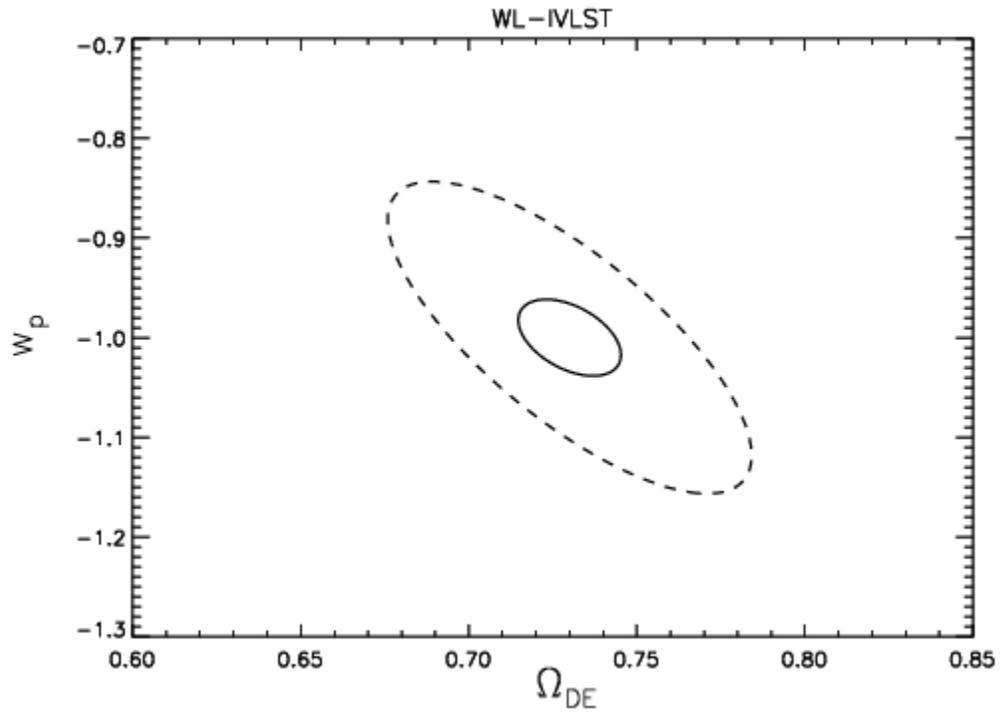

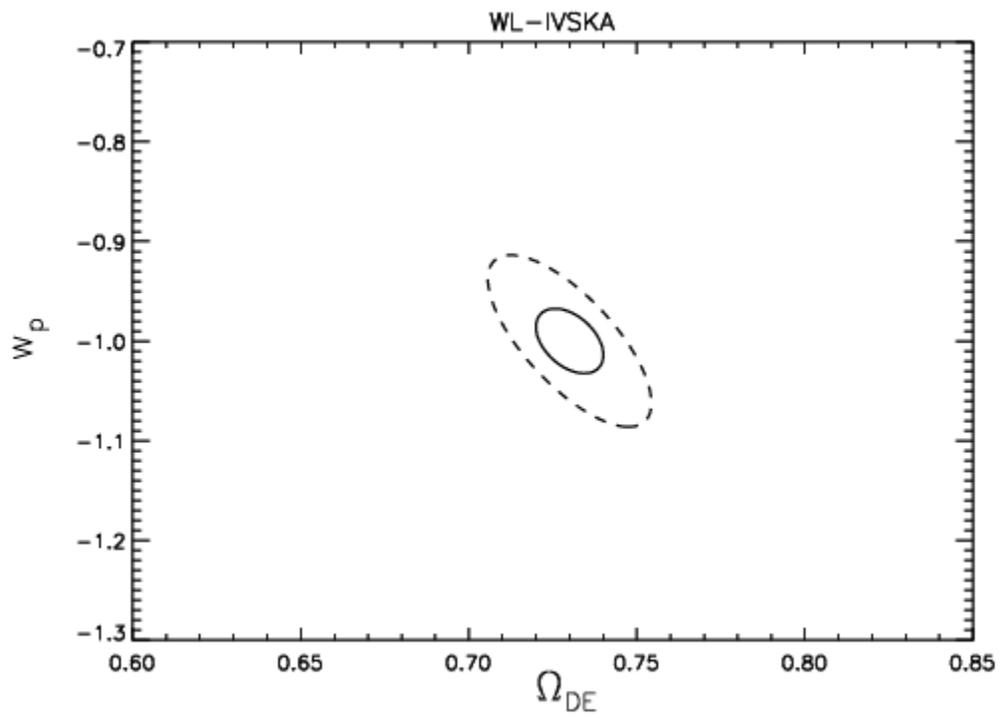

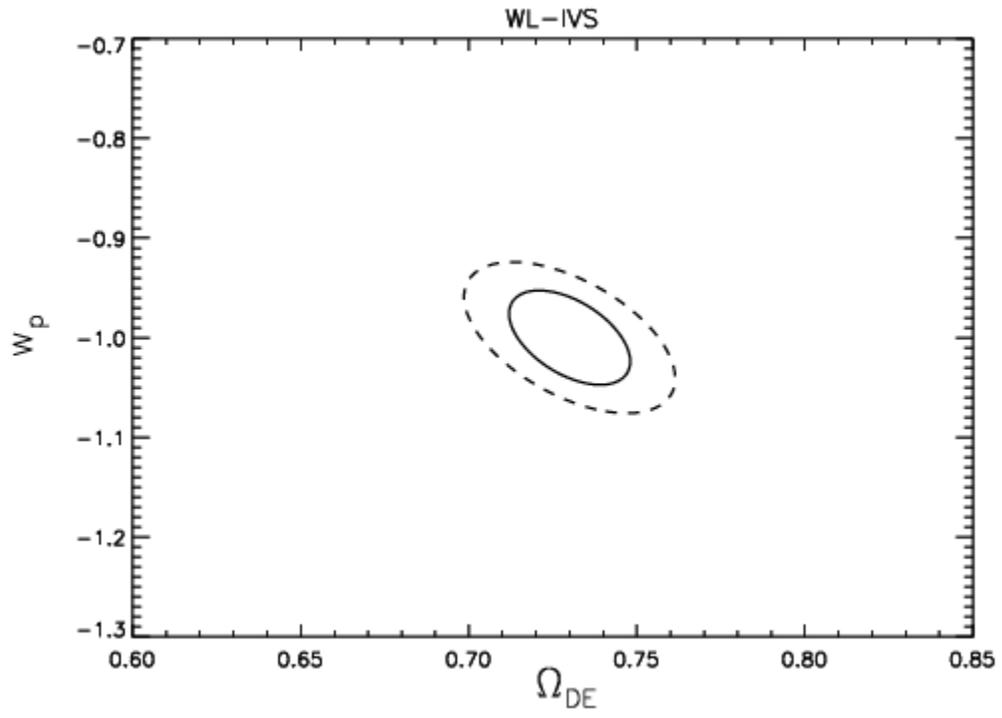

## **Results for models**

| MODEL | $\sigma(w_0)$ | $\sigma(w_a)$ | $\sigma(\Omega_{DE})$ | $a_p$ | $\sigma(w_p)$ | $[\sigma(w_a) \times \sigma(w_p)]^{-1}$ |
|---|---|---|---|---|---|---|
| Stage II | | | | | | |
| (CL-II+SN-II+WL-II) | 0.115 | 0.523 | 0.01 | 0.79 | 0.04 | 53.82 |
| BAO-IIIp-o | 0.911 | 3.569 | 0.06 | 0.76 | 0.26 | 1.06 |
| BAO-IIIp-p | 1.257 | 5.759 | 0.06 | 0.79 | 0.32 | 0.55 |
| BAO-IIIs-o | 0.424 | 1.099 | 0.04 | 0.63 | 0.11 | 8.04 |
| BAO-IIIs-p | 0.442 | 1.169 | 0.04 | 0.64 | 0.12 | 6.97 |
| BAO-IVLST-o | 0.489 | 1.383 | 0.04 | 0.65 | 0.09 | 7.78 |
| BAO-IVLST-p | 0.582 | 1.642 | 0.05 | 0.65 | 0.13 | 4.58 |
| BAO-IVSKA-o | 0.202 | 0.556 | 0.02 | 0.64 | 0.03 | 55.15 |
| BAO-IVSKA-p | 0.293 | 0.849 | 0.02 | 0.66 | 0.05 | 21.53 |
| BAO-IVS-o | 0.243 | 0.608 | 0.02 | 0.61 | 0.04 | 42.19 |
| BAO-IVS-p | 0.330 | 0.849 | 0.03 | 0.62 | 0.06 | 19.84 |
| CL-II | 1.089 | 3.218 | 0.05 | 0.67 | 0.18 | 1.76 |
| CL-IIIp-o | 0.256 | 0.774 | 0.02 | 0.67 | 0.04 | 35.21 |
| CL-IIIp-p | 0.698 | 2.106 | 0.05 | 0.67 | 0.08 | 6.11 |
| CL-IVS-o | 0.241 | 0.730 | 0.02 | 0.67 | 0.04 | 38.72 |
| CL-IVS-p | 0.730 | 2.175 | 0.05 | 0.67 | 0.07 | 6.23 |
| SN-II | 0.159 | 1.142 | 0.03 | 0.90 | 0.11 | 7.68 |
| SN-IIIp-o | 0.092 | 0.872 | 0.03 | 0.95 | 0.08 | 13.91 |
| SN-IIIp-p | 0.185 | 1.329 | 0.03 | 0.89 | 0.12 | 6.31 |
| SN-IIIs | 0.105 | 0.880 | 0.03 | 0.94 | 0.09 | 12.39 |
| SN-IVLST-o | 0.076 | 0.661 | 0.03 | 0.95 | 0.07 | 22.19 |
| SN-IVLST-p | 0.150 | 1.230 | 0.03 | 0.91 | 0.10 | 7.93 |
| SN-IVS-o | 0.074 | 0.683 | 0.02 | 0.93 | 0.05 | 27.01 |
| SN-IVS-p | 0.088 | 0.692 | 0.03 | 0.94 | 0.08 | 19.10 |
| WL-II | 0.560 | 1.656 | 0.05 | 0.67 | 0.12 | 4.89 |
| WL-IIIp-o | 0.189 | 0.513 | 0.02 | 0.64 | 0.05 | 42.96 |
| WL-IIIp-p | 0.277 | 0.758 | 0.03 | 0.65 | 0.07 | 19.55 |
| WL-IVLST-o | 0.055 | 0.142 | 0.01 | 0.63 | 0.02 | 453.60 |
| WL-IVLST-p | 0.187 | 0.495 | 0.02 | 0.64 | 0.06 | 32.04 |
| WL-IVSKA-o | 0.039 | 0.118 | 0.00 | 0.68 | 0.01 | 645.76 |
| WL-IVSKA-p | 0.195 | 0.723 | 0.01 | 0.73 | 0.03 | 39.84 |
| WL-IVS-o | 0.063 | 0.169 | 0.01 | 0.64 | 0.02 | 310.10 |
| WL-IVS-p | 0.103 | 0.249 | 0.01 | 0.60 | 0.03 | 131.72 |

| MODEL: (combined with Stage II) | $\sigma(w_0)$ | $\sigma(w_a)$ | $\sigma(w_p)$ | $\sigma(\Omega_{DE})$ | Figure of merit (Normalized to Stage II) |
|---|---|---|---|---|---|
| BAO-IIIp-o | 0.103 | 0.461 | 0.035 | 0.010 | 1.1 |
| BAO-IIIp-p | 0.109 | 0.494 | 0.036 | 0.011 | 1.1 |
| BAO-IIIs-o | 0.091 | 0.376 | 0.034 | 0.009 | 1.5 |
| BAO-IIIs-p | 0.094 | 0.393 | 0.034 | 0.010 | 1.4 |
| BAO-IVLST-o | 0.092 | 0.376 | 0.033 | 0.009 | 1.5 |
| BAO-IVLST-p | 0.099 | 0.427 | 0.034 | 0.010 | 1.3 |
| BAO-IVSKA-o | 0.071 | 0.222 | 0.023 | 0.007 | 3.7 |
| BAO-IVSKA-p | 0.082 | 0.305 | 0.029 | 0.008 | 2.1 |
| BAO-IVS-o | 0.069 | 0.210 | 0.025 | 0.007 | 3.6 |
| BAO-IVS-p | 0.078 | 0.279 | 0.030 | 0.008 | 2.2 |
| | | | | | |
| CL-IIIp-o | 0.081 | 0.292 | 0.026 | 0.007 | 2.4 |
| CL-IIIp-p | 0.100 | 0.420 | 0.034 | 0.010 | 1.3 |
| CL-IVS-o | 0.080 | 0.287 | 0.026 | 0.007 | 2.5 |
| CL-IVS-p | 0.098 | 0.405 | 0.033 | 0.010 | 1.4 |
| | | | | | |
| SN-IIIp-o | 0.071 | 0.349 | 0.025 | 0.009 | 2.1 |
| SN-IIIp-p | 0.095 | 0.453 | 0.027 | 0.010 | 1.5 |
| SN-IIIs | 0.077 | 0.369 | 0.026 | 0.009 | 2.0 |
| SN-IVLST-o | 0.062 | 0.311 | 0.022 | 0.008 | 2.7 |
| SN-IVLST-p | 0.089 | 0.426 | 0.027 | 0.010 | 1.6 |
| SN-IVS-o | 0.060 | 0.271 | 0.020 | 0.006 | 3.5 |
| SN-IVS-p | 0.070 | 0.328 | 0.023 | 0.008 | 2.5 |
| | | | | | |
| WL-IIIp-o | 0.073 | 0.261 | 0.030 | 0.007 | 2.4 |
| WL-IIIp-p | 0.085 | 0.336 | 0.033 | 0.009 | 1.7 |
| WL-IVLST-o | 0.043 | 0.116 | 0.015 | 0.005 | 11.0 |
| WL-IVLST-p | 0.076 | 0.292 | 0.032 | 0.009 | 2.0 |
| WL-IVSKA-o | 0.035 | 0.106 | 0.013 | 0.004 | 13.8 |
| WL-IVSKA-p | 0.083 | 0.336 | 0.026 | 0.006 | 2.1 |
| WL-IVS-o | 0.046 | 0.133 | 0.017 | 0.005 | 8.1 |
| WL-IVS-p | 0.056 | 0.161 | 0.024 | 0.007 | 4.8 |

# X. Dark-Energy Projects (Present and Future)

This inventory of present and future dark-energy observational projects is based on the white papers received by the DETF and on publicly available descriptions. It is necessarily incomplete as there are projects that have not reached these stages. The survey may be particularly incomplete for projects funded and executed outside the United States.

The targets for these projects are given as presented by their proposers. The DETF was not charged with the analysis of specific projects, and presents these figures without scrutiny or endorsement. We have not aimed to be complete. There are many other ongoing projects in addition to those presented here, reflecting the current interest and activity in this field.

## 1. Stage I

The case for dark energy currently comes from combinations of four types of observations: Type Ia supernovae, anisotropies of the cosmic microwave background radiation (combined with measurements of the Hubble constant or large-scale structure), weak lensing, and baryon acoustic oscillations.

**Type Ia Supernovae:** The primary evidence for the acceleration of the universe came originally from studies of Type Ia supernovae (Riess *et al.* 1998; Perlmutter *et al.* 1999). Type Ia SNe remain the strongest evidence for acceleration, though the need for dark energy can now be inferred from the other indicators. These studies have established that supernovae at high redshift are fainter than their local counterparts. Under the assumption of a flat universe, these results are consistent with a cosmology in which approximately one third of the universe is composed of matter, and two thirds dark energy. The most recent results are consistent with the existence of a cosmological constant, where $w = -1$ (Astier *et al.* 2006; Sollerman *et al.* 2006).

**CMB Anisotropies:** Studies of angular power spectrum of anisotropies in the cosmic microwave background radiation from WMAP, in combination with the Hubble Key Project or large-scale surveys of galaxies, have yielded a cosmological parameters in excellent agreement with the supernova results (Spergel *et al.* 2003; Bennett *et al.* 2003; Spergel *et al.* 2006), both for the one-year and three-year WMAP data.

**Weak Lensing:** The largest weak lensing surveys to date cover 50-100 square degrees in a single filter, and are dominated by statistical uncertainties (van Waerbeke *et al.* 2005, Jarvis *et al.* 2006). The CFHT Legacy Survey currently underway would survey 170 $deg^2$ in multiple colors, and has published preliminary results from partial coverage, yielding $w < -0.8$ (68% confidence) from weak-lensing data alone (Hoekstra et al. 2005). Jarvis et al. find $w = -0.89$ [+0.16,−0.21,95% CL] combining weak lensing, WMAP one-year data, and Type Ia SNe, under the assumption that $w$ is constant.

**Baryon Acoustic Oscillations:** Eisenstein et al. (2005) have used about 50,000 luminous red galaxies over 3800 $\deg^2$ from the Sloan Digital Sky Survey and made the first clear detection of the acoustic peak in the correlation function at a scale of 100$h^{-1}$ Mpc. This technique already this method provides geometric evidence for dark energy, particularly in combination with CMB data. The BAO analysis of SDSS would presumably be extended to the full survey area of about 8000 $\deg^2$ in the near future.

## Stage II

a. The **Canada-France-Hawaii Telescope Supernova Legacy Survey (CFHT-SNLS)** is a Canadian-French project using the 1-$\deg^2$ camera on the 3.5m CFHT telescope in Hawaii. The initial survey includes a "deep" search for SNe (which doubles as a small, deep weak-lensing survey), and a "wide" imaging weak lensing survey covering 170 $\deg^2$, both in multiple filters. The SNe survey expects to discover 700 SNe in the 0.2-0.9 redshift range.

b. The **ESSENCE** project is a multi-year ground-based supernova survey using the MOSAIC wide-field CCD camera on the Blanco 4-m telescope at the Cerro Tololo Inter-American Observatory (CTIO). The goal is to measure luminosity distances for a sample of about 200 SNIa at redshifts in the 0.2-0.8 redshift range.

χ. The **Sloan Digital Sky Survey-II (SDSS-II),** a three-year extension of the original SDSS, is using the wide-field, 2.5-meter telescope at Apache Point Observatory to undertake a supernova survey. Scanning the SDSS Southern equatorial stripe (about 2.5 deg wide by about120 deg long) over the course of three 3-month campaigns (Sept.-Nov. 2005-7), they are obtaining multi-band lightcurves for about 200 Type Ia supernovae in the redshift range $z$ =0.1−0.3.

d. The **Center for Astrophysics Supernova Program** obtains low dispersion spectra, UBVRI and recently JHK light curves for most supernovae brighter than 18[th] mag and north of about 20 degrees. To date, the sample of well-observed low-redshift supernovae totals almost 100. By the year 2007, the sample should double.

e. The **Nearby Supernova Factory** is an experiment being carried out using the large-area Yale-built QUEST camera at the 1.2m Samuel Oschin Schmidt Telescope at Palomar. The SN Factory is designed to collect 300 or more SNIa in the redshift range 0.03-0.08, each followed up by about 15 optical spectra (3400-10,000Å) spaced by about 3-4 day intervals at the Hawaii 2.2m telescope.

f.  The **Katzman Automatic Imaging Telescope (KAIT),** located at Lick Observatory on Mount Hamilton, is a 76-cm robotic telescope with a dedicated to searching and obtaining multicolor photometry for nearby supernovae. Since 1998, over 500 nearby supernovae have been discovered by KAIT, the largest nearby sample to date. Funding is provided by the National Science Foundation.

g.  The **Carnegie Supernova Project (CSP)** is using the Las Campanas 1-m, 2.5-m and 6.5-m telescopes to follow up several ongoing supernova search surveys (KAIT, CFHTLS, ESSENCE, SDSSII). The goal of the CSP is to obtain an I-band rest-frame Hubble diagram for approximately 200 supernovae. Multicolor (10-color) photometry, as well as optical spectroscopy is being obtained over the redshift range $0 < z < 0.1$, and near-infrared photometry is being obtained for supernovae in the range $0.1 < z < 0.7$.

h.  The Palomar **QUEST Survey** is a time variability sky survey using the 1.3m Samuel Oshin Schmidt Telescope with the Large Area ($4 \times 4$ deg$^2$) Yale-Indiana QUEST CCD camera. The survey is planned to last five years and cover 15,000 square degrees to a magnitude limit of about 21 with the aim of studying Type Ia and II supernovae as part of the Nearby Supernova Factory, for the dark energy part of its program.

i.  **HST Searches for High Redshift Supernovae:** The ACS HST Treasury program and continued later HST searches have been used to discover supernovae at $z > 1$. There are now 25 Type Ia supernovae known at $z > 1$, and if Hubble continues to function, in the next several years, the sample could be grown to more than 100.

j.  **PanSTARRS-1** is a 1.8m telescope survey being constructed at Haleakala in Hawaii. The telescope is funded by the Air Force, with science operations planned for the beginning of 2007. The telescope would have a prototype 1.4-billion-pixel CCD camera and six filters. The dark energy science goals potentially include supernovae, weak lensing, baryon-oscillation, and cluster surveys, with the exact allocation of survey resources currently undecided. It is a precursor experiment to PanSTARRS-4 (a stage III project described below).

k.  The **Parallel Imager for Southern Cosmological Observations (PISCO)** is being built for use on the 6.5-meter Magellan telescope for the purpose of obtaining simultaneous broadband images over a few arcminute field of view for photometric redshifts for upcoming Sunyaev-Zel'dovich surveys. The proposers expect the camera to be ready by December, 2006.

l. The **South Pole Telescope (SPT)** is a 10-meter submillimeter-wave telescope with a 1000-element bolometric focal plane array with channels at 90, 150, 220 and 270 GHz. It would conduct a deep, large solid angle (4000 square degree) galaxy-cluster survey using the Sunyaev –Zel'dovich effect. About 20,000 clusters with masses greater than $2 \times 10^{14}$ solar masses are expected to be discovered. The project is funded by the NSF Office of Polar Programs, and the survey is scheduled to start in spring 2007.

m. The **Atacama Cosmology Telescope (ACT)** would be located in the Atacama Desert in Chile. It would map 200 square degrees of the microwave sky at three frequencies (145 GHz, 220 Ghz and 265 GHz) at arcminute angular resolution over 100 square degrees. Clusters with masses greater than $3 \times 10^{14}$ solar masses would be discovered through the Sunyaev-Zel'dovich effect, and optical spectroscopic redshifts would be obtained for a subsample of 400 clusters. The millimeter bolometer array camera is composed of three $32 \times 32$ arrays of transition edge sensing bolometers. It is expected to be completed in 2006, with science observations from mid-2007 through 2008. It is funded by the NSF.

n. The **XMM Cluster Survey (XCS)** proposes to use the XMM EPIC camera to image about 500 deg$^2$ to discover thousands of clusters, 250 of which have redshifts $z > 1$. Optical photometric redshift data are being obtained from public archives, where possible (INT-Wide Field Survey, XMM-ESO Imaging Survey, UBVRI CFHT data, SDSS imaging). With a three-year time period, the proposers hope to optically image about 330 XMM cluster candidates.

o. The **Red-Sequence Cluster Survey 2 (RCS2)** is a shallow cluster-counting and weak-lensing survey in 3 filters over 1000 deg$^2$. The survey is underway at the Canada-France-Hawaii telescope.

**p.** The **Deep Lensing Survey (DLS)** is a deep BVRz′ imaging survey covering 20 deg$^2$ in five 2 deg by 2 deg fields, with a primary focus on weak lensing. It is being carried out at the CTIO 4-m with the Mosaic2 camera. As of May 2006, data-taking is almost complete, and preliminary photometric redshift and shape catalogs have been produced.

q. The **Kilo-Degree Survey (KIDS)** is a 4-band, 1500 deg$^2$ optical survey using the VLT Survey Telescope, with infrared follow-up on the VISTA telescope. This European project is focused on weak gravitational lensing but has potential for galaxy-cluster studies as well. It is pending approval by the ESO TAC, to begin observations in the very near future.

r. The **DEEP2 Galaxy Redshift Survey** of about 50,000 galaxies at $0.7 < z < 1.3$ over 3 $\text{deg}^2$ of sky is nearing completion at the Keck telescopes. This would facilitate an optically-based census of galaxy clusters and groups.

## 2. Stage III

a. The **Dark Energy Survey (DES)** would be a new 520 megapixel wide-field camera (2.2 square degrees) mounted on the 4-m Blanco Telescope of the Cerro Tololo Inter-American Observatory (CTIO) in Chile. This is a US-led collaboration with collaborators from the UK and Spain. The DES plans to obtain photometric redshifts in four bands. The planned survey area is 5,000 square degrees. The techniques to study dark energy would be baryon oscillations, clusters, supernovae, and weak lensing. With the Cluster technique they plan to exploit the clusters detected by the South Pole Telescope. The proponents plan a five-year survey and hope to start observing in 2009.

b. The **Hobby-Eberly Telescope Dark Energy Experiment (HETDEX)** aims to measure BAO over two areas, each 100 square degrees, using 500,000 galaxies over the redshift range $1.8 < z < 3.7$. The proposers expect that the instrument can be completed within 3.5 years of full funding, and that dark energy constraints would be provided by 2011.

c. The **Wide-Field Multi-Object Spectrograph (WFMOS)** is being designed for the Subaru 8-m telescope at Mauna Kea. It would have a field of view of about 1.5 degrees in diameter, and the capability of simultaneously obtaining spectra for 4,000-5,000 objects, or about 20,000 objects per night. A redshift range of $0.5 < z < 1.3$ for emission-line galaxies would be targeted, and $2.3 < z < 3.3$ for Lyman-break galaxies. WFMOS is a second-generation Gemini instrument, proposed as part of the 'Aspen' process. Its primary dark-energy goal is to measure baryon acoustic oscillations, as well as to measure thousands of supernovae. A precursor imaging survey would be required. It is expected by the proposers that building and commissioning would occur in 2010-2012, and the dark energy science survey in 2013-2016.

d. **Pan-STARRS-4** is a large optical/near-IR survey telescope to be sited on Mauna Kea in Hawaii, planned for 2009. It consists of an array of four 1.8m telescopes with a 7 degree field of view, each telescope equipped with a 1.4 billion-pixel CCD camera and six filters. The dark energy science goals include supernovae, weak lensing and clusters surveys. The supernova survey aims to discover supernovae to $z \sim 1$, covering approximately 1200 square degrees with a cadence of about 4 days. A $3\pi$-steradian survey would be undertaken, useful for weak lensing and cluster surveys. It is expected that the survey would continue for ten years. The telescopes and instruments are funded by the Air Force, but support for operations is not included.

e. A **One-Degree Imager (ODI)** at the WIYN 3.5m telescope at Kitt Peak is currently being planned, and is largely funded. A baseline survey covering 9 square degrees to a depth of z ~ 27 and Y ~ 27 mag, with a time cadence similar to LSST is planned. The details of an expanded survey are yet to be determined. The dark energy science goals include supernovae and weak lensing. It is expected to be commissioned in 2009.

f. The **One Thousand Points of Light Spectrograph** is a 1000-fiber spectrograph to be prototyped at Lawrence Berkeley Lab. The goal is to undertake a baryon oscillation experiment by surveying 1 million galaxies at $0.7 < z < 1.2$ on an existing 4-m class telescope, followed by 1 million galaxies at $2.3 < z < 3$ on a 10-m class telescope.

g. The **ALPACA** project proposes to install an 8-meter zenith-pointing liquid-mirror telescope in Chile to conduct a 1000 deg$^2$ survey of a 3-degree-wide strip circling the sky in five filters. The survey could begin in 2010 and would require three years. The survey would be well-suited to SN and weak-lensing observations, and like other multicolor imaging surveys would produce data of use for cluster counting and BAO measurement as well.

h. The **Cluster Imaging eXperiment (CIX)** is a 64-pixel four spectral band radiometer to be installed on the Large Millimeter Telescope (LMT). It aims to use high-resolution (12 arcsecond) Sunyaev-Zel'dovich images to measure masses, peculiar velocities and shapes of clusters, with the aim of improving the understanding of systematic errors in SZ surveys. Construction would be completed three years after the start of funding, and the proposal for the camera has been submitted to the NSF ATI program.

i. The **Cornell-Caltech Atacama Telescope (CCAT)** is a 25-m sub-millimeter and millimeter-wave telescope to be located in the Atacama Desert in Chile for high-angular-resolution thermal Sunyaev-Zel'dovich measurements. A range of low- to high-mass clusters would be followed up in detail to test SZ systematics and the relation between SZ flux and cluster mass. The proposers hope CCAT would be operational in 2012.

### 3. Stage IV

a. The **Large Synoptic Survey Telescope (LSST)** would have a newly constructed 8.4m telescope and a 3 Gigapixel camera. The survey would scan a hemisphere (approximately 20,000 deg$^2$) several times per month in six colors. It would reach galaxies in the redshift range $0.5 < z < 3$. It would study dark energy through baryon oscillations, supernovae, and weak-lensing techniques. The proponents hope to see first light in 2013, with science runs in 2014.

b. The **Joint Dark Energy Mission** (JDEM) is a DOE/NASA effort aimed toward a space mission to investigate dark energy. There are several proposals that have been advanced, three of which submitted white papers to the DETF: DESTINY, JEDI and SNAP. Other proposals may exist and may have submitted proposals for NASA's Mission Concept Design competition, but they have not been described in open publications or in confidential white papers submitted to the Task Force.

    i. The **Dark Energy Space Telescope (DESTINY)** is a proposed 2m-class space telescope, which aims to measure near-infrared grism spectrophotometry over the wavelength range $0.85\mu m < \lambda < 1.7\mu m$ for high-redshift supernovae. Continuous observations of several square degrees would yield an estimated 2500 SN1a with $0.5 < z < 1.7$ in two years.

    ii. The **Joint Efficient Dark-energy Investigation (JEDI)** is a proposed 2m space telescope with a one-degree field of view. It would have simultaneous imaging and multi-slit (microshutter array) spectroscopic capability to exploit the 0.8-4μm wavelength range for measurement of supernovae, baryon oscillations and weak lensing. It would discover 14,000 type Ia supernovae, and survey over 10,000 and 1,000 square degrees for baryon oscillations and weak lensing, respectively.

    iii. The **Supernova Acceleration Probe (SNAP)** is a 2m space telescope concept with a 0.7 square-degree field of view and optical and near-infrared imaging plus spectroscopy for the study of Type Ia supernovae, weak lensing, and baryon oscillations. About 2,000 extremely well-characterized, "Branch-normal" Type Ia supernovae (out of 10,000 total) would be discovered out to a redshift of 1.7. A 1,000 square-degree field would be covered for weak lensing and baryon oscillations, with a larger survey possible in an extended mission.

c. The **Square Kilometer Array (SKA)** is a proposed radio telescope with a collecting area of order one square kilometer, capable of operating at a wide range of frequencies and angular resolutions. Current baseline (goal) specifications give a frequency range of 100 MHz - 25 GHz (60 MHz - 35 GHz) and an angular resolution of better than 0.02 arcsec at 1 GHz, scaling with wavelength. Several dark energy experiments are planned for the SKA: a neutral hydrogen survey of about $10^9$ galaxies for the study of baryon oscillations; shear statistics for about $10^{10}$ continuum-detected galaxies for weak lensing; and a determination of the Hubble constant with about 1% accuracy from extragalactic maser sources. It is anticipated that construction of Phase 1 (10% of the collecting area) would begin in 2011, first science with Phase 1 would begin in 2014, and the full SKA would be operational in 2019.

d. **Cluster Surveys:**

   i. The **10K X-Ray Cluster Survey** would use a proposed x-ray telescope with large FOV mirrors to undertake a galaxy cluster survey over 10,000 $\text{deg}^2$ out to a redshift of $z \sim 1.5$. Optical photometric redshifts would be obtained to identify a sample for deeper follow-up x-ray observations of approximately 1,000 massive high-redshift clusters, for which masses would be determined. Optical spectroscopic redshifts with 6.5-m class telescopes would be obtained for clusters with x-ray follow-up.

   ii. A **NASA Medium-Explorer Mission** has been proposed for an x-ray telescope to undertake a survey of about 20,000 $\text{deg}^2$ out to $z \sim 1.5$ for a sample of about 100,000 galaxy clusters. A photometric redshift survey all of these objects is also planned, with a spectroscopic training set. The proponents say the mission could be ready to be flown around 2011.

   iii. **Constellation-X** would probe dark energy using X-ray observations of galaxy clusters in two different ways: 1) measurements of the X-ray gas mass fraction, with follow-up observations of the Sunyaev-Zel'dovich effect and 2) using the spectroscopic capability of Con-X to measure scaling relations between X-ray measurements and mass. Short (~1ks) exposures of 2,000 of the most massive clusters out to redshifts $0 < z < 2$ would be obtained, and deeper (20-40ks) exposures of the 250-500 most relaxed systems would be obtained to measure the X-ray gas mass fraction. It is proposed that about 10-15% of the available time over the first five years of the Con-X mission be aimed at dark energy studies.

e. **Other Projects:**

i. **The Giant Segmented Mirror Telescope (GSMT)** is a 30-meter class optical/near-infrared telescope, a public-private partnership. Currently there are two concepts under design, the Giant Magellan Telescope (GMT) and the Thirty-Meter Telescope (TMT), with planned completion dates in the middle of next decade. These telescopes would be useful for studies of supernovae, weak lensing experiments, galaxy clusters, and baryon oscillations.

ii. The **James Webb Space Telescope (JWST)** is a 6.5-meter deployable space telescope covering the wavelength range from 0.6 to 29 micrometers, planned for launch in 2013. It is being built by a NASA-led partnership with the European and Canadian space agencies. With cameras and spectrometers at all wavelengths, it would be useful for many types of dark-energy studies including supernovae, clustering, and weak lensing at high redshift.

# XII. Acknowledgments

We would like to acknowledge the great interest and cooperation of the community in putting together this report. More than fifty well thought out White Papers guided our thoughts and are reflected in our findings and recommendations.

The continual engagement and encouragement of the Astronomy and Astrophysics Advisory Committee and the High Energy Physics Advisory Panel, as well as the Agencies, kept our enthusiasm high.

A preliminary version of this report was read by Jonathan Bagger, Rachel Bean, Daniel Eisenstein, Keith Jahoda, Bob Kirshner, Rene Ong, and Adam Reiss. Their comments were very useful and led to an improved product.

Meetings of the Dark Energy Task Force were held at NSF Headquarters, Fermilab, UC Davis, and MIT. We would like to acknowledge the hospitality of those institutions and thank the people involved. We would also like to thank the people who took time from their busy schedule to speak to us at the meetings.

Thanks also to Mike Jarvis for sharing some calculations with us, and to Augusta Abrahamse and Brandon Bozek for technical assistance.

Finally, this report consumed a lot of time of many people. We would like to acknowledge the support and understanding of our family and loved ones during the ordeal; they share our pleasure that the task is completed.

# XIII. Technical Appendix

# 1 Our tools and methods

## 1.1 Fisher Matrix Overview

Here we review the Fisher Matrix methods used by the DETF. These methods are standard in many fields. First we consider a statistically simple case of a series of measurements with Gaussian error distributions. Suppose we measure the quantity $y$ when the remaining observables have the values $x$ and suppose we put the values of $x$ in bins $b=1,...B$. Suppose in addition that the data should be described by a function $f$ of the bin $b$ and some parameters $p$ and that the expected variance in bin $b$ is $\sigma^2_b$, then we can form

$$\chi^2 = \sum_{b=1}^{B} \sum_{i_b} \frac{(f_b(p) - y_{i_b})^2}{\sigma^2_b} \tag{1.1}$$

where $i_b$ labels the events in bin $b$. If the parameters $p$ give the true underlying distribution $\bar{p}$, then a Gaussian distribution of data values is:

$$P(y_{i_b}) \propto \exp(-\frac{1}{2}\chi^2) \tag{1.2}$$

The problem, however, is to estimate parameters $p$ given a realization of the data $y$. Using Bayes' theorem with uniform prior we have $P(p\,|\,y) \propto P(y\,|\,p)$ so that the likelihood of a parameter estimate can be described as a Gaussian with the same $\chi^2$, now viewed as a function of parameters. If we expand about the true values of the parameters, $p_i = \bar{p}_i + \delta p_i$, and average over realizations of the data,

$$\left\langle \chi^2(p) \right\rangle = \left\langle \chi^2 \right\rangle + \left\langle \frac{\partial \chi^2}{\partial p_j} \right\rangle \delta p_j + \frac{1}{2} \left\langle \frac{\partial \chi^2}{\partial p_j \partial p_k} \right\rangle \delta p_j \delta p_k ... \tag{1.3}$$

where the expectation values are taken at the true values $\bar{p}$. The mean value of the events in bin $b$ is indeed $f_b(\bar{p})$, so the second term vanishes. The distribution of errors in the measured parameters is thus in the limit of high statistics proportional to

$$\exp\left(-\frac{1}{2}\chi^2\right) \propto \exp\left(-\frac{1}{4} \left\langle \frac{\partial \chi^2}{\partial p_j \partial p_k} \right\rangle \delta p_j \delta p_k\right) = \exp\left(-\frac{1}{2} F_{jk} \delta p_j \delta p_k\right) \tag{1.4}$$

where the Fisher matrix is

$$F_{jk} = \sum_b \frac{N_b}{\sigma_b^2} \frac{\partial f_b}{\partial p_j} \frac{\partial f_b}{\partial p_k} \tag{1.5}$$

and $N_b$ is the average number of events in bin $b$. From this expression it follows that

$$\left\langle \delta p_j \delta p_k \right\rangle = \left( F^{-1} \right)_{jk} \tag{1.6}$$

In other words, the covariance matrix is simply the inverse of the Fisher matrix (and vice versa).

More generally, if one can create a probability $P(p_i \mid y_i)$ of the model parameters given a set of observed data, *e.g.* by Bayesian methods, then one can define the Fisher matrix components via

$$F_{ij} = -\left\langle \frac{\partial^2 \ln P}{\partial p_i \partial p_j} \right\rangle$$

and the Cramer-Rao theorem states that any unbiased estimator for the parameters will deliver a covariance matrix on the parameters that is no better than $F^{-1}$. The Fisher matrix therefore offers a best-case scenario for ones ability to constrain cosmology parameters given a set of observations.

If we want to use some other set of parameters $q$, the new Fisher matrix is simply

$$F'_{lm} = \sum_b \frac{N_b}{\sigma_b^2} \frac{\partial f_b}{\partial q_l} \frac{\partial f_b}{\partial q_m} = \sum_b \frac{N_b}{\sigma_b^2} \frac{\partial p_j}{\partial q_l} \frac{\partial p_k}{\partial q_m} \frac{\partial f_b}{\partial p_j} \frac{\partial f_b}{\partial p_k} = \frac{\partial p_j}{\partial q_l} \frac{\partial p_k}{\partial q_m} F_{jk} \equiv (\mathbf{M})^{\mathbf{T}} \mathbf{F} \mathbf{M} \tag{1.7}$$

using the usual summation convention on $j,k$.

## 1.2 Priors

A Gaussian prior with width $\sigma$ can be placed on the $i$th parameter by adding to the appropriate diagonal element of the Fisher matrix:

$$F_{kl} \rightarrow F_{kl} + \frac{\delta_{ki}\delta_{li}}{\sigma^2} \tag{1.8}$$

which can also be written as

$$\mathbf{F} \rightarrow \mathbf{F} + \mathbf{F}^P \tag{1.9}$$

where in this case $\mathbf{F}^P$ is an extremely simple matrix (with a single non-zero diagonal element).

If one wants to add a prior on some quantity that is not a single parameter of the Fisher matrix one can work in variables where it is and then transform the diagonal $\hat{\mathbf{F}}^P$ (where the hat indicates that the new variables are used) back into the working variables using the inverse of the transformation given in Eqn. (1.7). This generates a less trivial $\mathbf{F}^P$ given by

$$\mathbf{F}^P = \left(\hat{\mathbf{M}}^{-1}\right)^{\mathbf{T}} \hat{\mathbf{F}}^P \hat{\mathbf{M}}^{-1} \qquad (1.10)$$

### 1.3 Marginalization

On many occasions we need to produce a Fisher matrix in a smaller parameter space by marginalizing over the undesired "nuisance" parameters. This amounts to integrating over the nuisance parameters without assuming any additional priors on their values. There is a simple way to do this: Invert $\mathbf{F}$, remove the rows and columns that are being marginalized over, and then invert the result to obtained the reduced Fisher matrix.

### 1.4 Adding two data sets

If one calculates the Fisher matrices $\mathbf{F}^A$ and $\mathbf{F}^B$ for two independent data sets $A$ and $B$ one can find the Fisher matrix for the combined probability distribution by adding:

$$\mathbf{F}^{A+B} = \mathbf{F}^A + \mathbf{F}^B \qquad (1.11)$$

In general any marginalization over nuisance parameters must be done *after* summation of the two Fisher matrices. If, however, the nuisance parameters of $A$ are disjoint from those of $B$, then the two data sets have independent probability distributions over the set of nuisance parameters, and it is permissible to marginalize before summation.

The Fisher matrices produced by DETF to represent individual experiments have been marginalized over all parameters other than the eight in Eqn. (1.12). We can therefore legitimately sum these reduced matrices as long as we consider experiments with distinct nuisance parameters. In many cases of interest this is incorrect, *e.g.* when combining two different supernova experiments that may share nuisance parameters related to supernova evolution. Care must be exercised about such combinations.

### 1.5 The DETF Cosmological Parameters

Because our goal is to estimate the precision with which various cosmological parameters can be determined, the Fisher matrix provides exactly the tool needed: we simply invert it to find the expected uncertainties and covariances.

For many of our calculations a convenient set of parameters is

$$p_i \in \left\{\omega_k, \omega_m, \omega_{DE}, \omega_b, w_0, w_a, \delta_\zeta, n_s\right\} \qquad (1.12)$$

The $\omega_i$ are equal to $\Omega_i h^2$ for each component, i.e. their present-day mass-energy densities. The dark-energy equation of state follows $w(a)=w_0+(1-a)w_a$. The power spectrum of primordial scalar density fluctuations has slope $n_s$ and normalization $\delta_\zeta$ defined with the conventions of the WMAP publications (Verde et al. 2003). These parameters are "natural" because the Friedmann equation looks simple in terms of these parameters

$$h(a)^2 = \frac{\omega_m}{a^3} + \frac{\omega_r}{a^4} + \frac{\omega_k}{a^2} + \omega_{DE} \times \exp\left\{3\left(\left(1+w_0+w_a\right)\ln\left(\frac{1}{a}\right) - w_a\left(1-a\right)\right)\right\} \qquad (1.13)$$

where $h^2(a=1) = \omega_m + \omega_r + \omega_k + \omega_{DE}$, and current observations provide a prior constraint of $h = 0.72 \pm 0.08$ (Freedman *et al* 2001).

## 1.6 Fiducial Model

The fiducial model around which the parameters are perturbed has

$$
\begin{aligned}
\omega_{DE} &= 0.380 & \Omega_{DE} &= 0.722 \\
\omega_m &= 0.146 & \Omega_m &= 0.278 \\
\omega_k &= 0 & \Omega_k &= 0 \\
\omega_b &= 0.024 & H_0 &= 72.5 \\
w_0 &= -1 \\
w_a &= 0 \\
\delta_\zeta &= 0.87 \\
n_s &= 1 \\
\omega_r &= 4.16 \times 10^{-5} & \Omega_r &= 7.91 \times 10^{-5}
\end{aligned}
\tag{1.14}
$$

Note that $\omega_r$ is not a free parameter for our calculations but is fixed by the CMB temperature (and the standard assumption of three massless neutrinos).

## 1.7 Pivot Parameters

As discussed in Huterer and Turner, 2001 and Hu and Jain, 2004, the pivot point is the value of $a$ for which the uncertainty in $w(a)$ is least. If we minimize

$$\left\langle (\delta w_0 + (1-a)\delta w_a)^2 \right\rangle$$

we find that the pivot $a_p$ occurs when

$$1 - a_p = -\frac{\left\langle \delta w_0 \delta w_a \right\rangle}{\left\langle \delta w_a^2 \right\rangle}$$

so we can take as our parameters $w_a$ and

$$w_p = w_0 + (1-a_p)w_a$$

The "pivot" parameters are a linear transformation from $p = \{w_0, w_a\}$ space to $p = \{w_p, w_a\}$:

$$
M \equiv \frac{\partial p}{\partial q} = \begin{pmatrix} 1 & 1 - a_p \\ 0 & 1 \end{pmatrix}
\tag{1.15}
$$

The DETF figure of merit is inversely proportional to the area of the error ellipse in the $w_0 - w_a$ plane, that is to *det(F)*. Since the Fisher matrix in the $w_p$- $w_a$ variables is $F' = M^T F M$ and *det M=1*, it follows that the error ellipse in the $w_p$- $w_a$ plane has the same area.

# 2 Supernova data

The observables for SN data are apparent magnitudes $m_i$, corrected using light curve shapes or spectroscopic data to behave as standard candles with absolute magnitude $M$ so that

$$m_i = M + \mu(z_i),$$

The $\mu(z_i)$ for the set of measured redshifts $\{z_i\}$ are

$$\mu(z_i) = 5\log_{10}(d_L(z_i)) + 25 \tag{2.1}$$

$$d_L(z_i) = \frac{1}{a}\begin{cases} \frac{1}{\sqrt{|k|}}\sinh\left(\sqrt{|k|}\chi(z)_i\right) & k < 0 \\ \chi(z)_i & k = 0 \\ \frac{1}{\sqrt{|k|}}\sin\left(\sqrt{|k|}\chi(z)_i\right) & k > 0 \end{cases} \tag{2.2}$$

and

$$\chi(z_i) = \eta_0 - \eta(z_i) \equiv \int_{a_i}^{1} \frac{da}{a^2 H(a)} \tag{2.3}$$

with $\sqrt{|k|} \equiv H_0\sqrt{|\Omega_k|} = \frac{H_0}{h}\sqrt{|\omega_k|}$ in Mpc$^{-1}$.

The Fisher matrix is constructed by assigning Gaussian uncertainties of size $\sigma_i$ to the corrected apparent magnitude $m_i$ of each supernova. More commonly we will consider a set of bins in redshift centered on the set of mean redshifts $\{\overline{z_i}\}$ with $N_i$ SNe per bin. The observables are the mean values of $m$ in each bin but this time the statistical magnitude uncertainty per bin is reduced by a factor $\sqrt{N_i}$. The treatment of the systematic errors is unchanged by binning.

## 2.1 Statistical Errors

The peak luminosities of Type Ia supernovae vary, even after other observable features of the supernovae have been used to "standardize" the events. The uncertainty of the corrected apparent magnitudes due solely to variation in the properties of SNe is denoted as $\sigma_D$, to which we add in quadrature a measurement uncertainty $\sigma_m$. The assumed values for each supernova data model are listed in Section 3.

## 2.2 Systematic Errors

Various systematic effects can be represented by additional nuisance parameters, possibly applying a Gaussian prior to that parameter. The systematic-error nuisance parameters are marginalized away to reduce the Fisher matrix to the cosmological parameter set described in Section 1.5.

Absolute Magnitude:
The absolute magnitude $M$ of the SNe is a nuisance parameter in all SN calculations. This is equivalent to adding an additional parameter $\mu_{off}$ to the distance moduli: $\mu(z_i) \rightarrow \mu(z_i) + \mu_{off}$ . By incorporating the absolute magnitude into the definition of the distance moduli, we can simply consider the $\mu_i$ as observables rather than the apparent magnitudes. No prior is assigned to $\mu_{off}$ . A consequence of this is that SN measurements do not determine $h$, causing a degeneracy among $\omega_m$, $\omega_b$ and $\omega_k$. For SN measurements we can therefore take the basic parameter set to be $w_0$, $w_a$, $\Omega_{DE}$, and $\Omega_k$. Including $\omega_m$ would be redundant and would lead to a singular Fisher matrix.

Photo-z errors:
If the $\{z_i\}$ are determined photometrically then the resultant uncertainties in redshift are modeled in the following way: there is a nuisance parameter $\delta z_i$, which is the bias in the measurement of photo-z's in each bin. The observables are then modeled as $\mu_i = \mu(z_i + \delta z_i)$. A prior $\sigma_{z_i}$ is assigned to each of the nuisance parameters:

$$\sigma_z = \frac{\sigma_F (1+z)}{\sqrt{N_C}} \qquad (2.4)$$

Please note: $N_C$ represents the number of available spectra for calibration of photometric redshifts in the bin, and is **not** the number of detected SNe in the bin $N_i$. All our calculations use $N_C = 100$ and $\sigma_F \in \{0.01, 0.05\}$ .

Quadratic $\mu$ offset:
The peak luminosity of supernovae might show some z-dependent effects, for example due to evolution of the SN population or extinction properties, that are not fully corrected by recourse to light-curve or spectroscopic information. This is represented with two additional nuisance parameters $\mu^L$ and $\mu^Q$ to give the evolution a quadratic redshift dependence:

$$\mu(z_i) \rightarrow \mu(z_i) + \mu^L z_i + \mu^Q z_i^2 .$$

We apply equal, independent priors on each of these two parameters with

$$\sigma_{L/Q} \in \left\{ \frac{0.01}{\sqrt{2}}, \frac{0.03}{\sqrt{2}} \right\} . \qquad (2.5)$$

These two options are our "optimistic" and "pessimistic" cases respectively.

Step $\mu$ offset:

Each supernovae data model combines a collection of nearby supernovae with a collection of distant supernovae obtained from separate experiments. We allow for an offset between the photometric systems of the near and far samples. This is represented with as an additional nuisance parameter $\mu^S$ ("S" is for step). Specifically,

$$\mu(z_i) \rightarrow \mu(z_i) + \mu^S$$

for the *near sample only*. We apply a prior to $\mu^S$ with $\sigma_S = 0.01$ mag.

# 3 Specific SN models

## 3.1 Near Sample

All specified SN data sets (except for STAGE I) include a "near sample" of 500 SNe at $z \approx 0.025$. These SNe are always assumed to have spectroscopic redshifts and have the same $\sigma_D$ as the far data.

## 3.2 Stage II

The Table defines the bins and number of supernovae in each bin. The per-event error is given by adding in quadrature $\sigma_D$=0.15 and $\sigma_m$ ($\equiv \sigma_{bin}$) from the Table. The distribution is modeled on SNLS (Astier *et al.* 2006).

| zmax | 1 | 0.9 | 0.8 | 0.7 | 0.6 | 0.5 | 0.4 | 0.3 | 0.2 | 0.1 | 0.08 |
|------|-----|-----|------|------|------|-----|-----|-----|-----|------|------|
| zmin | 0.9 | 0.8 | 0.7 | 0.6 | 0.5 | 0.4 | 0.3 | 0.2 | 0.1 | 0.03 | 0.03 |
| N_bin | 68 | 104 | 111 | 104 | 100 | 86 | 68 | 43 | 11 | 4 | 500 |
| sigma_bin | 0.3 | 0.3 | 0.09 | 0.07 | 0.06 | 0.04 | 0.02 | 0.02 | 0.02 | 0.02 | 0.02 |

Note that the last bin is the "near sample" which overlaps (slightly) with the "deep" sample.

## 3.3 Stage IIIs

This models a spectroscopic experiment identical to Stage II, but with number of SNe per bin rescaled to give a total of 2001 SNe in the far sample:

| zmax | 1 | 0.9 | 0.8 | 0.7 | 0.6 | 0.5 | 0.4 | 0.3 | 0.2 | 0.1 | 0.08 |
|------|-----|-----|------|------|------|-----|-----|-----|-----|------|------|
| zmin | 0.9 | 0.8 | 0.7 | 0.6 | 0.5 | 0.4 | 0.3 | 0.2 | 0.1 | 0.03 | 0.03 |
| N_bin | 195 | 298 | 318 | 298 | 286 | 246 | 195 | 123 | 31 | 11 | 500 |
| sigma_bin | 0.3 | 0.3 | 0.09 | 0.07 | 0.06 | 0.04 | 0.02 | 0.02 | 0.02 | 0.02 | 0.02 |

### 3.4 Stage IIIp

The same distribution as Stage IIIs, but done with photo-z's.

| zmax | 1 | 0.9 | 0.8 | 0.7 | 0.6 | 0.5 | 0.4 | 0.3 | 0.2 | 0.1 | 0.05 |
|---|---|---|---|---|---|---|---|---|---|---|---|
| zmin | 0.9 | 0.8 | 0.7 | 0.6 | 0.5 | 0.4 | 0.3 | 0.2 | 0.1 | 0.05 | 0.01 |
| N_bin | 195 | 298 | 318 | 298 | 286 | 246 | 195 | 123 | 31 | 11 | 500 |

In this case the random errors are described by $\sigma_D$=0.12 mag, $\sigma_m$=0. Additional photo-z errors are modeled by the nuisance parameters discussed above, so that StageIIIp01 and StageIIIp05 are produced according to the value of $\sigma_F$ for the photo-z errors. The last bin is the "near sample" that does not accrue photo-z errors. Note that (for historical reasons) the last two bins are slightly different than for the Stage II and Stage IIs cases.

### 3.5 Stage IVLST

This high-statistics, photo-z experiment is taken to represent LST experiments.

| zmax | 1.2 | 1.1 | 1 | 0.9 | 0.8 | 0.7 | 0.6 | 0.5 |
|---|---|---|---|---|---|---|---|---|
| zmin | 1.1 | 1 | 0.9 | 0.8 | 0.7 | 0.6 | 0.5 | 0.4 |
| N_bin | #### | 20543 | 25505 | #### | 31940 | 31646 | 30379 | 29520 |
| Continued: | | | | | | | | |
| zmax | 0.4 | 0.3 | 0.2 | 0.1 | 0.05 | | | |
| zmin | 0.3 | 0.2 | 0.1 | 0.05 | 0.01 | | | |
| N_bin | #### | 25652 | 20419 | 7221 | 500 | | | |

In this case the random errors are described by $\sigma_D$=0.10 mag, $\sigma_m$=0. Additional photo-z errors are modeled by the nuisance parameters discussed above, so that StageIVLST01 and StageIVLST05 are produced according to the value of $\sigma_F$ for the photo-z errors. Again, the last bin is the "near sample" that does not accrue photo-z errors.

### 3.6 Stage IVS

This represents a supernova experiment carried out on JDEM.

| zmax | 1.7 | 1.6 | 1.5 | 1.4 | 1.3 | 1.2 | 1.1 | 1 | |
|---|---|---|---|---|---|---|---|---|---|
| zmin | 1.6 | 1.5 | 1.4 | 1.3 | 1.2 | 1.1 | 1 | 0.9 | |
| N_bin | 80 | 94 | 107 | 119 | 130 | 142 | 155 | 170 | |
| Continued: | | | | | | | | | |
| zmax | 0.9 | 0.8 | 0.7 | 0.6 | 0.5 | 0.4 | 0.3 | 0.2 | 0.08 |
| zmin | 0.8 | 0.7 | 0.6 | 0.5 | 0.4 | 0.3 | 0.2 | 0.1 | 0.03 |
| N_bin | 179 | 183 | 171 | 150 | 124 | 95 | 64 | 35 | 500 |

In this case the random errors are described by $\sigma_D$=0.10 mag, $\sigma_m$=0. This is similar to the simulation of SNAP found in astro-ph/0304509 but with more SNe in the near sample and with a smaller value of $\sigma_D$.

# 4 Baryon Oscillation data

## 4.1 The Formulas

We model errors on BAO data using the fitting formulae presented by Blake et al (2006). The observables are comoving angular diameter distance $d_A^{co}(z)$ and expansion rate $H(z)$, and the quantities $\ln\left(d_A^{co}(z)\right)$ and $\ln\left(H(z)\right)$ are the observables $y_{i_b}$ that appear in (1.1).

$$d_A = a^2 d_L \tag{4.1}$$

(where $d_L$ is defined in Eqn.(2.2)). The comoving angular diameter distance is

$$d_A^{co} = a^{-1} d_A = a d_L \tag{4.2}$$

To simulate BAO data we consider bins in redshift space of width $\Delta z_i$ centered on the grid $z_i$. For each value of $z_i$ we take $\ln\left(H(z_i)\right)$ and $\ln\left(d_A^{co}(z_i)\right)$ to be measured with errors

$$\sigma_d^i = x_0^d \frac{4}{3} \sqrt{\frac{V_0}{V_i}} f_{nl}(z_i) \tag{4.3}$$

and

$$\sigma_H^i = x_0^H \frac{4}{3} \sqrt{\frac{V_0}{V_i}} f_{nl}(z_i) \tag{4.4}$$

where the commoving survey volume in the redshift bin is, for a survey spanning solid angle $\Omega_{sky}$,

$$V_i = \frac{\left(d_A^{co}(z_i)\right)^2}{H(z_i)} \Omega_{Sky} \Delta z_i \tag{4.5}$$

and erasure of the baryon features by non-linear evolution is factored in using

$$f_{nl}(z_i) = \begin{cases} 1 & z > z_m \\ \left(\frac{z_m}{z_i}\right)^\gamma & z < z_m \end{cases} \tag{4.6}$$

We use the parameters

$$x_0^d = 0.0085$$

$$x_0^H = 0.0148$$

$$V_0 = \frac{2.16}{h^3} Gpc^3 \tag{4.7}$$

$$\gamma = 1/2$$

$$z_m = 1.4$$

The values of $z_i$ and $\Omega_{Sky}$ are specified by the particular survey being modeled.

Systematic errors in the BAO method are possible due to uncertainties in the theory of non-linear evolution and galaxy biasing. These are modeled (for both types of observable) as independent uncertainties in the (log of) the distance measures in each redshift bin:

$$\sigma^i \to \sqrt{\left(\sigma^i\right)^2 + \left(\sigma_S^i\right)^2} \tag{4.8}$$

with

$$\sigma_S^i = 0.01 \times \sqrt{\frac{0.5}{\Delta z_i}} \tag{4.9}$$

## 4.2 Photometric-redshift Surveys

If photo-z's are used, these formulae are slightly modified. The bin redshifts $z_i$ at which the distance measures are evaluated are, as for the SNe, taken to have biases $\delta z_i$, and we apply a Gaussian prior to each of these nuisance parameters with

$$\sigma_z = \frac{\sigma_F \left(1+z\right)}{\sqrt{N_C}} \tag{4.10}$$

Here $N_c$ is intended to represent the number of calibrating galaxies. In calculations we use $N_C = 1000$ and $\sigma_F \in \left\{0.01, 0.05\right\}$. Thus the assumed photo-z bias is, at z=1, in the range 0.003 to 0.0006.

The photometric-redshift survey also surrenders nearly all information along the line of sight, and degrades the accuracy transverse to the line of sight as well. Continuing to follow Blake et al (2006), we replace $x_0$ in Eqns. (4.3), (4.4) and (4.7) with

$$x_0^d = 0.0123 \times \sqrt{\frac{\sigma_F \left(1+z\right)}{\left(\frac{34.1 h^{-1} Mpc}{H^{-1}\left(z\right)}\right)}} \tag{4.11}$$

$$x_0^H = \infty$$

# 5 Specific BAO models

| Name | Omega_Sky (sq. deg) | zmin | zmax | number of bins | Photo-z*** |
|------|---------------------|------|------|----------------|------------|
| IIIa Part 1** | 2000 | 0.5 | 1.3 | 10 | no |
| IIIa Part 2 | 300 | 2.3 | 3.3 | 10 | no |
| IIIb | 4000 | 0.5 | 1.4 | 9 | yes |
| IVS (JDEM) | 10000 | 0.5 | 2 | 10 | no |
| IVG (SKA) | 20000 | 0.01 | 1.5 | 10 | no |
| IVG_2 (LST) | 20000 | 0.2 | 3.5 | 32 | yes |

** Stage III BAO is comprised of two survey parts. These are combined in one Fisher matrix.

*** "yes" here means photo-z systematic errors are implemented as described.

# 6 Clusters

## 6.1 Statistical Errors

The number density of rare clusters as a function of their mass $M$ is exponentially sensitive to the linear density field and hence its rate of growth as well as linearly sensitive to the volume element. Cosmological $N$-body simulations can accurately calibrate this mass function and suggest that the differential comoving number density of clusters can be approximated by

$$\frac{dn}{d \ln M} = 0.3 \frac{\rho_m}{M} \frac{d \ln \sigma^{-1}}{d \ln M} \exp\left[ -\left| \ln \sigma^{-1} + 0.64 \right|^{3.82} \right], \qquad (6.1)$$

Where $\sigma^2(M;z) \equiv \sigma_R^2(z)$, the linear density field variance in a region enclosing $M = 4\pi R^3 \rho_m / 3$ at the mean matter density today $\rho_m$. Given this exponential sensitivity, dark energy projections for clusters also depend sensitively on the amplitude of structure today $\sigma_8(z=0)$ and we have assumed a value of 0.91 throughout.

Take as the data the number of clusters $N_i$ in the $i$th pixel. The pixels are defined by their angular and redshift extent as well as selection criteria for the clusters in the sample. These selection criteria can be characterized as a mass selection function $P_i(M)$ such that

$$\bar{N}_i = V_i \int d(\ln M) P_i(M) \frac{dn}{d \ln M}. \qquad (6.2)$$

Statistical fluctuations in the number counts arise from Poisson shot noise and sample variance due to the large scale structure of the universe. The covariance matrix of the counts becomes

$$C_{ij} \equiv \left\langle \left( N_i - \bar{N}_i \right)_i \left( N_j - \bar{N}_j \right) \right\rangle = \delta_{ij} \bar{N}_i + S_{ij}, \tag{6.3}$$

where the sample variance is

$$S_{ij} = b_i \bar{N}_i b_j \bar{N}_j \int \frac{d^3 k}{(2\pi)^3} W_i^*(\mathbf{k}) W_j(\mathbf{k}) P(k). \tag{6.4}$$

$W_i(\mathbf{k})$ is the Fourier transform of the pixel window normalized such that $\int d^3 x W_i(\mathbf{x}) = 1$. Here $b_i$ is the average bias of the selected clusters

$$b_i = \frac{V_i}{\bar{N}_i} \int \frac{dM}{M} \frac{dn_i}{d \ln M} b(M; z_i). \tag{6.5}$$

Cosmological *N*-body simulations calibrate the bias as a function of mass. We adopt

$$b(M; z) = 1 + \frac{a_c \delta_c^2 / \sigma^2 - 1}{\delta_c} + \frac{2 p_c}{\delta_c \left[ 1 + \left( a_c \delta_c^2 / \sigma^2 \right)^{p_c} \right]}, \tag{6.6}$$

with $a_c = 0.75$, $p_c = 0.3$, and $\delta_c = 1.69$.

The Fisher matrix of the number counts is then

$$F_{\mu\nu} = \sum_{ij} \frac{\partial \bar{N}_i}{\partial p_\mu} \left( \mathbf{C}^{-1} \right)_{ij} \frac{\partial \bar{N}_j}{\partial p_\nu}, \tag{6.7}$$

where $p_\nu$ contains both cosmological parameters and nuisance parameters that define the mass selection.

## 6.2 Mass Selection

The mass selection function $P_i(M)$ carries uncertainties since the mass is not a direct observable. Observable quantities such as the SZ flux decrement, X-ray temperature and gas mass have uncertainties in their mean relationship to cluster mass as well as an unknown distribution around the mean. Finally instrumental effects and inaccurate modeling of contaminating point sources can distort the selection especially near threshold. Characterizing the mass selection function is the main challenge for extracting the dark energy information in the number counts.

To assess the importance of characterizing the mass selection function we introduce two nuisance parameters per redshift interval of $\Delta z = 0.1$

$$P_i(M) = \frac{1}{2} \text{Erfc} \left[ \frac{\ln M_{th}(z) - \ln M}{\sqrt{2 \sigma_{\ln M}^2(z)}} \right] \tag{6.8}$$

that models an unknown mean and Gaussian variance in the mass observable relations.

Priors on these mass selection parameters can come from hydrodynamic cosmological simulations, hydrostatic equilibrium solutions involving a measured $X$-ray temperature profile, gas mass from the $X$-ray surface brightness, weak lensing measurements of individual clusters, and statistical weak lensing calibration through the cluster-mass correlation function. Low and intermediate redshift observations currently suggest that these mass selection parameters are known to better than the 10% level.

Priors can also come from the counts data themselves in a form of "self-calibration." Self-calibration may be especially useful for the high redshift clusters where multiple wavelength measurements are more difficult to obtain. There are at least two sources of information for self-calibration: the spatial clustering of the clusters and their abundance as a function of the observable. Given that cosmological simulations predict both the shape of the mass function and the bias as a function of mass, the statistical properties of the number counts will only be consistent for the correct mass selection function. The full Fisher matrix of the counts as a function of spatial position and observable therefore contains more information than we represent in Eqn. (6.7). Since this information is mainly useful for calibrating the mass-observable relation, we absorb it into priors on the nuisance parameters. Our pessimistic projections reflect the efficacy of self-calibration. Note however that if the mass selection is fixed by external calibration, this additional information in the counts can enhance the dark energy constraints themselves.

## 6.3 Mass Outliers

Given the steepness of the mass function near threshold, non-Gaussian tails in the mass-observable relations can easily contaminate the counts from the low mass end. We do not attempt to model these effects in detail but instead provide a rough translation of constraints based on our Gaussian distribution for a more general case.

In order for tails in the mass selection function not to overwhelm the dark energy signal, uncertainties in the mass selection function at some $M << M_{th}$ must be controlled at roughly the level

$$P_i(M) < \sigma(w_p)\left(\frac{M}{M_{th}}\right)^\alpha,$$ (6.9)

where $\alpha$ is the effective slope of the mass function $dn/d\ln M$ between $M$ and $M_{th}$. For example, near $M_{th} = 10^{14.2} h^{-1} M_\odot$ and a typical $z = 0.7$, $\alpha = 2$. Thus the selection function for clusters that are 1/3 of the threshold must yield less that $\sim$1% contamination in order to measure $w_p$ to $\sim$10%.

Examples of effects that might produce outliers include unsubtracted point source contamination and projection effects. Note however that point sources would affect different cluster observables in different ways. For example for SZ-selected clusters, point contamination fills in the decrement and scatters rare high mass clusters down in effective mass whereas for *X*-ray-selected clusters, AGN can bring the abundant low mass clusters into the sample unless the instrument resolution is better than approximately 10". Conversely, *X*-ray-selected clusters are the most robust against chance projections of structure that can cause low mass clusters to masquerade as high mass clusters whereas lensing-selected clusters will be limited in their utility by these projection effects. Thus the approximately 20000 5-$\sigma$ shear selected clusters from LST or SNAP may not reflect the number of clusters that can be used for dark energy studies.

### 6.4 Fiducial Surveys

We model a generic cluster count survey that has mass selection function with a *constant* threshold mass $M_{th}$ and *constant* scatter of $\sigma_{\ln M} = 0.25$. This is *not* expected to model an *X*-ray or *SZ* selection in detail so that the results should be taken qualitatively not quantitatively. We are mainly interested in making a correspondence between a level of degradation in $\sigma(w_p)$ vs statistical errors, $N = \sigma(w_p)/\sigma_{statistical}(w_p)$, and control over the mass selection. Our pessimistic projection reflects control over the selection that should conservatively be available through self-calibration or extrapolation of current data provided the selection function varies slowly with mass and redshift. Our optimistic projection reflects levels that are potentially achievable through multi-wavelength observations and detailed modeling of the clusters. The number of clusters in each category should be taken as an arbitrary normalizing point since there is a currently unknown trade-off between decreasing statistical errors and increasing systematic errors by including more objects identified in the survey. Statistical errors will scale mainly with $N_{clusters}^{-1/2}$ and our results can be adjusted accordingly as estimates of the trade-off become more refined.

Stage II:

200 deg$^2$; $M_{th} = 10^{14}h^{-1}M_\odot$; $z_{max} = 2$; 4000 to 5000 clusters.

Pessimistic: $N = 3$ or $\sigma(\ln M_{th}) = 0.27$ and $\sigma(\sigma_{\ln M}^2) = 0.27$ per $\Delta z = 0.1$.

Optimistic: $N = \sqrt{2}$ or $\sigma(\ln M_{th}) = 0.08$ and $\sigma(\sigma_{\ln M}^2) = 0.08$ per $\Delta z = 0.1$.

Stage III:

4000 deg$^2$; $M_{th} = 10^{14.2}h^{-1}M_\odot$; $z_{max} = 2$; 30,000 clusters.

Pessimistic: $N = 3$ or $\sigma(\ln M_{th}) = 0.14$ and $\sigma(\sigma_{\ln M}^2) = 0.14$ per $\Delta z = 0.1$.

Optimistic: $N = \sqrt{2}$ or $\sigma\left(\ln M_{th}\right) = 0.02$ and $\sigma\left(\sigma^2_{\ln M}\right) = 0.02$ per $\Delta z = 0.1$.

Stage IV:

20000 deg$^2$; $M_{th} = 10^{14.4} h^{-1} M_\odot$; $z_{max} = 2$; 30,000 clusters.

Pessimistic: $N = 3$ or $\sigma\left(\ln M_{th}\right) = 0.11$ and $\sigma\left(\sigma^2_{\ln M}\right) = 0.11$ per $\Delta z = 0.1$.

Optimistic: $N = \sqrt{2}$ or $\sigma\left(\ln M_{th}\right) = 0.016$ and $\sigma\left(\sigma^2_{\ln M}\right) = 0.016$ per $\Delta z = 0.1$.

# 7 Weak Lensing

## 7.1 Overview

This Appendix details the method for predicting dark-energy constraints from future weak-lensing experiments. The theory and forecasting methods for weak lensing are still evolving rapidly (Mirlada-Escude 1991; Blandford et al.1991; Kaiser 1992; Jain and Seljak 1997; Hu 1999; Jain and Taylor 2003; Bernstein and Jain 2004; Song and Knox 2004; Takada and Jain 2004; Zhang *et al.* 2005; Knox *et al.* 2005; Bernstein (2006); Huterer et al. 2006) so the DETF calculations can be considered only a snapshot. Most, but not all, of the statistical information and systematic effects believed to be important have been included in this analysis: shear-shear 2-point correlations are analyzed, assuming use of tomography to exploit redshift information; similarly for galaxy-shear 2-point correlations. Systematic errors that are treated include redshift-dependent shear calibration biases; photo-$z$ biases; intrinsic galaxy shape correlations; intrinsic shape-density correlations; and uncertainties in the theoretical power spectrum due to baryonic physics. Bispectrum information is known to be important, but not included here (Takada and Jain 2004); uncertain structure in the photo-$z$ distributions beyond simple biases are also known to be important, but not included here (Ma *et al.* 2006). To date there are no analyses in the literature that treat all these elements simultaneously.

## 7.2 Lensing Fisher Matrix

There are three fields of interest: $\kappa$ is the convergence that would be inferred from the (E-mode) shear pattern of source galaxies at this redshift shell; $m$ is the mass overdensity; and $g$ is an *estimate* of the mass overdensity that is derived from the galaxy distribution. Note that this need not be the galaxy distribution itself, but more likely will involve an attempt to assign halos to galaxies or groups.

All three are properly functions in the 3d continuum space $(\theta, \varphi_s)$. To render our problem discrete, we decompose the angular variables into spherical harmonic coordinates $lm$, and slice the depth variable into a finite set of shells indexed by $k$. Each depth bin is idealized by a set of galaxies confined to a thin shell at redshift $z_k$ with comoving angular diameter distance $D_k = D_A(z_k)$. Taking the limit of a large number of shells should converge back to the continuum limit.

We will drop the spherical-harmonic indices $lm$ from our quantities for brevity, leaving our three fields described by vectors $\kappa$, $\mathbf{m}$, and $\mathbf{g}$ over the distance bins. To make a Fisher matrix, we need a likelihood function $L(\kappa, \mathbf{m}, \mathbf{g})$, which we then must marginalize over the unobservable $\mathbf{m}$. This probability can be expressed as

$$L(\kappa, \mathbf{m}, \mathbf{g}) = L(\kappa \mid \mathbf{m}) L(\mathbf{m}, \mathbf{g}). \tag{7.1}$$

Note the implicit assumption here that likelihoods for different spherical harmonics are independent.

The first likelihood term is straightforward, if we split the convergence into the deterministic term from gravitational lensing, and a stochastic term from the intrinsic orientations of galaxies:

$$\kappa = \kappa_{\text{lens}} + \kappa_{\text{intrinsic}}, \tag{7.2}$$

$$\kappa_{\text{lens}} = \mathbf{Am}, \tag{7.3}$$

$$A_{s\ell} = \frac{3\omega_m \Delta\chi_\ell D_\ell}{2a_\ell} \begin{cases} \left(1 - D_\ell/D_s\right)\left(1 - \omega_K D_\ell D_s/2\right)(1 + f_s) & s > \ell \\ 0 & s \le \ell \end{cases} \tag{7.4}$$

Here we take $\omega_{[mK]} = \Omega_{[mK]} h^2$, measure distance in units of $c/H_{100} = 2998$ Mpc, and approximate the effect of curvature to first order in $\omega_K$. The comoving radial thickness of the lens shell is $\Delta\chi = \Delta D(1 + \omega_K D^2/2)$. We've also introduced the systematic-error variables $f_s$, which describe a multiplicative (calibration) error on the shear measured on galaxies in source shell $s$. The calibration error is assumed to be scale-independent.

The intrinsic galaxy shapes are usually assumed to have scale-independent variance of $N_s = \sigma_\gamma^2/n_s$, where $n_s$ is an effective number of sources on shell $s$. We allow, however, for the possibility that galaxies have intrinsic alignment with each other and potentially with the local mass distribution, such that $\langle \kappa_{\text{intrinsic},k} \mid m_k \rangle = A_{kk} m_k$. A non-zero shear-mass correlation will produce the "GI" systematic effect described by Hirata and Seljak (2004). We must allow this term to depend upon angular scale $l$ and redshift. Similarly, there may exist intrinsic shape correlations that do not correlate with the local density, so we must consider $\text{Var}(\kappa_s) = N_s$ potentially to have slight scale-dependent departures from the fiducial $\sigma_\gamma^2/n_s$. Leaving $N_s(l)$ as a free parameter will allow us to marginalize over the intrinsic alignments termed "II" by Hirata and Seljak (2004).

The random shape noise will be Gaussian to high accuracy, with a covariance matrix that is diagonal over multipole and redshift indices. The intrinsic correlations will likely be weak, and will not cross distance bins, so the likelihood function of the total convergence will be Gaussian to good approximation at all scales, given by:

$$L(\kappa\,|\,\mathbf{m}) = \left|2\pi\mathbf{N}\right|^{-1/2}\exp\left\{-\frac{1}{2}\left[(\kappa-\mathbf{Am})^T\,\mathbf{N}^{-1}(\kappa-\mathbf{Am})\right]\right\}. \tag{7.5}$$

Here $\mathbf{N} = \mathrm{diag}(N_s)$, and we have placed the intrinsic-correlation terms $A_{kk}$ into the otherwise empty diagonal of the geometry matrix $\mathbf{A}$.

### 7.3 Mass/galaxy likelihood

Next we need to determine the joint likelihood $L(\mathbf{m},\mathbf{g})$. We start with the assumption that the relation is local in redshift, and our redshift shells are thick enough that each shell is independent of every other. We define the bias and correlation

$$B \equiv \left\langle\frac{\langle m\,|\,g\rangle}{g}\right\rangle = \frac{\langle gm\rangle}{\langle g^2\rangle} = \frac{C_{gm}}{C_g}, \tag{7.6}$$

$$r \equiv \frac{\langle gm\rangle}{\sqrt{\langle g^2\rangle\langle m^2\rangle}} = \frac{C_{gm}}{\sqrt{C_g C_m}}. \tag{7.7}$$

Note that this is the bias of the *mass* for a fixed *galaxy* estimator; the usual bias parameter $b$ specifies the converse, and is given by $b = r^2/B$.

Next we assume that the likelihood function for $g$ and $m$ is Gaussian, with each multipole independent. This assumption is much less secure than the assumption that $L(\kappa\,|\,\mathbf{m})$ is Gaussian, so may limit the regime of applicability. With the Gaussian assumption, we have at each multipole

$$L(m,g) = \left[(2\pi)^2 C_u C_g\right]^{-1/2}\exp\left\{-\frac{1}{2}\left[\frac{(m-Bg)^2}{C_u}+\frac{g^2}{C_g}\right]\right\} \tag{7.8}$$

$$C_u \equiv (1-r^2)C_m \tag{7.9}$$

$C_u$ is the portion of the mass variance which is uncorrelated with the mass estimator $g$.

### 7.4 Likelihood Function

Multiplying Equations (7) and (10), then integrating over $\mathbf{m}$ gives the Gaussian distribution

$$L(\kappa,\mathbf{g}) = \left|2\pi\mathbf{K}\right|^{-1/2}\left|2\pi\mathbf{C_g}\right|^{-1/2}\exp\left\{-\frac{1}{2}\left[\mathbf{g}^T\,\mathbf{C_g}^{-1}\mathbf{g}+(\kappa-\mathbf{ABg})^T\,\mathbf{K}^{-1}(\kappa-\mathbf{ABg})\right]\right\}, \tag{7.10}$$

$$\mathbf{K} = \mathbf{N}+\mathbf{AC_u}\mathbf{A}^T. \tag{7.11}$$

Here the matrices $\mathbf{N}, \mathbf{C_g}, \mathbf{C_u}, \mathbf{B}$ are $\mathrm{diag}(N_s)$, etc., and we recall that this likelihood applies to a single harmonic $lm$. $\mathbf{K}$ is the covariance matrix for $\kappa$ if the mass estimators $g$ are held fixed, *i.e.* only the components of the mass distribution that are uncorrelated with $g$ are considered stochastic.

## 7.5 Lensing Fisher Matrix

From this multivariate Gaussian distribution we may derive a Fisher matrix using the formulae for zero-mean distribution given, *e.g.*, by Tegmark *et al.* (1997). We also multiply this by the number of spherical harmonic modes in our bin of width $\Delta l$ to get

$$F_{ij} = l\Delta l\, f_{\text{sky}} \operatorname{Tr}\left[\mathbf{C_g}^{-1}\mathbf{C_{g,i}}\mathbf{C_g}^{-1}\mathbf{C_{g,j}} + \mathbf{K}^{-1}\mathbf{K}_{,i}\mathbf{K}^{-1}\mathbf{K}_{,j} + 2\mathbf{K}^{-1}(\mathbf{AB})_{,i}\,\mathbf{C_g}(\mathbf{AB})_{,j}^T\right], \quad (7.12)$$

with the commas in the subscripts denoting differentiation. The Fisher information nicely separates into three parts: the first is the information that can be gleaned from the variances of the mass estimator $g$, *i.e.*, the galaxy power spectrum. The second term is Fisher information that would arise from adjusting parameters to minimize the $\chi^2$ in the fit of $\kappa$ to the estimated mass $g$, with the values of $g$ taken as fixed and the matrix $\mathbf{K}$ taken as a known covariance for the $\kappa$ values. The third term is information gleaned from the covariance of the $\kappa$ residuals to this fit, and looks just like the Fisher matrix for pure shear power-spectrum tomography, except that the relevant mass power spectrum in this term is $\mathbf{C_u}$, the power that is uncorrelated with the galaxies, not the (larger) total power of $m$.

We take the parameters of this Fisher matrix to be, most generally:

- $C_u, C_g$, and $B$ for each of the $Z$ redshift bins and $L$ bins in $l$, for a total of $3LZ$ parameters.
- The intrinsic-correlation parameters $A_{kk}$ and $N_s$ at each redshift and angular scale, another $2LZ$ free parameters;
- Cosmological parameters $\omega_k$ and $\omega_M$;
- The angular-diameter distances $D_k$ ($Z$ free parameters);
- The shear-calibration errors $f_s$ ($Z$ free parameters);
- The scale factor $a_k$ at each redshift shell ($Z$ free parameters), or equivalently a redshift estimation error $\delta z_k$ with $1/a_k = 1 + z_k + \delta z_k$.

Note that $\Delta\chi_k$ is expressible in terms of $D_k$, $D_{k\pm 1}$, and $\omega_k$. With these parameters, all of the derivatives required in Equation (7.14) are simple, sparse matrices.

## 7.6 Power Spectrum Prior

For a given cosmological theory we should be able to predict the 3-dimensional power spectrum $P_\delta(k, z)$ of mass overdensity. This prediction, however, may have some finite uncertainty $\sigma_{\log P}$ due to the difficulties of predicting non-linear growth, especially at high $k$ and low $z$ where baryonic physics may be important. We incorporate the theoretical knowledge/ignorance into the Fisher matrix by noting that the Limber approximation implies

$$C_m = C_u + B^2 C_g = P_\delta(l/D, z)D^2\Delta\chi \quad (7.13)$$

at each bin in $l$ and $z$. We therefore define a power-spectrum prior likelihood for the bin at $l$ to be

$$-\log L_{\mathrm{PS}} = \frac{1}{2} \sum_k \frac{\left[ \log(C_u + B^2 C_g)_k - \log(D_k^2 \Delta\chi_k) - \log P_\delta(l/D_k, z_k + \delta z_k) \right]^2}{\sigma_{\log P}^2(l/D_k, z_k)} \quad (7.14)$$

From this likelihood we may produce a Fisher matrix. We will assume that the theory uncertainty $\sigma_{\log P}$ is independent of all parameters, so we need only know the derivatives of the logarithms in the numerator of the sum. The parameters of the Fisher matrix will clearly include those in the lensing Fisher matrix, namely $\{C_u, B, C_g, D, \delta z, \omega_k\}$, but must also include any further parameters that will affect the power spectrum $P_\delta$. This would of course include dark-energy parameters, or any parameters of extensions to General Relativity.

In our current implementation we make the assumption that the full non-linear power spectrum $P_{NL}(k, z)$ is a function solely of the linear power spectrum $P_L$ at the same era. The linear power spectrum is in turn the product of the spectrum at an initial epoch (*e.g.* recombination) and a growth factor $G$ since that epoch. All dark-energy dependence of the non-linear spectrum is thence absorbed into a vector $\mathbf{G}$ of growth-function values $G_k$ at galaxy shell $k$. In General Relativity the linear growth factor is scale-independent; we could generalize to a scale-dependent growth factor given a suitable non-GR gravity theory.

We adopt a power-law primordial power spectrum $P_\phi = A_s(k/k_0)^{n_s-1}$, where $k_0 = 0.05$ Mpc$^{-1}$ by the WMAP convention. The transfer function is a function of $\omega_M$ and $\omega_B$ (ignoring neutrinos); we use the formulation of Eisenstein and Hu (1999). The non-linear power spectrum can hence be expressed as $P_{\mathrm{NL}}(l/D_k, G_k; \omega)$ where we take the shorthand $\omega = \{\omega_M, \omega_B, \omega_k, n_s, A_s\}$ for the set of cosmological parameters that are independent of any parameterization of dark energy.

Peacock and Dodds (1996) propose a mapping from the linear to non-linear power spectrum. We instead use the Smith *et al.* (2003) prescription, in which the non-linear power spectrum at redshift $z$ is determined by the linear power spectrum at $z$ but also requires the matter density $\Omega_m(z)$ (for nearly-flat models). We ignore this subtlety, fixing $\Omega_m(z)$ at the value for the fiducial cosmology, so as to preserve the simplicity that all dark-energy dependence is implicit through $G$. We emphasize that this simplification could be abandoned for a more complex description of the non-linear power spectrum if desired, at the cost of additional Fisher-matrix parameters.

The power-spectrum theory prior is now a Fisher matrix over the same parameters as the lensing Fisher matrix, with the addition of the linear-spectrum parameters $\omega_B$, $n_S$, and $A_s$, plus the growth factor vector $\mathbf{G}$ over the redshift shells.

## 7.7 Marginalization

Once we have summed the lensing Fisher matrix with the power-spectrum-theory Fisher matrix for a given $l$ bin, we can marginalize over the parameters $\{\mathbf{C_u}, \mathbf{B}, \mathbf{C_g}\}$ at all redshifts shells. We note that when $\sigma_{\log P}$ is small, the power-spectrum-theory Fisher matrix can be large, causing roundoff error in the subsequent marginalization. It is numerically advantageous to combine the addition of the prior and the marginalization into a single operation. The power-spectrum Fisher matrix can be expressed as $\mathbf{D}^T \Sigma^{-1} \mathbf{D}$, where $\Sigma = \mathrm{diag}(\sigma_{\log P}^2)$, and $\mathbf{D}$ is a matrix of the derivatives of the (log) power spectra with respect to the Fisher-matrix parameters. If we divide the Fisher parameters into the subvectors A that we wish to keep and B that will be marginalized away, then our summed, marginalized Fisher matrix is

$$\mathbf{F}' = \left[ \left( (\mathbf{F} + \mathbf{D}^T \Sigma^{-1} \mathbf{D})_{AA}^{-1} \right) \right]^{-1} \tag{7.15}$$

$$= F_{AA} - F_{AB} F_{BB}^{-1} F_{BA} \tag{7.16}$$

$$+ (D_A - F_{AB} F_{BB}^{-1} D_B)(\Sigma + D_B^T F_{BB}^{-1} D_B)^{-1} (D_A - F_{AB} F_{BB}^{-1} D_B)^T. \tag{7.17}$$

This form avoids roundoff error as $\sigma_{\log P} \to 0$.

In the cases $r = C_g = 0$ and $\Sigma = 0$, we also may use this expression to recover the usual Fisher matrix for power-spectrum tomography.

## 7.8 Power Spectrum Uncertainties

Zhan and Knox (2004) quantify the effect of baryonic pressure support on the observed $C_l$ lensing spectrum using a halo model. White (2004) estimates the effect of baryonic cooling with a similar approach. These two effects have opposite signs but roughly similar amplitudes. A numerical simulation by Jing *et al.* (2006) incorporates both effects and confirms the amplitude of the baryonic effects on the power spectrum. Hu Zhan has kindly provided a tabulation of $\delta \log P_b(k, z)$, the fractional change in power due to the hot-baryon effect. We will assume here that $\sigma_{\log P}(k, z) = E_{\mathrm{PS}} \, \delta \log P_b(k, z)$, with the logic that the accuracy of future power-spectrum calculations will be some fraction of the total baryonic effect.

A rough fit to the Zhan data is

$$\delta \log P_b(k, z) = 0.012 \begin{cases} 1 + 5 \log_{10}(k/k_1) & k > k_1 \\ (k/k_1)^{1+a} & k < k_1 \\ k_1 \equiv a^{-2.6} \ \mathrm{Mpc}^{-1} \end{cases} \tag{7.18}$$

When $\sigma_{\log P} > 1$, we simply marginalize over $\{C_u, C_g, B\}$ without applying any power-spectrum prior information. The Fisher matrix then becomes equivalent to that used in Bernstein (2006) for analysis of pure cross-correlation cosmography.

## 7.9 Projection onto Dark Energy

After summing the lensing and power-spectrum-prior Fisher matrices, and marginalizing over the shell powers and biases, we next eliminate the other $l$-dependent nuisance parameters $N_k$ and $A_{kk}$ associated with intrinsic galaxy alignments. At present we apply no external prior constraints on these systematic variables, *i.e.* we presume they must be completely self-calibrated from the lensing data.

The Fisher matrices for the different $l$ bins may now be summed. Because the power-spectrum information is degraded at highly non-linear $(k, z)$, we have less concern over the choice of maximum permissible $l$ for the analysis. The non-Gaussian nature of power at high $l$ is of less concern because the Fisher matrix is automatically using only cross-correlation information at high $l$, and the measurement uncertainties are primarily traceable to shear noise—which remains Gaussian at all $l$—rather than the behavior of the mass fluctuations.

We may now apply any desired prior to the shear-calibration factors $f_k$, then marginalize these away. What remains is a Fisher matrix over the parameter vector $\{\omega, D_k, G_k, \delta z_k\}$. At this point one may add prior constraints on the redshift biases $\delta z_k$. This Fisher matrix is independent of the choice of dark-energy or gravity model, as long as the gravity model preserves scale-independent linear growth and maintains the GR form for light deflection.

We next may constrain any model for dark energy or gravity which is described by additional cosmological parameters $Q_i$. The dark-energy model will predict the distances $D_k = D_A(z_k + \delta z_k; \omega, Q_i)$ and similarly the growth factors $G_k = G(z_k + \delta z_k; \omega, Q_i)$. We may therefore project the Fisher matrix onto a new set of variables $\{\omega, Q_i, \delta z_k\}$, then marginalize over redshift errors to obtain the final Fisher matrix over the cosmological parameters $\omega$ and the dark energy parameters.

## 7.10 A Note on Priors

Our formalism allows for prior constraints on systematic-error nuisance parameters which represent functions of $z$, namely $f_k$ and $\delta z_k$. We assume that the priors are Gaussian in $f_k$ and $\log a_k$, with standard deviations $\sigma_f$ and $\sigma_{\log a}$.

In order for our results to be stable under a change in $z$ step size, we must scale these priors by the inverse square root of the bin width $\Delta z$. We specify input values $E_f$ and $E_{\log a}$ which refer to the values on shells of width $\Delta \log a = 0.15$; then each shell gets an equal prior uncertainty

$$\sigma_{\log a} = E_{\log a}\left(\frac{\Delta \log a}{0.15}\right)^{1/2},$$

$$\sigma_f = E_f\left(\frac{\Delta \log a}{0.15}\right)^{1/2}.$$

In other words we are assuming that these systematic errors, which are functions of $z$ in the continuum limit, have a fixed variance per unit redshift. Alternatively we could adopt the approach of Huterer *et al.* (2006), and project the systematic errors onto the coefficients of a (truncated) series of orthogonal functions of $z$.

Similarly there are systematic-error priors on several quantities that are functions of both $z$ and angular scale $l$, namely the power-spectrum theoretical uncertainties and the intrinsic-correlation strengths. If we choose finite priors, we must rescale them by the square roots of both the $z$ and the $l$ bin widths. We use bins logarithmically spaced in $l$, and refer all systematic errors to a standard bin of width 0.5 dex. We specify an input $E_{\mathrm{PS}}$ and then adopt

$$\sigma_{\log P}(k,z) = E_{\mathrm{PS}} \left( \frac{\Delta \log a}{0.15} \frac{\Delta \log l}{0.5 \log 10} \right)^{1/2} \delta \log P_\delta(k,z). \qquad (7.19)$$

A statement of the level of systematic error on a $z$-dependent quantity is always meaningless without some accompanying description of the averaging width or functional form to which it applies.

### 7.11  Fiducial Case

The fiducial $\Lambda$CDM cosmology common to all DETF models is presumed. The Eisenstein and Hu (1999) transfer function and Smith *et al.* (2003) recipe for non-linearity define the power spectrum $P_\delta(k,z)$. The fiducial value of $C_m$ follows from Equation (7.15) given a choice of redshift binning. We set the fiducial $B = 1$ at all redshifts and scales—this choice does not affect the results. Of more importance is the choice of fiducial correlation coefficient $r$ between mass and its estimator. We take $r = 0.5$ unless otherwise noted, which leads to fiducial $C_u = (1 - r^2)C_m$ and $C_g = r^2 C_m$.

Our discretized Fisher matrix should converge toward the true continuum value as the shell width goes to zero. In practice we find convergence of parameter accuracies for $\Delta \log a \leq 0.1$, sufficiently large that our assumption of uncorrelated redshift shells is valid for the multipoles $l > 100$ which provide most of the information.

## 8  PLANCK priors

The Planck Fisher matrix is initially calculated for a flat $\Lambda$CDM Universe with an adiabatic power-law primordial scalar perturbation power spectrum. Tensor and isocurvature perturbations are assumed to be zero, as are neutrino masses. Reionization is assumed to occur in a step process at the redshift of $z_{ri}$ which we marginalize over. We also let the primordial Helium mass fraction, $Y_P$ float and marginalize over that. After calculating the Planck Fisher matrix in this space (which has $\Omega_k = 1 + w_0 = w_a = 0$ ) we then transform it to the full standard DETF parameter space.

We model the Planck dataset as foreground-free maps of CMB temperature and polarization over 80% of the sky with homogeneous white noise. Each map has been smoothed with a circular Gaussian beam. The beam size and noise levels are listed in the Table.

| Experiment | $l_{max}^{T}$ | $l_{max}^{E,B}$ | $\nu$ (GHz) | $\theta_b$ | $\Delta_T$ ($\mu K$) | $\Delta_P$ ($\mu K$) |
|---|---|---|---|---|---|---|
| Planck | 2000 | 2500 | 100 | 9.2' | 5.5 | $\infty$ |
| | | | 143 | 7.1' | 6 | 11 |
| | | | 217 | 5.0' | 13 | 27 |

Table 1. Experimental specifications. The $\theta_b$ are the full-width at half maximum of the beam profiles. The $\Delta_T$ and $\Delta_P$ are temperature and polarization (Stokes parameters $Q$ and $U$ ) noise standard deviations in a pixel of area $\theta_b^2$.

Although Planck has frequency bands from 30 up to 850 GHz we ignore these in our analysis. We are crudely taking foregrounds into account by assuming that over 80% of the sky these outer channels can be used to remove the foreground contributions to the central channels without significant increase in the noise in the central channels, and further by assuming that the other 20% of the sky is irretrievably contaminated by galactic emission.

We use the formalism for calculating the Fisher matrix given CMB temperature and polarization maps as laid out in Zaldarriaga *et al.* 1997 and Bond *et al.* 1997. The important quantities are the auto and cross angular power spectra of the two fields, temperature (T) and E-mode polarization (E), and their derivatives with respect to the cosmological parameters.

We discard polarization information at $l < 30$ because our assumptions about foregrounds are almost certainly too optimistic for polarization on large angular scales. These low $l$ values are important for constraining the optical depth, $\tau$, to Thomson scattering by electrons in the reionized inter-galactic medium. Forecasts that include foreground modeling (Tegmark et al. 2000) indicate that Planck can determine $\tau$ to $\pm 0.01$ so we include the appropriate prior on $z_{ri}$ in order to achieve $\sigma(\tau) = 0.01$. Were we to drop the prior, but include the polarization data at $l < 30$ we would find $\sigma(\tau) = 0.005$ which is not only better than what we expect because of foreground contamination, but also better than we can expect due to uncertainty in the shape of the reionization history (Holder et al. 2003).

We discard temperature data at $l > 2000$ to reduce our sensitivity to contributions from "patchy" reionization (Knox et al. 1998, Santos et al. 2003) and residual point source contamination. For similar reasons we discard polarization data at $l > 2500$.

We first calculate the Fisher matrix assuming a flat universe with a cosmological constant. That is, we fix $w_0 = -1$, $w_a = 0$ and $\Omega_K = 0$. This is because there is a strict geometric degeneracy (at least for $l \gtrsim 20$) between these three parameters and $\Omega_X$. We do the calculation in this manner to make sure the degeneracy is not artificially broken. It may appear that we are artificially breaking it by fixing three of the parameters, but we fix that later by putting the degeneracy back in by hand. Our parameter space is

$$Y = \{n_s, \ln A_s, \omega_m, \omega_b, \theta_s, Y_p, z_{\mathrm{ri}}\}$$

where $\theta_s$ is the angular size of the sound horizon, which we use instead of $\Omega_\Lambda$, $Y_p$ is the primordial fraction of nuclear mass in Helium-4, $z_{\mathrm{ri}}$ is the redshift of reionization, assumed to occur instantaneously. The Helium-4 mass fraction and reionization redshift are nuisance parameters that we marginalize over. The first five are the five parameters (of the eight that we care about) that are well-determined by the CMB. Varying $w_0$, $w_a$, $\Omega_k$, and $\Omega_X$ in a manner that leaves the above parameters fixed will not change the CMB observables, except at very low $l$ where there is large cosmic variance.

We calculate the Fisher matrix in the $Y$ parameter space, marginalize over $Y_p$ and $z_{\mathrm{ri}}$ and then transform the resulting 5-dimensional Fisher matrix to the eight-dimensional $X$ parameter space where

$$X = \{w_0, w_a, \Omega_X, \omega_m, \omega_b, n_s, \ln A_s\}$$

With some parameter re-ordering, the Jacobian for this transformation is mostly trivial; *i.e.,* most of the Jacobian is the identity matrix. The only non-trivial parts are the derivatives of $\theta_s$ with respect to all the $X$ parameters except for $n_s$ and $\ln P$ (since these derivatives are zero). These derivatives we calculate numerically by finite difference.

By following the above procedure we have restored and rigorously enforced the geometric degeneracy.

Our treatment of the CMB is conservative in the sense that we have ignored lensing signals (particularly the lensing-induced polarization B modes) and low $l$ signals (ISW) that are sensitive to the dark energy and could potentially provide us with more information about the dark energy. Including the ISW effect would not significantly change the DETF figure of merit forecasts for any probe, or combination of probes, with figure of merit greater than about 10. To exploit the dark-energy dependence of the lensing-induced B modes in a significant manner requires higher resolution and sensitivity than Planck (Acquaviva and Baccigalupi 2005).

# 9  Other technical issues

## 9.1  Combining issues

As discussed in Section 0, we combine model data sets by adding Fisher matrices produced individually for each data model in our eight dimensional cosmological parameter space.  This leads us to neglect several effects that should be studied in further work.

### 9.1.1  Nuisance Parameters

Some data models may have common nuisance parameters.  Ideally these data models should be combined in the higher dimensional parameter space that includes the common nuisance parameters before marginalizing down to the eight standard parameters.  As an illustration, when we construct Stage 3 and Stage 4 figures of merit "normalized to stage 2" we include Fisher matrices from all Stage 2 data in all single and combined cases.  This means, for example, that supernovae form Stage 2 and Stage 4 might be combined, but only after what are essentially the same evolution nuisance parameters are separately marginalized out.  We've investigated this particular effect and found that it typically leads to errors of no worse than 10% in the figure of merit, although in one case we found a 25% error.   This effect could also be important for other combinations and further investigations may even lead to significant new strategies for controlling systematic errors (see for example Zhan, 2006).

### 9.1.2  Duplication of the supernova near sample

Each of our supernova data models includes a "near sample."  When two SN data models are added using our methodology the near sample is included twice.  We've checked and found that a more careful calculation that makes sure the near sample only appears once do not lead to significant changes to the figure of merit.  The main reason for this is that in our data models the uncertainties on the far sample are inherited by the near sample.  (Whether this is an accurate model of a realistic program of SN studies deserves further scrutiny.)  Thus when combing two or more SN data models there is one near sample with smaller errors which dominates over the contributions from any other near sample.

## 9.2  Growth and transfer functions

Part of the construction of the weak lensing and cluster data models includes modeling the perturbation spectrum growth in the linear regime using standard transfer function techniques.  These techniques involve approximating the evolution of perturbations after some redshift ($z \approx 10$) as being scale independent, and also assumes that $\delta \propto 1/a$ around $z \approx 10$  to allow for the transition between two domains of approximation (see for example Dodelson 2004).  For clusters we additionally assume that the cluster abundance depends on the linear power spectrum in a known and universal manner

Our investigations show that in some cases we model such high quality data that the above approximations are not sufficiently valid to produce precise determinations of our figure of merit.  While we estimate that most single weak lensing data models to have 10-20% uncertainties in the figure of merit due to this issue, in some cases the error could approach 50%. We conclude that a more sophisticated treatment that does not depend on the above approximations would be required to achieve a full assessment of these corrections to the DETF calculations. But we note that for the combined data the overall impact of these approximations is reduced and we do not anticipate that these corrections could change our main findings.

## 10 Bibliography for the Appendix

# XV.    Logistical Appendix

1.  Membership

2.  Charge Letters

3.  Meeting dates, locations, and agendas

4.  White Papers received by the DETF

5.  Transmittal Letter

## Membership

Dr. Rocky Kolb (Chair)
Fermi National Accelerator Laboratory

Dr. Andreas Albrecht
University of California, Davis

Dr. John Huth
Harvard University

Dr. Gary Bernstein
University of Pennsylvania

Dr. Marc Kamionkowski
California Institute of Technology

Dr. Robert Cahn
Lawrence Berkeley National Laboratory

Dr. Lloyd Knox
University of California, Davis

Dr. Wendy Freedman
Carnegie Observatories

Dr. John Mather
NASA-GSFC

Dr. Jacqueline Hewitt
Massachusetts Institute of Technology

Dr. Suzanne Staggs
Princeton University

Dr. Wayne Hu
University of Chicago

Dr. Nicholas Suntzeff
Texas A&M University

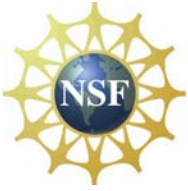

National Science Foundation
and the
National Aeronautics and Space Administration

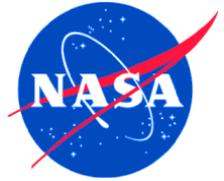

Professor Garth Illingworth
University of California Santa Cruz
Lick Observatory
Santa Cruz, California 95064

Dear Dr. Illingworth:

This letter is to request that the Astronomy and Astrophysics Advisory Committee (AAAC), in cooperation with the High Energy Physics Advisory Panel (HEPAP) of NSF and DOE, establish a Dark Energy Task Force as a joint sub-committee to advise NSF, NASA and DOE on the future of dark energy research.

Background and Purpose

The NRC Report *Connecting Quarks with the Cosmos*[2] poses eleven science questions for the new century, among which the nature of dark energy is identified as "probably the most vexing." The report outlines a near-term program to constrain the properties of dark energy, which includes the measurement of the apparent brightness of Type Ia supernovae as a function of redshift, the study of the number density of galaxies and clusters of galaxies as a function of redshift, and the use of weak gravitational lensing to study the growth of structure in the universe. The report also recommends the construction of two wide-field telescopes, one in space and one on the ground, to measure much larger numbers of supernovae with control of systematics and to map gravitational lensing over large scales.

In response, an NSTC interagency working group has established a federal strategy for approaching the dark energy question (see *The Physics of the Universe*[3]). The recommended triple-pronged strategy covers measurements of weak lensing, Type Ia supernovae and studies of the Sunyaev-Zel'dovich effect, primarily through a ground-based large survey telescope (LST), a space-based Joint Dark Energy Mission (JDEM) and coordinated ground-based CMB and space-based x-ray observations of galaxy clusters.

The joint Dark Energy Task Force (DETF) will help the agencies to identify actions that will optimize a near- and intermediate-term dark energy program and ensure rapid progress in the development and implementation of a concerted effort towards understanding the nature of dark energy.

The DETF is asked to advise the agencies on the optimum near- and intermediate-term programs to investigate dark energy and, in cooperation with agency efforts, to advance the justification, specification and optimization of LST and JDEM.

The DETF is asked to address the following areas:

1. Summarize the existing program of funded projects by projected capabilities, systematics, risks, required developments and progress-to-date.

2. Where possible, similarly summarize proposed and emergent approaches and techniques for dark energy studies; that is, characterize these approaches and techniques by the added value the projected capabilities would provide to the investigation of dark energy.

3. Identify important steps, precursors, R&D and other projects that are required in preparation for JDEM, LST and other existing or planned experiments.

4. If possible, identify any areas of dark energy parameter space that the existing or proposed projects fail to address.

The DETF is not constituted, nor has available time, to review individual proposals to determine their technical feasibility or likelihood of meeting performance goals. Rather, in addressing the areas above the DETF is asked to advise on the coverage of parameter space, to identify potential knowledge gaps that would preclude informed decisions about projects, to identify unnecessary or duplicated efforts, and to quantify the sensitivity of the determination of dark energy parameters to experimental performance goals such as sky coverage, number of objects, image quality or other requirements. The DETF should also comment on areas where expanded theoretical or modeling activity would be of significant benefit.

**Reporting**

The DETF Chair is responsible for preparing the final report, in consultation with all DETF members. In accordance with FACA rules, this report will be discussed independently at the first meetings of the AAAC and the HEPAP following completion of the report, before formal presentation to the agencies. We request that the DETF prepare their report for submission to the committees with a target date of December 2005.

We thank you for your efforts and wish you success in this important endeavor.

Sincerely,

Michael S. Turner                          Anne Kinney
Assistant Director, Directorate for        Director, Universe Division

Mathematical and Physical Sciences
National Science Foundation

Science Mission Directorate
National Aeronautics and Space
Administration

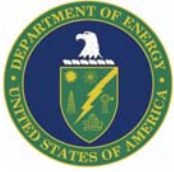

U.S. Department of Energy
and the
National Science Foundation

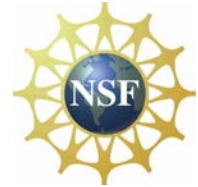

Professor Frederick Gilman
Carnegie Mellon University
5000 Forbes Avenue
Pittsburgh, Pennsylvania 15213

Dear Dr. Gilman:

This letter is to request that the High Energy Physics Advisory Panel (HEPAP), in
cooperation with the Astronomy and Astrophysics Advisory Committee (AAAC) of NSF
and NASA, establish a Dark Energy Task Force as a joint sub-committee to advise NSF,
NASA and DOE on the future of dark energy research.

Background and Purpose

The NRC Report *Connecting Quarks with the Cosmos*[4] poses eleven science questions
for the new century, among which the nature of dark energy is identified as "probably the
most vexing." The report outlines a near-term program to constrain the properties of dark
energy, which includes the measurement of the apparent brightness of Type Ia
supernovae as a function of redshift, the study of the number density of galaxies and
clusters of galaxies as a function of redshift, and the use of weak gravitational lensing to
study the growth of structure in the universe. The report also recommends the
construction of two wide-field telescopes, one in space and one on the ground, to measure
much larger numbers of supernovae with control of systematics and to map gravitational
lensing over large scales.

In response, an NSTC interagency working group has established a federal strategy for
approaching the dark energy question (see *The Physics of the Universe*[5]). The
recommended triple-pronged strategy covers measurements of weak lensing, Type Ia
supernovae and studies of the Sunyaev-Zel'dovich effect, primarily through a ground-
based large survey telescope (LST), a space-based Joint Dark Energy Mission (JDEM)
and coordinated ground-based CMB and space-based x-ray observations of galaxy
clusters.

The joint Dark Energy Task Force (DETF) will help the agencies to identify actions that
will optimize a near- and intermediate-term dark energy program and ensure rapid progress in the development and implementation of a concerted effort towards understanding the nature of dark energy.

---

[4] http://www.nap.edu/books/0309074061/html/
[5] www.ostp.gov/html/physicsoftheuniverse2.pdf

Charge to the Task Force

The DETF is asked to advise the agencies on the optimum near- and intermediate-term programs to investigate dark energy and, in cooperation with agency efforts, to advance the justification, specification and optimization of LST and JDEM.
The DETF is asked to address the following areas:

5. Summarize the existing program of funded projects by projected capabilities, systematics, risks, required developments and progress-to-date.

6. Where possible, similarly summarize proposed and emergent approaches and techniques for dark energy studies; that is, characterize these approaches and techniques by the added value the projected capabilities would provide to the investigation of dark energy.

7. Identify important steps, precursors, R&D and other projects that are required in preparation for JDEM, LST and other existing or planned experiments.

8. If possible, identify any areas of dark energy parameter space that the existing or proposed projects fail to address.

The DETF is not constituted, nor has available time, to review individual proposals to determine their technical feasibility or likelihood of meeting performance goals. Rather, in addressing the areas above the DETF is asked to advise on the coverage of parameter space, to identify potential knowledge gaps that would preclude informed decisions about projects, to identify unnecessary or duplicated efforts, and to quantify the sensitivity of the determination of dark energy parameters to experimental performance goals such as sky coverage, number of objects, image quality or other requirements. The DETF should also comment on areas where expanded theoretical or modeling activity would be of significant benefit.

**Reporting**

The DETF Chair is responsible for preparing the final report, in consultation with all DETF members. In accordance with FACA rules, this report will be discussed independently at the first meetings of the AAAC and the HEPAP following completion of the report, before formal presentation to the agencies. We request that the DETF prepare their report for submission to the committees with a target date of December 2005.

We thank you for your efforts and wish you success in this important endeavor.

Sincerely,

Robin Staffin
Associate Director, Office of High Energy
Physics
Office of Science
U.S. Department of Energy

Michael S. Turner
Assistant Director, Directorate for
    Mathematical and Physical Sciences

National Science Foundation

cc: K. Turner, SC-20
    P. K. Williams, SC-20

G. W. Van Citters, NSF–AST
J. Dehmer, NSF–PHY
K. Erb, NSF–OPP

**DETF face-to-face meetings:**

March 22–23, 2005 at the National Science Foundation, Arlington, VA

June 30–July1, 2005 at the Fermi National Accelerator Laboratory, Batavia, IL

September 29-30, 2005 at the National Science Foundation, Arlington, VA

October 19–21, 2005 at the University of California, Davis, CA

December 7–8, 2005 at the Massachusetts Institute of Technology, Cambridge, MA

**AGENDA**

**Dark Energy Task Force**
**22–23 March 2005**

*National Science Foundation*

4201 Wilson Blvd., Arlington, VA
Stafford I Bldg., Room 1235

*Tuesday, 22 March 2005*

| | | |
|---|---|---|
| 8:30 – 9:00 | *Coffee and Conversation* | |
| 9:00 – 9:15 | Welcome and Introductions | R. Kolb |
| | <u>Charge from the Agencies:</u> | |
| 9:15 – 9:30 | National Science Foundation | W. Van Citters |
| 9:30 – 9:45 | National Aeronautics and Space Administration | M. Salamon |
| 9:45 – 10:00 | U.S. Department of Energy | K. Turner |
| 10:00 – 10:30 | Discussion with the Agencies | |
| 10:30 – 11:00 | | |
| | Break | |
| 11:00 – 12:00 | Committee Discussion—Scope and Procedure | R. Kolb |
| 12:00 – 1:00 | | M. Turner |
| | Box Lunch and Discussion | |
| 1:00 – 1:30 | Discussions with AAAC Chair and HEPAP Chair | G. Illingworth; F. Gilman |
| 1:30 – 2:15 | Review of Dark Energy Theory | A. Albrecht |
| | <u>Review of Current and Emerging Approaches:</u> | |
| 2:15 – 2:45 | | R. Kolb |
| | *Identify 1ˢᵗ-Order Matrix of Approaches vs. Projects* | |
| 2:45 – 3:15 | | |
| | Break | |
| 3:15 – 4:00 | | W. Freedman |
| | *Type Ia Supernovae* | |
| 4:00 – 5:00 | Committee Discussion | |
| 5:00 | | |
| | Adjourn for the Day | |
| TBD | | |
| | *Committee Dinner* | |

Wednesday, 23 March 2005

8:30 – 9:00
                Coffee and Conversation

9:00 – 10:00      JDEM SDT Activities           C. Bennett
10:00 – 10:40   Committee Discussion

10:40 – 11:00
                Break

11:00 – 11:30   Identify Mechanisms for Community Input   R. Kolb
11:30 – 12:00   Outline Report and Select Writing Groups   R. Kolb

*12:00 – 1:00*     Lunch

                                                                  R. Kolb

*1:00 – 2:00*     *Plan Future Meetings*
2:00
                Adjourn

Dark Energy Task Force
29–30 September 2005

***National Science Foundation***
Room 595, Stafford II Building

*Thursday, 29 September 2005*

| | |
|---|---|
| 9:00 – 9:15 | Report from the Chair |
| 9:15 – 10:15 | Discussion of level 0 findings |
| 10:15 – 10:30 | Discussion of parameters to be used by technique working groups |
| 10:30 – 10:45 | Break |
| 10:45 – 12:15 | Working groups meet to complete reports on current status of dark-energy techniques |
| 12:15 – 1:15 | Lunch |
| 1:15 – 2:15 | Reports on current status from technique working groups |
| 2:15 – 3:15 | *Revision of current status reports or move on to "next step" goals* |
| 3:15 – 3:30 | Break |
| 3:30 – 5:30 | Working groups continue to discuss "next step" goals |
| 5:30 | Adjourn for the Day |
| 6:00 | *Optional committee dinner, TBD* |

Friday, 30 September 2005

| | |
|---|---|
| 9:00 – 9:30 | Planning for October meeting at UC Davis |
| 9:30 – 10:30 | Discussion of DETF report outline |
| 10:30 – 10:45 | Break |
| 10:45 – 12:30 | Continued work of technique subgroups |
| *12:30 – 1:30* | Lunch |
| *1:30 – 3:00* | *Wrap-up discussion* |
| 3:00 | Adjourn |

Dark Energy Task Force
30 June –1 July 2005
Fermi National Accelerator Laboratory
Room 1 East, Wilson Hall

*Thursday, 30 June 2005*

| | | |
|---|---|---|
| 9:00 – 9:10 | Welcome and Announcements | R. Kolb |
| 9:10 – 9:40 | Committee Discussion | |
| 9:40 – 10:20 | JDEM concept: DESTINY | J. Morse |
| 10:20 – 10:50 | Break | |
| 10:50 – 11:30 | JDEM concept: JEDI | A. Crotts |
| 11:30 – 12:00 | *Update on JDEM SDT Activities* | A. Albrecht |
| 12:00 – 1:00 | Lunch | |

Reconvene in Room 9 South East

| | | |
|---|---|---|
| 1:00 – 1:40 | JDEM concept: SNAP (videoconference) | S. Perlmutter/M. Levi/E. Linder |
| 1:40 – 1:50 | *Return to Room 1 East* | |
| 1:50 – 2:30 | LST concept: LSST | T. Tyson |
| 2:30 – 3:10 | LST concept: Pan-STARRS | G. Magnier |
| 3:10 – 3:40 | Dark Energy Survey (DES) | J. Frieman |
| 3:40 – 4:10 | Break | |

*Techniques and Calibrations:*

| | | |
|---|---|---|
| 4:10 – 4:30 | *Clusters* | J. Mohr |
| 4:30 – 4:50 | Weak Lensing | G. Bernstein |
| 4:50 – 5:10 | *CMB Studies* | W. Hu |
| 5:10 – 5:20 | *Baryon Oscillations* | L. Knox |
| 5:20 – 5:30 | *Redshifted 21-cm Measurements* | J. Hewitt |
| 5:30 – 6:00 | Committee Discussion | |
| 6:00 | Adjourn for the Day | |
| 7:00 | *Committee Dinner, Chez Leon* | |

Friday, 1 July 2005

| | |
|---|---|
| 9:00 – 9:30 | Planning for October Meeting |
| 9:30 – 10:30 | Discussion of White Papers |
| 10:30 – 11:00 | |
| | Break |
| 11:00 – 11:30 | Input of International Projects |
| 11:30 – 12:00 | Report Structure |
| *12:00 – 1:00* | Lunch |
| *1:00 – 3:00* | *Writing Assignments* |
| 3:00 | |
| | Adjourn |

Dark Energy Task Force Whitepapers

Beckwith, Andrew
projectbeckwith2@yahoo.com
Proposal for using mix of analytical work with data analysis of early CMB data obtained
from the JDEM NASA - DOE Investigation

Riess, Adam
ariess@stsci.edu
Dark Energy Evolution from HST and SNe Ia at z > 1

Allen, Steve
swa@stanford.edu
Probing Dark Energy with Constellation X

Clarke, Tracy
tclarke@ccs.nrl.navy.mil
Laying the Groundwork for Cluster Studies: The Long Wavelength Array

Kaiser, Nick
kaiser@ifa.hawaii.edu
The Pan-STARRS Project

Baltay, Charles
cahrles.baltay@yale.edu
The Palomar QUEST Variability Survey

Vikhlinin, Alexey
avikhlinin@cfa.harvard.edu
Probing Dark Energy with Cluster Evolution in a 10,000 Square Degree ("10K")

Thompson, Rodger
thompson@as.arizona.edu
A Molecular Probe of Dark Energy

Miller, Chris
cmiller@ctio.noao.edu
The XMM Cluster Survey (XCS)

Aldering, Greg
aldering@panisse.lbl.gov
The Nearby Supernova Factory

Jahoda, Keith
keith@milkyway.gsfc.nasa.gov
An X-ray Galaxy Cluster Survey for Investigations of Dark Energy

Pritchet, Chris
pritchet@uvic.ca
The Supernova Legacy Survey -- SNLS

Peoples, John
peop@fnal.gov
Dark Energy Survey

Schlegel, David
djschlegel@lbl.gov
Baryon Oscillations with the "One Thousand Points of Light Spectrograph"

Golwala, Sunil
golwala@caltech.edu
Supplementing Thermal Sunyaev-Zeldovich Effect Surveys with CCAT High-Angular-Resolution Follow-Up

Wang, Yun
wang@nhn.ou.edu
Joint Efficient Dark-energy Investigation (JEDI): a Candidate Implementation of the NASA-DOE Joint Dark Energy Mission (JDEM)

Crotts, Arlin
arlin@astro.columbia.edu
ALPACA: Advanced Liquid-mirror Probe of Asteroids, Cosmology and Astrophysics

Gerke, Brian
bgerke@berkeley.edu
Measuring the Growth of Structure with Spectroscopically Identified Galaxy Groups and Clusters

Dey, Arjun
dey@noao.edu
Dark Energy and Cosmic Sound: w(z) Surveys with the Gemini/Subaru Wide-Field Multi-Object Spectrograph

Kirshner, Robert
rkirshner@cfa.harvard.edu
Supernova Observations: Foundation of the Accelerating Universe

Lamb, Don
lamb@oddjob.uchicago.edu
A Gamma-Ray Burst Mission to Investigate the Properties of Dark Energy

Kaiser, Mary Beth
kaiser@pha.jhu.edu
DETF CalibrationWhite Paper: ACCESS - Absolute Color Calibration Experiment for Standard Stars

Bailyn, Charles
bailyn@astro.yale.edu
The WIYN One-Degree Imager and Associated Yale Survey

Meyer, Stephan
meyer@uchicago.edu
The Cluster Imaging eXperiment (CIX) and the Importance of Large Single Dish Sub-MM Measurements

Adelberger, Eric
eric@npl.washington.edu
Laboratory Tests of Gravity at the Dark Energy Length Scale

Gebhardt, Karl
gebhardt@astro.as.utexas.edu
HETDEX: Hobby-Eberly Telescope Dark Energy Experiment

Perlmutter, Saul
saul@lbl.gov
Pre-JDEM Dark Energy Studies with SNe Ia

Perlmutter, Saul
saul@lbl.gov
Probing Dark Energy via Weak Gravitational Lensing with the SuperNovaAcceleration Probe (SNAP)

Perlmutter, Saul
saul@lbl.gov
Supernova Acceleration Probe: Studying Dark Energy with Type Ia Supernovae

Perlmutter, Saul
saul@lbl.gov
SNAP/JDEM Overview

Cordes, Jim
cordes@astro.cornell.edu
Dark Energy Experiments with the Square Kilometer Array

Carlstrom, John
jc@hyde.uchicago.edu
The South Pole Telescope

Stubbs, Chris
cstubbs@fas.harvard.edu
The ESSENCE Project: A Supernova Survey Optimized to Constrain the Equation of State of the Cosmic Dark Energy

Tyson, Tony
tyson@physics.ucdavis.edu
The Large Synoptic Survey Telescope (LSST)

Stubbs, Chris
cstubbs@fas.harvard.edu
Photometric Redshifts of SZ Clusters with Targeted Multiband Imaging

Morse, Jon
jon.morse@asu.edu
DESTINY: Dark Energy Space Telescope

Dark Energy Task Force Whitepapers (Theory)

Bean, Rachel
rbean@astro.cornell.edu
Insights into Dark Energy: Interplay Between Theory and Observation

Daniel Eisenstein
eisenste@cmb.as.arizona.edu
Theoretical Investigations To Support the Acoustic Oscillation Method

Gus Evrard
evrard@umich.edu
Dark Energy Studies: Challenges To Computational Cosmology

David Weinberg
dhw@astronomy.ohio-state.edu
Cosntraining dark Energy with the Dark Energy Survey: Theoretical Challenges

Eddie Baron
baron@nhn.ou.edu
Quantitative Spectroscopy of Supernovae for Dark Energy Studies

Eric Linder
evlinder@lbl.gov
Light & Dark: the Clear and Present Role of the CMB for Dark Energy

Eric Linder
evlinder@lbl.gov
Baryon Acoustic Oscillations for Dark Energy

Bill Baker
bill.baker@furham.edu
Dark Energy as a Geometric Effect

Karl Gebhardt
gebhardt@astro.as.utexas.edu
HETDEX Hobby-Eberly Telescope Dark Energy Experiment

Michael Pierce
mpierce@uwyo.edu
Characterizing Dark Energy Through Nano-Arcsecond Astrometry of Quasars

Henk Hoekstra
hoekstra@uvic.ca

Theory Requirements for Cosmic Shear Studies

Nick Gorkavyi
gorkavyi@gist.us

Dark Energy and Acceleration of the Universe with Variable Gravitational Mass

Rudy Schild
rschild@cfa.harvard.edu

White Paper on the Dark Energy Problem

Daniel Holtz
deholz@cfcp.uchicago.edu

Using Gravitational Wave Standard Sirens

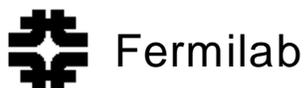 **Fermilab**

Fermi National Accelerator Laboratory
Particle Astrophysics Center
P.O.Box 500 - MS209
Batavia, Illinois • 60510

June 6, 2006

Dr. Garth Illingworth
Chair, Astronomy and Astrophysics Advisory Committee
Dr. Mel Shochet
Chair, High Energy Physics Advisory Panel

Dear Garth, Dear Mel,

I am pleased to transmit to you the report of the Dark Energy Task Force.

The report is a comprehensive study of the dark energy issue, perhaps the most compelling of all outstanding problems in physical science. In the Report, we outline the crucial need for a vigorous program to explore dark energy as fully as possible since it challenges our understanding of fundamental physical laws and the nature of the cosmos.

We recommend that program elements include
1. Prompt critical evaluation of the benefits, costs, and risks of proposed long-term projects.
2. Commitment to a program combining observational techniques from one or more of these projects that will lead to a dramatic improvement in our understanding of dark energy. (A quantitative measure for that improvement is presented in the report.)
3. Immediately expanded support for long-term projects judged to be the most promising components of the long-term program.
4. Expanded support for ancillary measurements required for the long-term program and for projects that will improve our understanding and reduction of the dominant systematic measurement errors.
5. An immediate start for nearer term projects designed to advance our knowledge of dark energy and to develop the observational and analytical techniques that will be needed for the long-term program.

Sincerely yours, on behalf of the Dark Energy Task Force,

*Rocky*

Edward Kolb
Director, Particle Astrophysics Center, Fermi National Accelerator Laboratory
Professor of Astronomy and Astrophysics, The University of Chicago